\documentclass[final,5p,times,twocolumn,authoryear]{elsarticle}
\usepackage{graphicx}
\usepackage{subcaption}
\usepackage{amssymb,amsmath}
\usepackage{pdflscape}
\usepackage{adjustbox}
\usepackage[T1]{fontenc}
\usepackage{url}
\usepackage{hyperref}
\usepackage{etoolbox}
\usepackage{booktabs}
\usepackage{lipsum}
\usepackage{epsfig}
\usepackage{txfonts}
\usepackage{natbib}
\usepackage{xcolor}
\usepackage{tabularx}
\usepackage{footnote}
\usepackage{soul}
\usepackage{multirow}
\usepackage{multicol}
\usepackage{mathpazo}
\usepackage{lineno,hyperref}
\usepackage{textcomp}
\usepackage[utf8]{inputenc}
\usepackage{booktabs}

\usepackage{amsthm}

\journal{High Energy Astrophysics}
\UseRawInputEncoding

\begin{document}

\begin{frontmatter}
\title{Spectral-Regime Overlap and Transition-like Behavior in the Blazar Population from Multi-Instrument X-ray and TeV Observations}

\author[label1]{Javaid Tantry\corref{cor1}}
\ead{javaidtantray9@gmail.com}

\address[label1]{Department of Physics, University of Kashmir, Srinagar 190006, India}

\cortext[cor1]{Corresponding author: Javaid Tantry}

\author[label1]{Naseer Iqbal}

\begin{abstract}

Blazars are traditionally divided into two major subclasses--BL Lac objects and flat-spectrum radio quasars (FSRQs)--commonly associated with radiatively inefficient and efficient accretion onto supermassive black holes, respectively. Increasing observational evidence, however, suggests that several sources exhibit intermediate or transition-like spectral behavior that challenges strictly discrete classification schemes. Since the photon index directly traces the shape and evolution of the non-thermal emission spectrum, it provides an effective diagnostic of changes in jet energetics, particle acceleration, and radiative processes across different blazar states. By combining data from \textit{Swift-XRT}, \textit{Swift-BAT}, \textit{NuSTAR}, \textit{ROSAT}, \textit{Chandra}, \textit{XMM-Newton}, NICER, \textit{AstroSat}, \textit{TeVcat}, and VizieR compilation (ROSAT/ASCA/BeppoSAX/EXOSAT/Einstein), we examine the distribution and evolution of photon indices across multiple blazar subclasses.

The analysis reveals broad overlap regions in photon-index space linking EHBL, HBL, IBL, LBL, and FSRQ-like populations. Across  X-ray instruments, the intermediate spectral regime is predominantly concentrated around $\Gamma_{\mathrm{X}} \approx 2$, typically spanning $\Gamma_{\mathrm{X}} \sim 1.5$--$2.2$. Multi-epoch observations further demonstrate substantial intra-source spectral variability, including continuous stochastic spectral evolution in Mrk~421 and pronounced state-dependent variability in OJ~287. Analysis revealed sources exhibit strong spectral-state changes with $\Delta \Gamma_{\mathrm{X}} > 0.5$, while intermediate spectral-regime occupancy reaches $\sim 22$--$43\%$ depending on the instrument. Several sources repeatedly occupy spectral regions connecting traditionally separated subclasses, supporting their interpretation as candidate transition-like blazars. The observed overlap in $\Gamma_{\mathrm{X}}$ distributions broadly supports the conventional blazar classification framework while indicating that the subclass boundaries are not strictly discrete.

Overall, the results favor a framework in which blazar subclasses represent overlapping and dynamically evolving spectral-regime populations rather than strictly fixed categories. The X-ray and TeV photon index appears to provide a useful empirical tracer of long-term spectral evolution  across the BL Lac--FSRQ continuum, supporting a phenomenological picture in which long-term spectral variability reflects changes in jet emission and radiative processes across the blazar population.

\end{abstract}


\end{frontmatter}

\section{Introduction}
\label{introduction}
Blazars are the most luminous and variable class of radio-loud AGN, with relativistic jets oriented within a few degrees of the line of sight\cite{urry1995unified}.
The defining features of blazars include strong $\gamma$-ray emission, a high degree of polarization, and frequent variability in both flux and spectra \cite{ 2020ApJ...892..105A, Zhang_2021}. These properties arise from relativistic jets that efficiently accelerate particles, producing non-thermal radiation that dominates the objects' broadband emission \cite{1995MNRAS.273..583D, urry1995unified}.
Their broadband emission is characterised by a double-humped spectral energy distribution (SED): a lower hump produced by synchrotron radiation from relativistic jet electrons, and a higher hump from inverse Compton (IC) scattering — either of internal synchrotron photons (synchrotron self-Compton; SSC) or external photon fields (external Compton; EC) from the accretion disc, broad-line region (BLR), or dusty torus \cite{Dermer_2009, 2013ApJ...763..134F}.
A further sub-classification by synchrotron peak frequency $\nu_s$ separates sources with low-synchrotron peak (LSP; $\nu_s < 10^{14}$~Hz), intermediate-synchrotron peak (ISP; $10^{14} < \nu_s < 10^{15}$~Hz), and high-synchrotron peak (HSP; $\nu_s > 10^{15}$~Hz) sources~\cite{urry1995unified}. However, this classification scheme is subject to two key limitations. First, it relies on single-epoch or time-averaged observations of intrinsically variable sources, such that objects may exhibit observational
properties associated with both FSRQ-like and BL
Lac-like states over time. \citep{10.1111/j.1365-2966.2011.20044.x, 2019MNRAS.484L.104P}. Second, the adoption of a fixed boundary at EW $= 5$ Å is increasingly inconsistent with evidence for a continuous blazar sequence rather than discrete subclasses \citep{10.1046/j.1365-8711.1998.01828.x, 10.1093/mnras/stx806}. These limitations are underscored by long-term multiwavelength observations, which reveal sources that defy static classification, producing the so-called  \textit{transition blazars} or Changing look blazars (Clbs).
\section{Transition blazars or Changing look blazars (clbs)}
A growing population of AGN exhibits properties intermediate between or alternating between BL Lacs and FSRQs \citep{2013MNRAS.432L..66G,Xiao_2022,2024A&A...685A.140R}, which pose a significant challenge to the standard AGN unification model \citet{Peña-Herazo_2021}. The transition blazars may undergoing accretion mode transitions between a radiatively efficient Shakura-Sunyaev disc (FSRQ phase) and an advection-dominated accretion flow (ADAF) \citep{Cavaliere_2002, 2014ARA&A..52..529Y}. When the accretion rate drops below the critical Eddington threshold, EC is suppressed, SSC begins to dominate, and the SED may evolve toward BL
Lac-like spectral characteristics.
Alternatively, Changing look blazars (CLBs) are exceptionally rare objects whose transition between FSRQs and BL Lacs — marked by the appearance or disappearance of broad emission lines — has been explained by several scenarios. The broad-line equivalent width (EW) may be diluted by strong, beamed jet continuum variability driven by changes in jet power or bulk Lorentz factor \citep{Vermeulen_1995,2009A&A...496..423B,Giommi:2011sn, Pasham_2019}. Alternatively, weak radiative cooling can allow the non-thermal continuum to overwhelm the broad lines \citet{10.1111/j.1365-2966.2012.21554.x}. At high redshifts ($z > 0.7$), the strong H$\alpha$ line shifts out of the optical window, potentially misclassifying FSRQs as BL Lacs \citet{10.1093/mnras/stv573}. Additionally, observational factors such as signal-to-noise ratio and spectral resolution can affect line detection and thus classification \citep{Peña-Herazo_2021}. Their defining observational signatures
include — the optical emission-line equivalent width, the $\gamma-ray$ photon index, the Compton dominance parameter, and the accretion disk luminosity.

\subsection{The Dominance of $\gamma$-Ray Studies and the X-ray Gap}
The $\gamma$-ray band has been the primary hunting ground for transition blazars over the past decade, principally because Fermi-LAT has provided continuous, unbiased, all-sky monitoring of AGN in the 0.1--300\,GeV energy range since 2008. The $\gamma$-ray photon index $\Gamma_\gamma$ separates BL\,Lacs (mean $\Gamma_\gamma \approx 1.9$) from FSRQs (mean $\Gamma_\gamma \approx 2.4$) in the 4LAC Catalog, with a clear spectral gap between the two populations. Transition blazars systematically occupy the intermediate range $\Gamma_\gamma \approx 2.0$--$2.2$, the $\gamma$-ray transition zone identified in the  study \citet{10.1093/mnras/stag542}. Using a Gaussian Mixture Model applied to 3,934 Fermi-LAT AGN, transition sources are shown to cluster at the boundary between the two populations in both spectral index and Compton dominance, supporting the use of the photon index as a potential transition diagnostic. Changing look blazars have been documented through multi-year $\gamma$-ray light curve analysis combined with optical spectroscopy. The blazar OQ\,334 underwent a documented oscillation between FSRQ-to-BL \, Lac-to-FSRQ between 2018 and 2023, with the $\gamma$-ray photon index satisfying $\Gamma_\gamma \gtrsim 2.2$ in FSRQ phases, $2.0 < \Gamma_\gamma < 2.2$ in transition phases, and $\Gamma_\gamma \lesssim 2.0$ in BL \, Lac phases \citet{2024A&A...685A.140R}. The confirmed transition blazar S5 \, 1803+784 showed $\Gamma_\gamma$ evolving from $1.65 \pm 0.41$ to $2.48 \pm 0.09$ in four distinct Fermi epochs \citep{TANTRY2025100372}, with spectral evolution spanning the BLLac-like to FSRQ-like regimes.
In 2024, the FSRQ OP313 underwent a remarkable flare with gamma-ray flux increasing 60-fold, peaking in less than two days. Multiwavelength observations revealed prolonged activity spanning 100 days, with the source transitioning from an FSRQ-like to BL Lac-like state. This transition featured an increased synchrotron peak frequency and the disappearance of broad-line emission. Post-flare analysis showed synchrotron-inverse Compton decoupling and possible magnetic field decrease, suggesting variations in electron density and accretion rate within the shock front \citep{Zhang_2025}. OJ\,287 has recently been shown to exhibit significant flux variability across optical, UV, X-ray, and $\gamma$-ray bands, with strong correlations among optical, UV, and X-ray but no significant correlation with $\gamma$-rays. The source displays a flux-dependent transitional behaviour: quiescent states show hard X-ray spectra dominated by inverse Compton emission, intermediate states show mixed contributions, and flaring states show soft X-ray spectra dominated by synchrotron emission \citep{10.1093/mnras/staf1781}.\\
In contrast to the extensive $\gamma$-ray literature, the X-ray domain has received relatively little attention as a transition diagnostic. The X-ray domain therefore represents
a significant gap in transition-like blazar studies. The 0.3--80\,keV band is the only energy range that directly straddles the IC/synchrotron crossover for ISP/IBL sources, encoding the relative dominance of both emission mechanisms simultaneously in a single spectral index \citep{giommi2019open,2022MNRAS.514.3179M}. While $\gamma$-ray studies capture the high-energy IC hump, they cannot resolve the precise point at which IC and synchrotron contributions become equal — a boundary naturally encoded in the X-ray photon index $\Gamma$.
Prior X-ray studies have highlighted the diagnostic importance of X-ray spectral properties in blazars. \citet{2014Ap&SS.352..207Y} reported strong anti-correlations between $\alpha_X$, the Doppler factor, and the $\gamma$-ray spectral index for 245 \textit{Fermi} blazars. In addition, long-term \textit{Swift}-XRT and \textit{NuSTAR} observations of 1ES\,0229+200 (2008--2024) revealed transitional X-ray spectral behavior, where a spectral upturn near $\sim$25 keV during low states suggests a transition between the synchrotron and inverse-Compton emission components. These results further indicate that the position of the X-ray spectrum within the broadband SED evolves with flux state \citep{2025A&A...703A.150W}. Nevertheless, a comprehensive multi-instrument  observational evidence of the X-ray transition regime in  blazars remains sparse. In particular, the existence of a common transition region near $\Gamma_{\mathrm{X}} \approx 2.0$ across observations from \textit{NuSTAR}, \textit{Swift}-XRT/BAT, \textit{ROSAT}, \textit{NICER}, and \textit{AstroSat} has not yet been systematically explored. The present work aims to address this observational gap.

\subsection{This Work}
We adopt instrument-dependent boundaries reflecting energy-band 
sensitivity: \textit{Swift}-XRT (0.3--10 keV): 
$\Gamma_{\rm intermediate} = 1.5$--$2.0$; 
\textit{NuSTAR} (3--79 keV): 
$\Gamma_{\rm intermediate } = 1.6$--$2.2$. 
These shifts reflect spectral curvature effects when sampling 
different portions of the synchrotron--IC crossover.
The specific goals are: (i) to identify the X-ray transition zone in each dataset; (ii) to test whether the X-ray photon index ($\Gamma_{\mathrm{X}}$) serves as a consistent, 
model-independent tracer of spectral-state variability across multiple 
instruments and blazar subclasses; (iii) to identify transition-like sources through cross-class spectral boundary violations; (iv) to quantify multi-epoch spectral variability; and (v) to validate findings with AstroSat/XMM-Newton/NICER observations.

\section{Observations and Data}
\subsection{NuBlazar: NuSTAR Hard X-ray Catalog (3--79 keV)}
The NuBlazar Catalog \citep{2022MNRAS.514.3179M} provides the first systematic hard X-ray spectral database of blazars observed by \textit{NuSTAR} \citep{2013ApJ...770..103H}. The \textit{NuSTAR} 3--79~keV bandpass is uniquely positioned to capture the high-energy tail of the synchrotron peak in HBL blazars, the rising synchrotron self-Compton (SSC) component in LBL objects, and the critical transition region between these two Spectral Energy Distribution (SED) humps. Compiled from the \textit{NuSTAR} public archive through September 2021, the Catalog includes 253 observations of 126 distinct blazars. 

\begin{figure*}[p] 
\centering
\includegraphics[width=\textwidth]{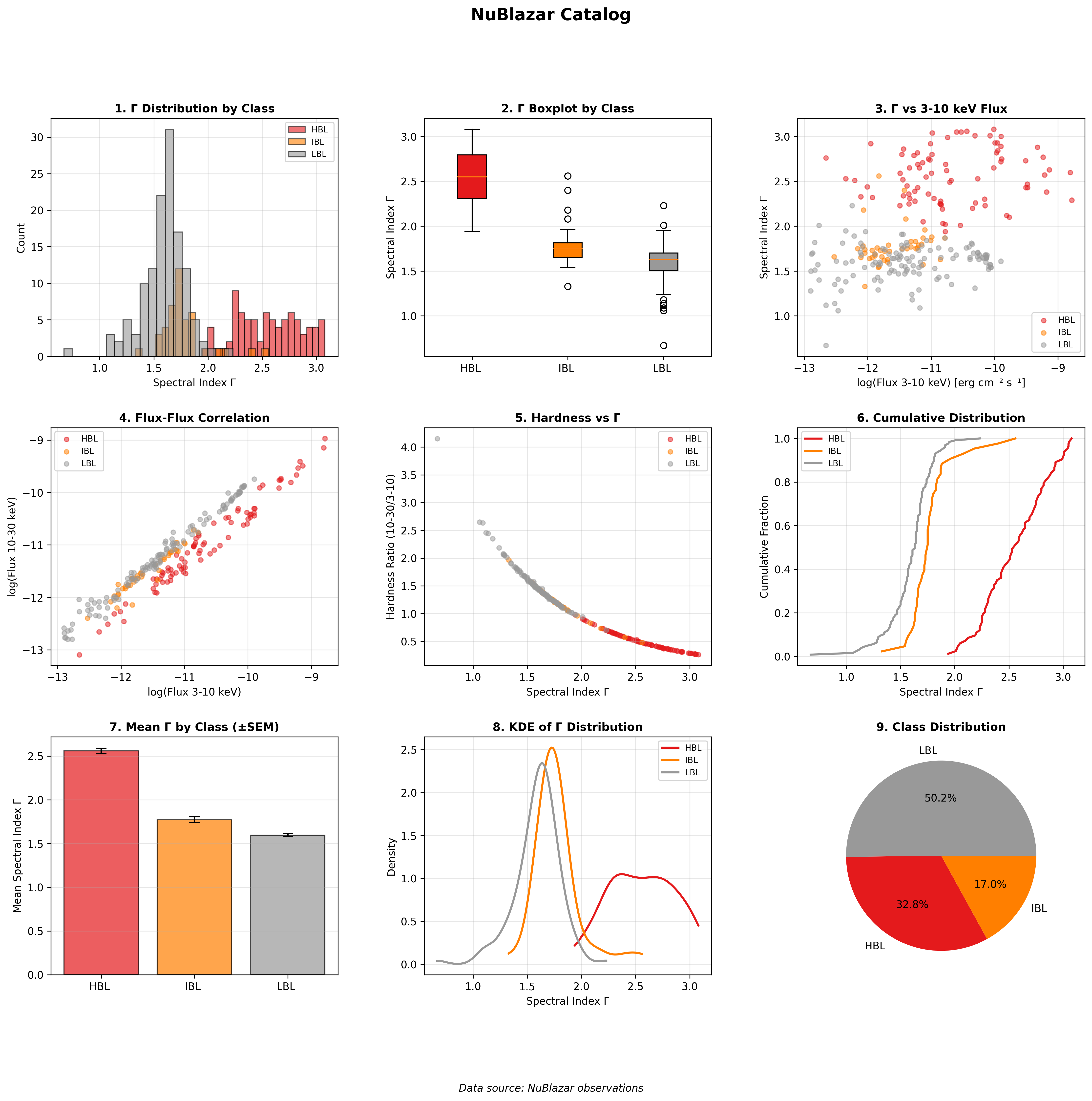}
\caption{Multi-panel overview of NuBlazar catalog properties for HBL, IBL, and LBL populations observed with NuSTAR. The figure summarizes spectral-index distributions, flux correlations, hardness-ratio behavior, cumulative and kernel-density distributions, and class fractions, highlighting clear spectral and population differences among the blazar subclasses.}
\label{fig:nublazar_dashboard}
\end{figure*}

Analysis identifies 57 observations (22.5\%) occupying photon-index  ranges ($\Gamma_{\mathrm{X}} = 1.6$--$2.2$) intermediate between the LBL-dominated ($\Gamma_{\mathrm{X}} < 1.6$) and HBL-dominated ($\Gamma_{\mathrm{X}} > 2.2$) distributions in the NuBlazar sample as shown in Figure~\ref{fig:nublazar_dashboard}
These observations are further grouped into approximate intermediate spectral subdivisions:
\begin{itemize}
    \item \textbf{LBL $\to$ IBL-like Transition:} 15 observations (e.g., BL Lac, TXS~0506+056) characterized by $\Gamma_{\mathrm{X}} = 1.70$--$1.90$ and Hardness Ratios (HR) of $1.00$--$1.30$.
    \item \textbf{ IBL-like intermediate  regime:}: 32 observations (e.g., TXS~0506+056, OJ~287) with $\Gamma_{\mathrm{X}} = 1.75$--$2.20$ and $\mathrm{HR} = 0.85$--$1.20$, centered within the  intermediate spectral range.
    \item \textbf{IBL $\to$ HBL-like Transition:} 10 observations (e.g., 1ES~0229+200, PKS~1441+25) showing steeper indices ($\Gamma_{\mathrm{X}} = 2.00$--$2.40$) and lower HR ($0.70$--$0.90$).
\end{itemize}
The remaining 196 observations reside within the characteristic LBL or HBL spectral domains, whereas sources occupying intermediate photon-index regimes
may represent blazars exhibiting enhanced spectral variability
or overlap between traditional spectral population are summarized in Table~\ref{tab:nustar_revised}.
Previous studies have shown that HBL blazars generally exhibit steeper X-ray spectra than LBL sources, with characteristic average photon indices of $\langle \Gamma \rangle = 2.56 \pm 0.30$ and $\langle \Gamma_{\mathrm{X}} \rangle = 1.58 \pm 0.22$, respectively \citep{2022MNRAS.514.3179M}.

\begin{table*}[htbp]
\centering
\caption{Unified NuSTAR Hard X-ray Blazar Classification with Intermediate Spectral-Regime Subdivision}
\label{tab:nustar_revised}
\renewcommand{\arraystretch}{1.2}

\begin{tabular}{p{3.8cm} c c c p{6.5cm}}
\toprule

\textbf{Class / Spectral Regime} &
\textbf{Observations} &
\textbf{Fraction (\%)} &
\textbf{$\Gamma$ Range} &
\textbf{Representative Sources / Notes} \\

\midrule

\multicolumn{5}{l}{\textbf{Primary Hard X-ray Spectral Classification}} \\

\midrule

LBL-like region &
83 &
32.8 &
$\Gamma < 1.6$ &
Hard-spectrum dominated population \\

Intermediate spectral region &
109 &
43.1 &
$1.6 \leq \Gamma \leq 2.2$ &
Intermediate-regime population \\

HBL-like region &
61 &
24.1 &
$\Gamma > 2.2$ &
Soft-spectrum dominated population \\

\midrule

\multicolumn{5}{l}{\textbf{Intermediate Spectral-Regime Subdivision}} \\

\midrule

LBL-like / Intermediate regime &
30 &
-- &
1.70--1.90 &
BL~Lac, TXS~0506+056, PKS~0521$-$36, PG~2209+184 \\

Intermediate-dominated regime &
28 &
-- &
1.80--2.10 &
TXS~0506+056, 1H~0323+342, OJ~287, S5~0716+714 \\

Intermediate / HBL-like regime &
11 &
-- &
2.00--2.40 &
1ES~0229+200, OJ~287, 3C~264, PKS~0625$-$354 \\

\midrule

\textbf{Total} &
\textbf{253} &
\textbf{100} &
-- &
-- \\

\bottomrule
\end{tabular}

\vspace{0.2cm}

\begin{minipage}{0.95\textwidth}
\footnotesize
\textbf{Note:}
The classifications represent empirical hard X-ray spectral regimes derived from NuSTAR photon-index distributions. The terms ``hard-spectrum'', ``intermediate-spectrum'', and ``soft-spectrum'' refer to phenomenological spectral behavior within the NuSTAR energy band and should not be interpreted as definitive physical subclass transformations. Due to long-term spectral variability, some sources may occupy multiple intermediate spectral-regime intervals.
\end{minipage}

\end{table*}

\subsection{Swift-XRT Blazar Catalog (0.3--10 keV)}
We analyzed 31,068 observations of 65 blazars from the \textit{Swift}-XRT archive, obtained through the Open Universe blazar monitoring programme \citep{2021MNRAS.507.5690G}. Following established conventions \citep{comastri1997soft,2001yCat..33750739D,Kadler2005,giommi2019open}, we adopt provisional classification boundaries at $\Gamma_{\mathrm{X}} = 1.5$ (LSP/ISP boundary) and $\Gamma_{\mathrm{X}} = 2.0$ (ISP/HSP boundary) as empirical reference points for our analysis.

It is important to emphasize that these boundaries are phenomenological thresholds derived from historical X-ray spectral studies rather than physically motivated dividing lines. Our analysis examines whether these conventional boundaries correspond to genuinely distinct source populations or instead reflect gradual transitions within a continuous spectral sequence.

\begin{figure*}[htbp]
\centering
\includegraphics[width=\textwidth]{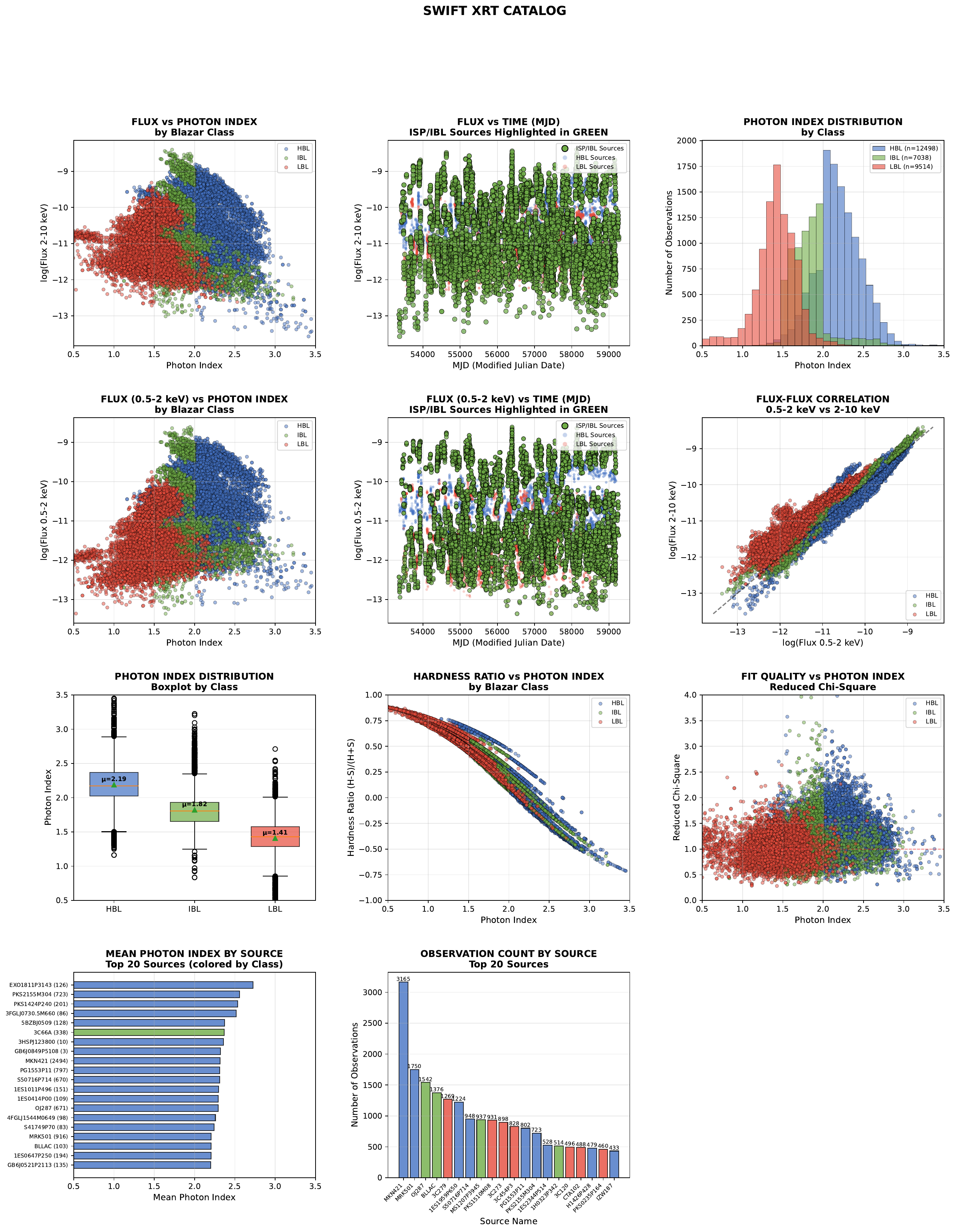}
\caption{Multi-panel overview of Swift-XRT blazar properties for HBL, IBL, and LBL populations. The figure summarizes spectral-index distributions, flux variability, flux--flux correlations, hardness-ratio behavior, and spectral-fit properties, highlighting clear class-dependent trends and long-term X-ray variability across the Swift-XRT sample.}
\label{fig:giomi}
\end{figure*}

As demonstrated in Figure~\ref{fig:giomi}, the observed distributions do not form sharply separated clusters but instead exhibit substantial overlap across these nominal boundaries. This overlap is particularly pronounced in the intermediate spectral region ($1.5 \leq \Gamma_{\mathrm{X}} \leq 2.0$), where approximately $26\%$ of all observations reside--a fraction far too large to represent merely transitional scatter between well-separated populations.

The continuity of the flux--photon index distribution, combined with the systematic wandering of individual sources across these boundaries over multi-year timescales, provides strong observational evidence that the LSP/ISP/HSP classification captures different spectral states of an underlying continuous blazar sequence rather than fundamentally distinct source populations.

These classifications do not appear as completely isolated clusters but instead occupy partially continuous regions in the flux–index plane.  observe a systematic decrease in the $2$--$10$~keV flux from LSP to HSP, accompanied by a corresponding increase in the photon index. This continuous shift indicates that the traditional blazar subclasses are snapshots of a broad observational continuum. The intermediate ISP class is particularly significant, as it marks the regime where the synchrotron peak crosses the \textit{Swift}-XRT bandpass. Consequently, ISP sources serve as direct observational evidence of the transition in dominant cooling mechanisms from synchrotron to inverse Compton (IC) radiation.

\begin{figure*}[htbp]
\centering
\includegraphics[width=\textwidth]{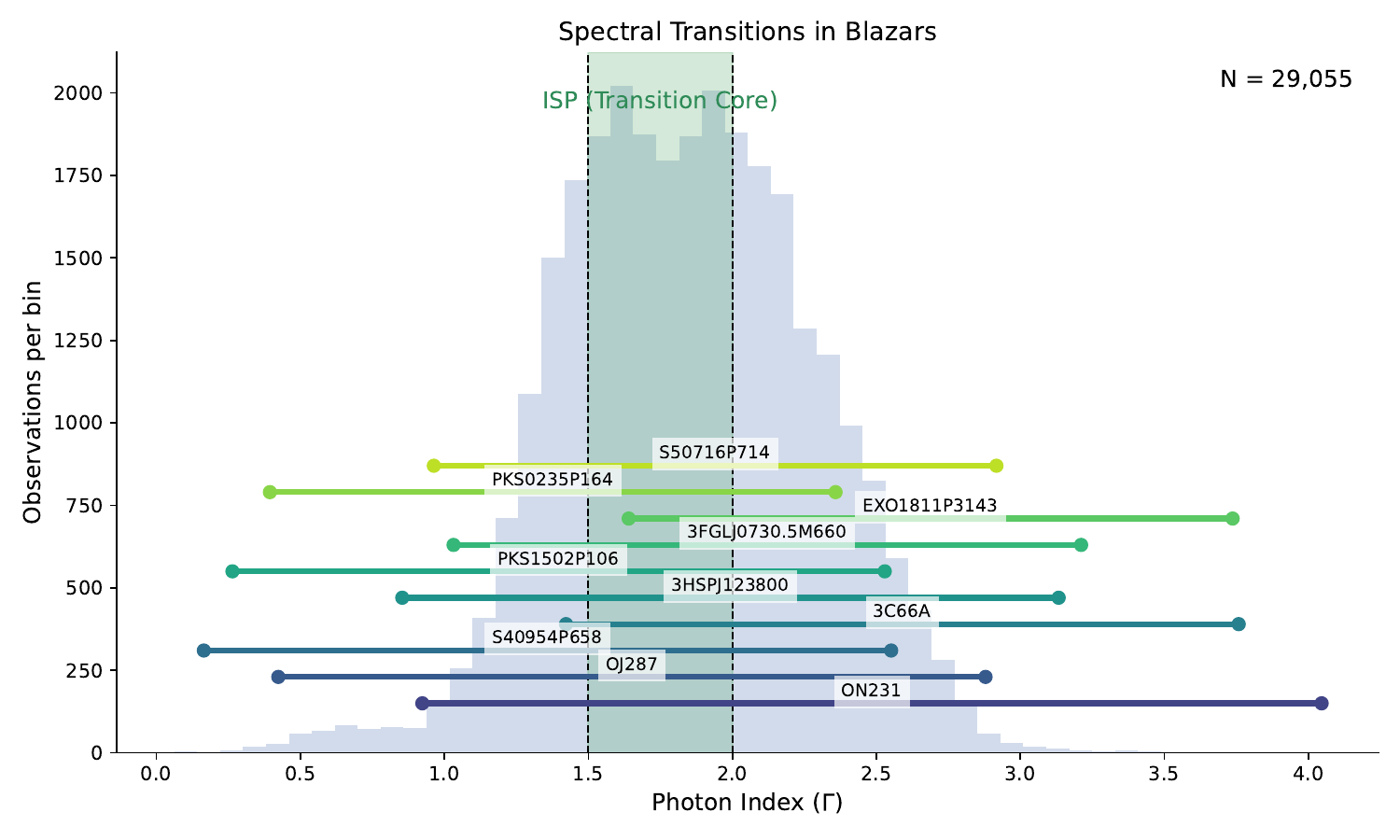}
\caption{Spectral transition distribution of blazars from the \textit{Swift}-XRT dataset. The histogram shows the X-ray photon index ($\Gamma_{\rm X}$) distribution for 29,055 observations. The shaded region highlights the ISP regime ($1.5 < \Gamma_{\rm X} < 2.0$), which acts as the transition core. Horizontal lines represent individual sources, showing their spectral range ($\Gamma_{\rm X,min}$ to $\Gamma_{\rm X,max}$), demonstrating transitions across traditional blazar subclasses.}
\label{fig:transition_plot}
\end{figure*}

Long-term light curves spanning 2005 to 2025 demonstrate that individual sources frequently wander across these nominal class boundaries. This spectral wandering suggests that the LSP/ISP/HSP classification reflects a transient spectral state driven by dynamic changes in the synchrotron peak frequency during outburst or quiescent phases. Multi-epoch observations enable the measurement of intra-source spectral variability ($\Delta \Gamma_{\mathrm{X}}$), identifying ``strong transition'' sources , with large spectral-index variability. The distribution of these observations across subclasses and the most active transition-like sources are summarized in Table~\ref{tab:swift_revised}. The presence of a large population within the ISP ``transition zone'' $\Gamma_{\mathrm{X}} \approx 1.5$--$2.0$) reinforces the paradigm of a dynamic, evolving blazar sequence as shown in figure \ref{fig:transition_plot}.

\begin{table*}[htbp]
\centering
\caption{Swift-XRT Blazar Catalog: Candidate Transition-like Sources and Spectral-Regime Distribution}
\label{tab:swift_revised}
\begin{tabular}{lccccccc}
\toprule
\multicolumn{8}{c}{\textbf{Candidate Transition-like Blazars}} \\
\midrule
\textbf{Source} &
\textbf{$N_{\rm obs}$} &
\textbf{Year$_{\rm start}$} &
\textbf{Year$_{\rm end}$} &
\textbf{$\Gamma_{\rm min}$} &
\textbf{$\Gamma_{\rm max}$} &
\textbf{$\Delta\Gamma$} &
\textbf{Classification} \\
\midrule

OJ~287 &
1556 & 2005 & 2021 & 1.86 & 2.55 & 0.69 &
Likely \\

3C~66A &
361 & 2007 & 2021 & 1.70 & 3.70 & 2.00 &
Strong \\

ON~231 &
352 & 2007 & 2021 & 1.80 & 4.10 & 2.30 &
Strong \\

EXO~1811+3143 &
355 & 2009 & 2021 & 1.90 & 3.80 & 1.90 &
Strong \\

PKS~1502+106 &
178 & 2008 & 2021 & 1.50 & 2.53 & 1.03 &
Likely \\

BL~Lac &
1376 & 2005 & 2021 & 1.70 & 2.30 & 0.60 &
Likely \\

S5~1803+784 &
226 & 2006 & 2021 & 1.60 & 2.10 & 0.50 &
Likely \\

3C~279 &
1269 & 2006 & 2020 & 1.00 & 2.08 & 1.08 &
Strong \\

PKS~1510$-$08 &
931 & 2006 & 2019 & 0.54 & 2.03 & 1.49 &
Likely \\

PKS~2155$-$304 &
723 & 2005 & 2020 & 2.08 & 3.09 & 1.01 &
Strong \\

H~1426+428 &
479 & 2005 & 2020 & 1.54 & 2.27 & 0.74 &
Likely \\

PKS~0235+164 &
460 & 2005 & 2016 & 0.39 & 2.36 & 1.97 &
Likely \\

PKS~0208$-$512 &
401 & 2005 & 2020 & 0.75 & 2.54 & 1.79 &
Likely \\

1ES~0033+595 &
397 & 2005 & 2020 & 1.16 & 2.12 & 0.96 &
Likely \\

B3~1633+382 &
389 & 2007 & 2018 & 0.84 & 2.22 & 1.38 &
Likely \\

PKS~0548$-$322 &
370 & 2005 & 2020 & 1.33 & 2.15 & 0.82 &
Likely \\

\midrule
\multicolumn{8}{c}{\footnotesize
$\Delta\Gamma = \Gamma_{\rm max} - \Gamma_{\rm min}$} \\
\midrule

\multicolumn{8}{c}{\textbf{Observation-level Distribution across Spectral Regimes}} \\
\midrule

\textbf{Spectral Region} &
\multicolumn{2}{c}{\textbf{Observations ($N$)}} &
\textbf{Fraction (\%)} &
\multicolumn{4}{c}{\textbf{Photon-index Regime}} \\
\midrule

Hard-spectrum region &
\multicolumn{2}{c}{$\sim$17,000} &
55 &
\multicolumn{4}{c}{Hard ($\Gamma < 1.5$)} \\

Intermediate-spectrum region &
\multicolumn{2}{c}{$\sim$8,000} &
26 &
\multicolumn{4}{c}{Intermediate ($1.5 \leq \Gamma \leq 2.0$)} \\

Soft-spectrum region &
\multicolumn{2}{c}{$\sim$6,000} &
19 &
\multicolumn{4}{c}{Soft ($\Gamma > 2.0$)} \\

\midrule

\textbf{Total} &
\multicolumn{2}{c}{31,068} &
\multicolumn{4}{c}{100} & \\

\bottomrule
\end{tabular}

\vspace{0.2cm}

\begin{minipage}{0.95\textwidth}
\footnotesize
\textbf{Note:}
The classifications represent empirical X-ray spectral-regime variability derived from long-term Swift-XRT monitoring observations. The terms ``hard-spectrum'', ``intermediate-spectrum'', and ``soft-spectrum'' refer to phenomenological photon-index regimes and should not be interpreted as definitive physical subclass transformations. Candidate transition-like sources are identified through statistically significant spectral excursions across multiple photon-index regimes.
\end{minipage}

\end{table*}

\subsection{Swift-BAT Hard X-ray Catalog}
\begin{figure*}[htbp]
\centering
\includegraphics[width=\textwidth]{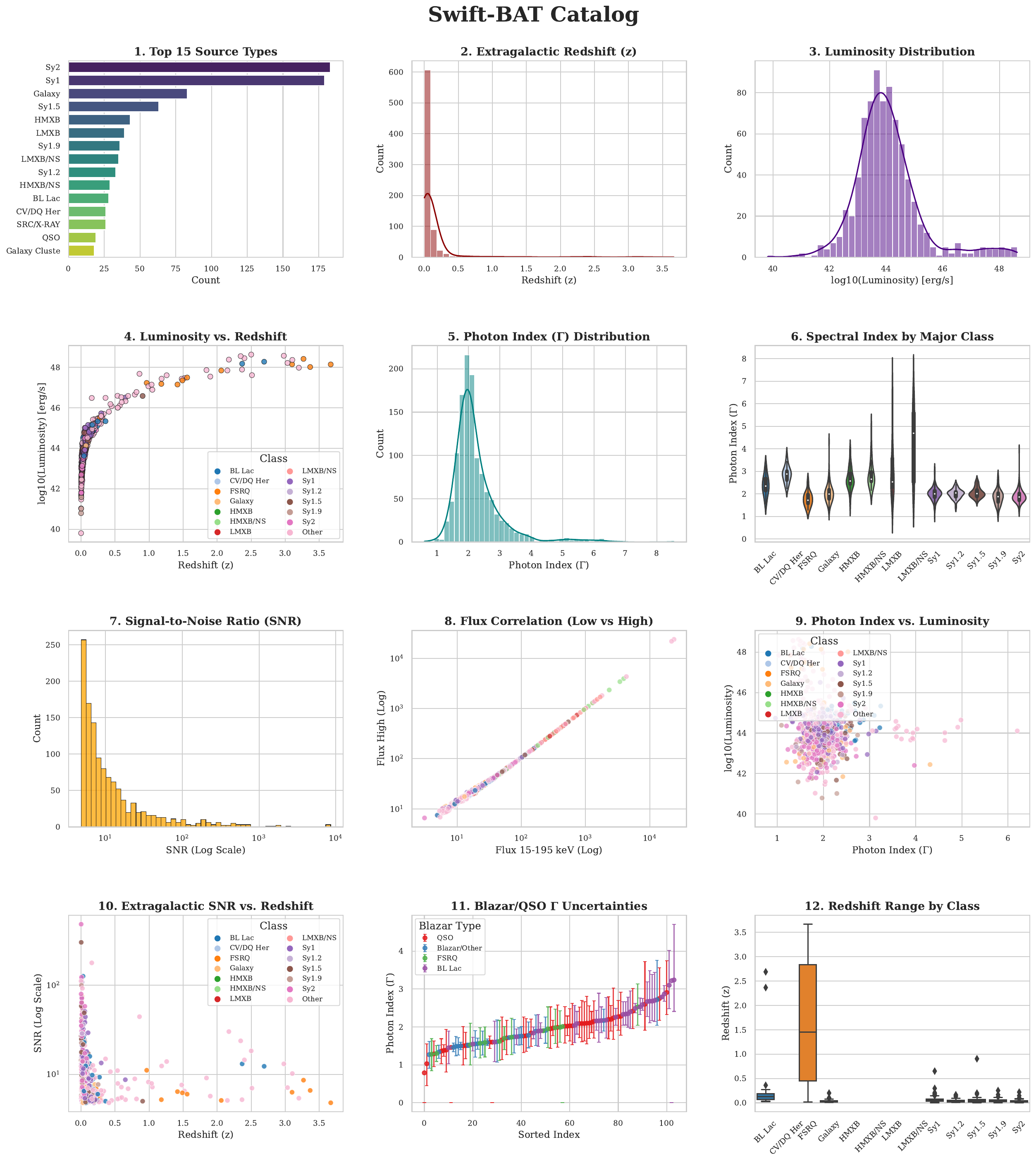}
\caption{Comprehensive overview of the Swift-BAT source population, showing the statistical, spectral, luminosity, and redshift properties of major source classes. The panels summarize source-type distributions, luminosity and redshift trends, photon-index behavior, signal-to-noise characteristics, and flux correlations, highlighting the diversity of spectral and observational properties across the Swift-BAT catalog.}
\label{fig:bat}
\end{figure*}

This work Utilizes the \textit{Swift} Observatory BAT 70-month survey \citep{Baumgartner_2013}, which provides a uniquely unbiased, penetrating view of the active universe across the $14$--$195$~keV band. Unlike optical or soft X-ray surveys, the BAT Catalog is relatively unaffected by line-of-sight obscuration, offering a clearer diagnostic of the intrinsic central engine. As demonstrated in our Figure~\ref{fig:bat}. The local hard X-ray universe ($z < 0.2$) is predominantly populated by Seyfert galaxies and Galactic X-ray binaries. Conversely, the high-redshift regime is dominated by Flat-Spectrum Radio Quasars (FSRQs) and Quasi-Stellar Objects (QSOs). This stark division is a classic manifestation of Malmquist bias in a flux-limited survey; at cosmological distances, only the most intrinsically luminous sources---specifically those with relativistically beamed jets--exceed the detection threshold. Our findings are broadly consistent with the Unified
Model of AGN. In the BAT band,  observe a continuum of source classes that are several hard X-ray spectral properties despite differing
in orientation, obscuration, and jet contribution. Seyfert 1 and Seyfert 2 galaxies exhibit comparable hard X-ray photon index ($\Gamma_{\mathrm{X}}$) distributions, confirming that the circumnuclear dusty torus becomes largely transparent above $15$~keV. While FSRQs share similar accretion disk properties with local Seyferts, they represent the extreme beamed population. The observed disparities in luminosity and spectral hardness underscore that the $15$--$195$~keV emission in FSRQs is dominated by inverse-Compton scattering within the relativistic jet, which effectively overpowers the underlying coronal emission characteristic of Seyfert galaxies. Detailed classifications and spectral ranges for these 1210 sources are summarized in Table~\ref{tab:bat_summary}.

\begin{table*}[htbp]
\centering
\caption{Swift-BAT AGN Catalog: Spectral-population Summary in the 15--195\,keV Band (1210 Sources)}
\label{tab:bat_summary}

\renewcommand{\arraystretch}{1.3}

\begin{tabular}{l l c c cc}
\toprule

\textbf{Population} &
\textbf{Representative Source Types} &
\textbf{$N$} &
\textbf{Fraction (\%)} &
\multicolumn{2}{c}{$\Gamma$ Range} \\

\cmidrule(lr){5-6}

& & & & \textbf{Min} & \textbf{Max} \\

\midrule

\multicolumn{6}{c}{\textbf{Extragalactic AGN Populations}} \\

\midrule

Seyfert 1 &
Sy1, Sy1.x, Sy1.2, Sy1.5, Sy1.8 &
380 &
31.4 &
1.2 &
3.0 \\

Seyfert 2 &
Sy2, Sy1.9 &
260 &
21.5 &
1.3 &
3.5 \\

QSO / Quasar &
QSO, Quasar, LPQ, HPQ &
50 &
4.1 &
1.3 &
2.8 \\

LINER &
LINER &
15 &
1.2 &
1.5 &
2.5 \\

NLRG / Radio Galaxy &
NLRG, Radio Galaxy &
10 &
0.8 &
1.5 &
2.2 \\

Galaxy / AGN &
Galaxy, AGN, GPAIR, XBONG &
200 &
16.5 &
1.4 &
3.0 \\

\midrule

\multicolumn{6}{c}{\textbf{Blazar-related Populations}} \\

\midrule

BL Lac Objec&&

29 &
2.4 &
1.45 &
3.24 \\

FSRQ-like Blazars &&

32 &
2.6 &
1.27 &
2.72 \\

\midrule

\multicolumn{6}{c}{\textbf{Galactic / Stellar Populations}} \\

\midrule

HMXB / LMXB / XRB &
HMXB, LMXB, XRB, SFXT &
170 &
14.1 &
1.5 &
6.5 \\

CV / Symbiotic &
CV, CV/DQ Her, CV/AM Her, Symb/WD &
40 &
3.3 &
2.0 &
4.5 \\

Other Galactic &
Pulsar, SNR, Nova, Star, AXP &
35 &
2.9 &
1.3 &
5.0 \\

\midrule

\textbf{Total} &
-- &
\textbf{1210} &
\textbf{100} &
-- &
-- \\

\bottomrule
\end{tabular}

\vspace{0.2cm}

\begin{minipage}{0.95\textwidth}
\footnotesize
\textbf{Note:}
The quoted photon-index ranges represent observed hard X-ray spectral distributions in the Swift-BAT energy band and should not be interpreted as strict physical subclass boundaries. Significant overlap between populations is expected because of variability, absorption effects, instrumental selection biases, and energy-dependent spectral behavior.
\end{minipage}

\end{table*}

\subsection{Rossi X-ray Timing Explorer (RXTE) Meta-Analysis}
\begin{figure*}[htbp]
\centering
\includegraphics[width=\textwidth]{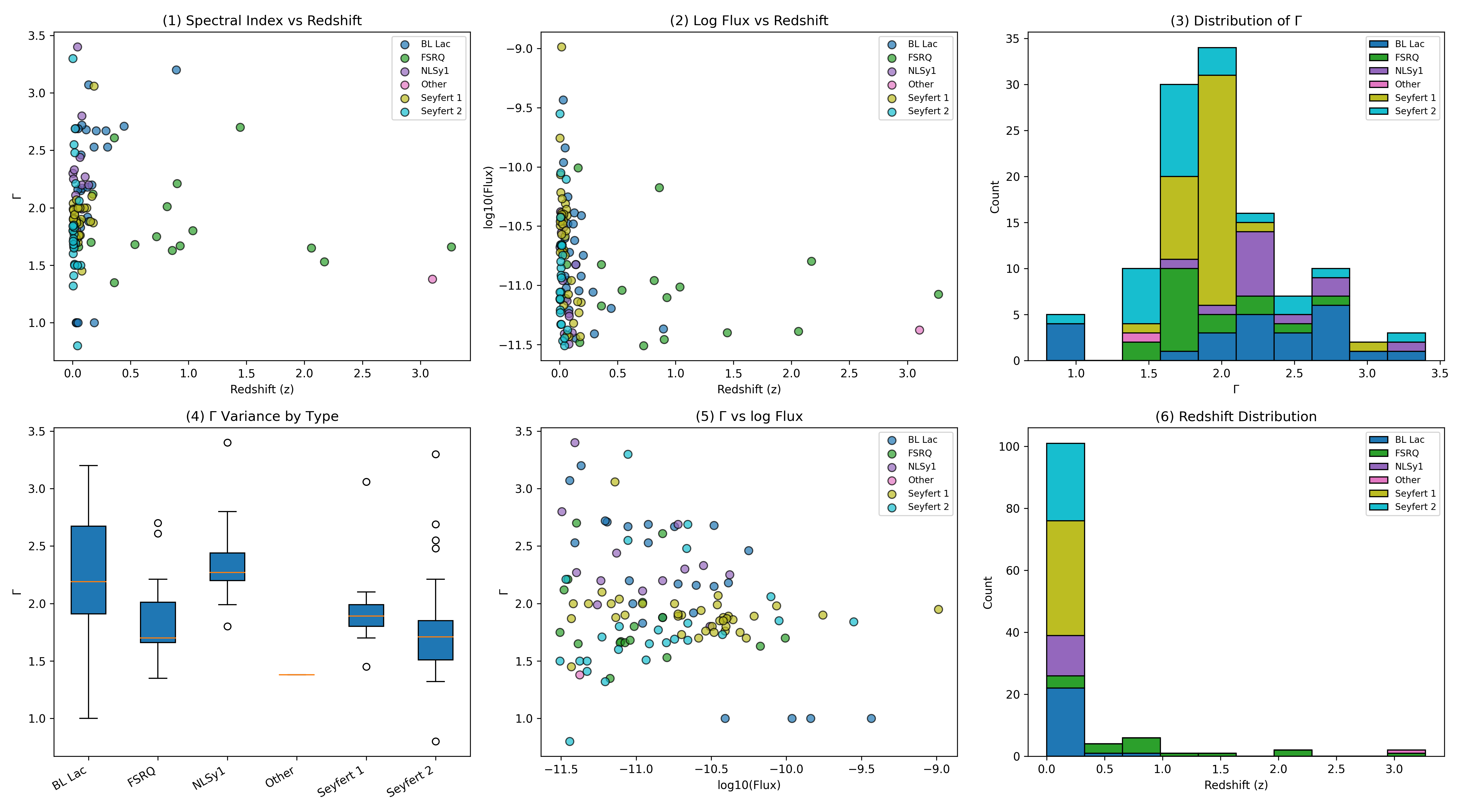}
\caption{Distribution of spectral index ($\Gamma_{\mathrm{X}}$), flux, and redshift for different source classes (BL Lac, FSRQ, NLSy1, Seyfert 1, Seyfert 2, and others). The panels show class-dependent trends in $\Gamma_{\mathrm{X}}$, flux, and redshift distributions, highlighting the diversity of spectral and observational properties across the sample.}
\label{fig:rxte}
\end{figure*}

To investigate the relationship between thermal accretion-driven emission and non-thermal jet processes, I utilize the RXTE AGN spectral Catalog \citep{Rivers_2013}. This dataset provides long-term average 2--10 keV spectral parameters for 100 bright AGNs, offering a standardized framework to compare Seyfert galaxies, Narrow-Line Seyfert 1s (NLSy1s), and blazars without instrument-specific bias. The photon index ($\Gamma_{\mathrm{X}}$) serves as the primary diagnostic for the underlying physical regime, with $\Gamma_{\mathrm{X}} \approx 2.0$ acting as a critical observational dividing line. As summarized in Table~\ref{tab:rxte_spectral_regimes}, indices satisfying  ($\Gamma_{\mathrm{X}} > 2.0$) typically indicate thermal dominance, where the X-ray output is governed by the accretion disk and a compact hot corona. Conversely, harder spectra ($\Gamma_{\mathrm{X}}< 2.0$) signify either non-thermal dominance-characteristic of synchrotron self-Compton emission in relativistic jets or significant intrinsic obscuration. Crucially, the RXTE data reveals that the observed photon-index distributions suggest partial spectral continuity between several AGN populations. The dynamic spectral evolution of these objects is evident in intermediate cases, such as Seyfert 1.2 and 1.5, which do not cluster in isolated classes but demonstrate evolutionary or orientation-dependent migration between regimes. The most profound insight from the RXTE sample is the spectral overlap between NLSy1 galaxies and BL Lac objects as shown in  Figure~\ref{fig:rxte}
.Despite their distinct optical and jet signatures, both classes occupy a shared parameter space defined by steep photon indices ($\Gamma_{\mathrm{X}} > 2.0$). This overlap confirms that an AGN’s spectral identity is often a transient manifestation of its current accretion efficiency. During periods of high-Eddington accretion, an NLSy1 and a blazar can exhibit near-identical X-ray signatures, suggesting that AGN spectral
classifications may reflect partially overlapping
observational states .
\begin{table*}[htbp]
\centering
\caption{RXTE AGN Catalog: Empirical Spectral Regimes and Photon-Index Distributions}
\label{tab:rxte_spectral_regimes}
\begin{tabular}{lccccccc}
\toprule
\textbf{Source Type} & \textbf{N} & \textbf{Fraction (\%)} & \textbf{$\Gamma_{\rm min}$} & \textbf{$\Gamma_{\rm max}$} & \textbf{Mean $\Gamma$} & \textbf{Spectral Regime} & \textbf{Spectral Evolution} \\
\midrule

Seyfert 1 & 35 & 28.5 & 1.70 & 2.33 & 1.89 & Intermediate-spectrum & Intermediate-to-soft variability \\

Seyfert 2 & 28 & 22.8 & 1.32 & 3.30 & 1.86 & Intermediate-spectrum & Broad intermediate-to-soft excursions \\

\midrule
\multicolumn{8}{c}{\textit{\textbf{Transition-like Spectral Populations}}} \\
\midrule

BL Lac & 20 & 16.3 & 1.83 & 3.20 & 2.15 & Soft-spectrum & Broad soft-spectrum variability \\

FSRQ & 14 & 11.4 & 1.35 & 2.21 & 1.79 & Hard-spectrum & Hard-to-intermediate spectral range \\

NLSy1 & 10 & 8.1 & 2.11 & 3.40 & 2.47 & Very soft-spectrum & Extreme soft-spectrum variability \\

QSO & 3 & 2.4 & 1.45 & 2.00 & 1.77 & Hard-spectrum & Moderate hard-spectrum evolution \\

NLRG & 2 & 1.6 & 1.84 & 1.84 & 1.84 & Intermediate-spectrum & Stable intermediate-spectrum regime \\

LINER & 1 & 0.8 & 1.73 & 1.73 & 1.73 & Intermediate-spectrum & Stable intermediate-spectrum regime \\

ULIRG & 1 & 0.8 & 1.50 & 1.50 & 1.50 & Hard-spectrum & Stable hard-spectrum regime \\

Other & 9 & 7.3 & -- & -- & -- & Mixed/Unclassified & -- \\

\midrule
\textbf{Total} & \textbf{123} & \textbf{100} & -- & -- & -- & -- & -- \\

\bottomrule
\end{tabular}

\vspace{0.2cm}

\begin{minipage}{0.95\textwidth}
\footnotesize
\textbf{Note:} The spectral classifications refer to empirical X-ray photon-index regimes derived from RXTE observations and should not be interpreted as definitive physical state transitions. Hard-, intermediate-, and soft-spectrum classifications are phenomenological descriptions based on observed photon-index distributions.
\end{minipage}

\end{table*}

\subsection{Chandra AKARI NEP-Deep Population Analysis}
We extend  spectral analyses to the \textit{Chandra} AKARI NEP-Deep Catalog \citep{2024A&A...689A..83M}, which encompasses 27 galactic objects, 57 Type I AGNs, and 131 additional active galactic nuclei. For this population-level study, the hardness ratio (HR) is utilized as a robust proxy for spectral shape and absorption, defined as:
\begin{equation}
HR = \frac{F_{2-7~\text{keV}} - F_{0.5-2~\text{keV}}}{F_{2-7~\text{keV}} + F_{0.5-2~\text{keV}}}.
\end{equation}
In the absence of individualized photon index modeling for faint sources, HR provides a critical diagnostic for identifying obscuration and emission regimes. Out of 369 analyzed sources, the population exhibits a clear bimodal distribution: a soft-spectrum population ($HR \leq -0.2$; 138 sources) likely dominated by thermal or stellar processes, and a harder population ($HR > 0.3$; 146 sources) interpreted as candidate obscured AGN, where photoelectric absorption suppresses soft X-ray emission .

\begin{figure*}[htbp]
 \centering
 \includegraphics[width=\textwidth]{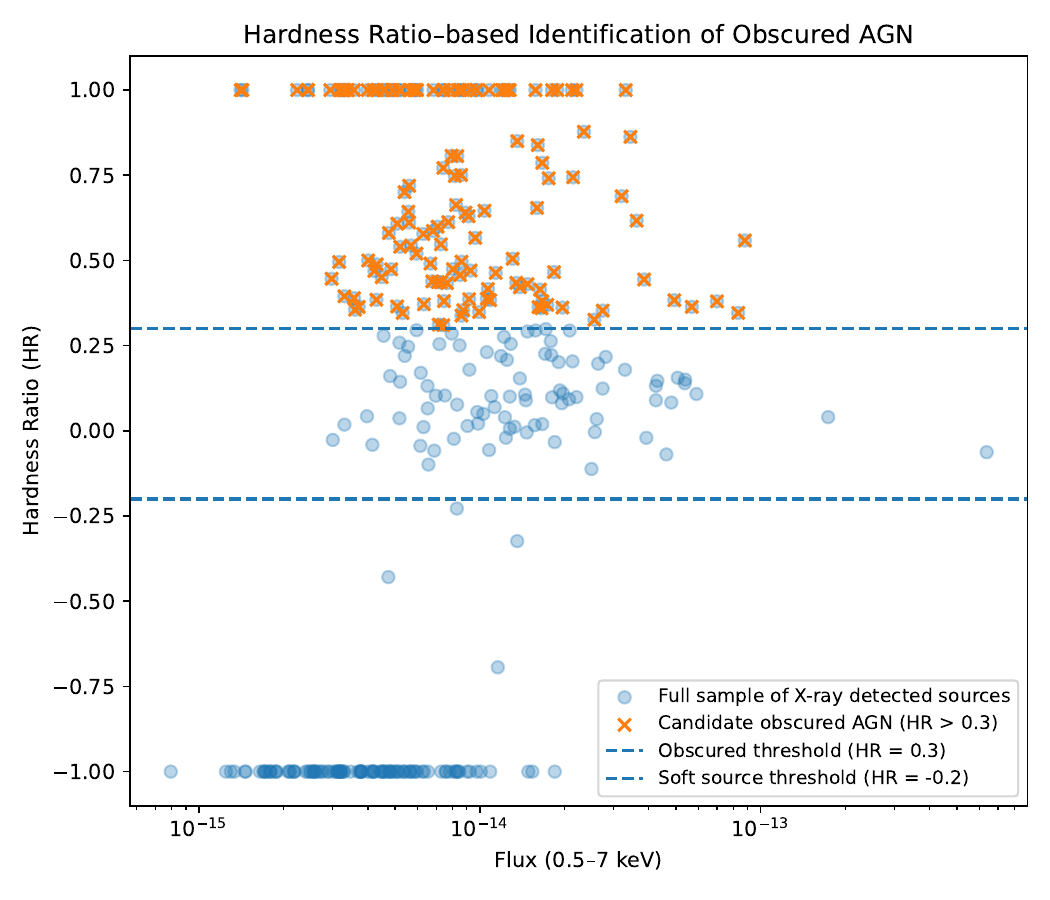}
 \caption{Hardness ratio (HR) as a function of 0.5--7 keV X-ray flux for the Chandra source sample. The figure highlights the separation between soft and hard populations, with candidate obscured AGN predominantly occupying the high-HR region, demonstrating the usefulness of HR as a diagnostic of X-ray obscuration.}
 \label{fig:chanpdf}
\end{figure*}

As illustrated in Figure~\ref{fig:chanpdf}, the hardness ratio appears independent of the total X-ray flux (0.5--7~keV), suggesting that spectral hardness is governed primarily by intrinsic source properties rather than flux-dependent selection effects. Identify a well-defined transition region in hardness ratio space ($-0.2 < HR \leq 0.3$) that separates these two distinct populations. This observational boundary is physically consistent with the state-transition spectral-overlap behavior reported in individual blazars, such as the corona-to-jet transition observed in OT~081 \citep{Ding_2023}. By employing HR as a proxy, we extend the statistical view of spectral state transitions to a larger, fainter sample of X-ray sources where detailed spectral fitting is unfeasible,suggesting that similar
spectral-overlap behavior may occur across diverse
AGN populations.

\subsection{Historical Multi-Satellite Archives (VizieR)}
\begin{figure*}[htbp]
\centering
\includegraphics[width=0.8\textwidth]{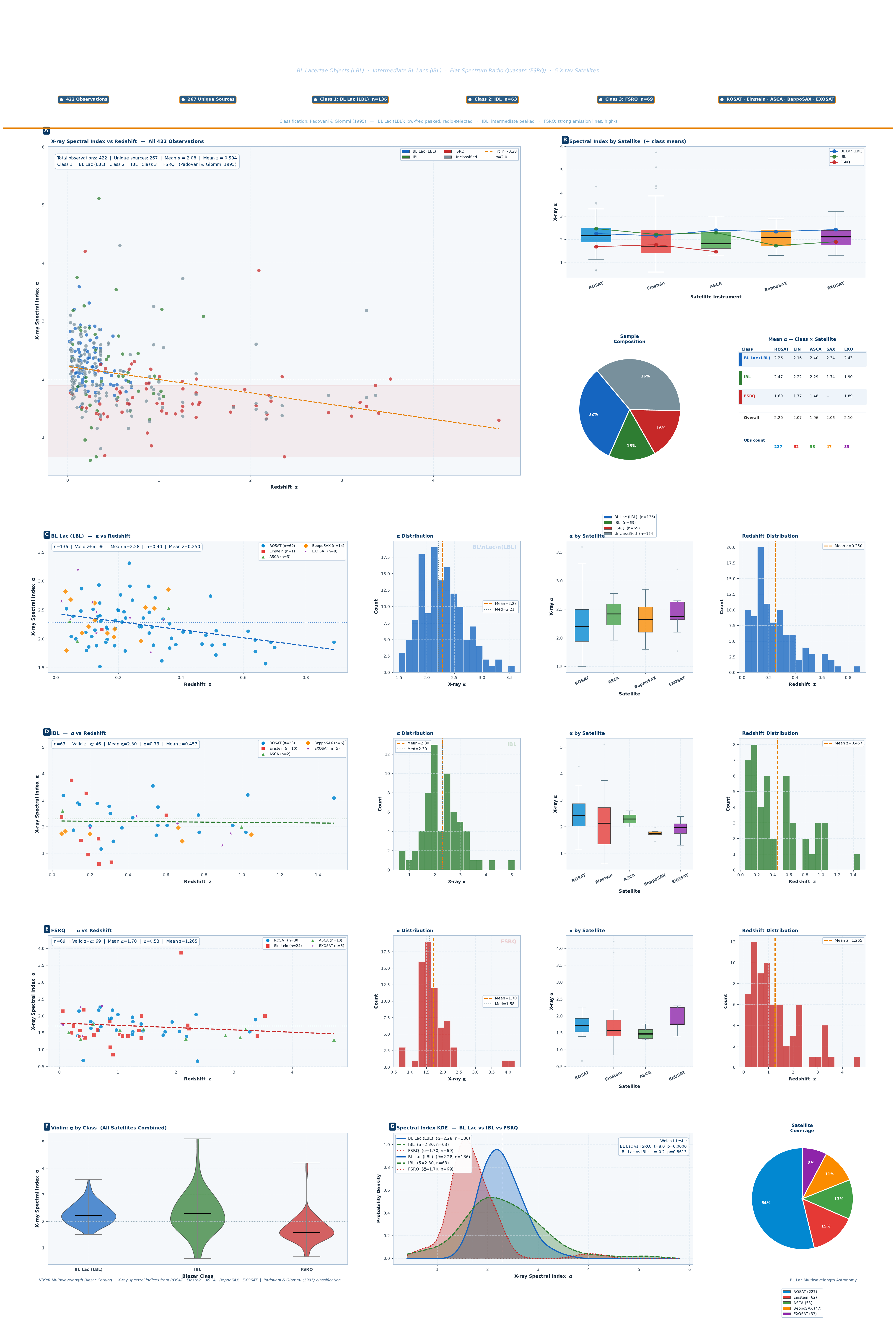}
\caption{Vizier Catalog}
\label{fig:vizier}
\end{figure*}

The historical multi-instrument VizieR compilation further supports the presence of broad spectral-regime continuity across the blazar population (Table~\ref{tab:vizier_revised}). The combined distributions derived from ROSAT, Einstein, ASCA, BeppoSAX, and EXOSAT observations reveal systematic differences between HBL-, IBL-, and FSRQ-like populations while still exhibiting substantial overlap in photon-index space. As shown in Figure~\ref{fig:vizier}, HBL populations preferentially occupy softer X-ray spectral regimes with a mean photon index near $\Gamma \sim 2.4$, whereas FSRQ populations remain concentrated toward harder spectra with mean values near $\Gamma_{\mathrm{X}}\sim 1.7$. Intermediate BL~Lac populations occupy broad distributions centered near $\Gamma_{\mathrm{X}} \sim 2.3$, bridging the harder FSRQ-like and softer HBL-like regimes. The distribution analyses further demonstrate that the spectral populations are not sharply separated but instead form partially overlapping continuous distributions. The statistical comparison shown in Figure~\ref{fig:vizier} indicates a significant separation between BL~Lac/HBL-like and FSRQ-like populations, whereas the IBL distributions remain broadly consistent with intermediate spectral behavior. The redshift-dependent distributions additionally show that FSRQ populations preferentially extend toward higher redshift systems with systematically harder X-ray spectra, while HBL-like populations remain concentrated at lower redshifts with softer synchrotron-dominated continum.
Importantly, the persistence of similar spectral-regime behavior across five independent historical X-ray missions strongly suggests that the observed overlap regions are not solely artifacts of a single instrument or survey strategy. Instead, the consistency of the distributions across multiple archival datasets supports the interpretation that intermediate spectral-regime occupation represents a persistent observational characteristic of the blazar population.

\begin{table*}[htbp]
\centering
\scriptsize
\caption{Vizier Multi-Instrument Catalog: Hard X-ray Spectral Regimes of Blazar Populations (423 Sources)}
\label{tab:vizier_revised}
\begin{tabular}{lcccccccl}
\toprule
\multirow{2}{*}{\textbf{Class}} &
\multirow{2}{*}{\textbf{Population}} &
\multirow{2}{*}{\textbf{N}} &
\multirow{2}{*}{\textbf{Fraction (\%)}} &
\multicolumn{3}{c}{\textbf{Spectral Index ($\Gamma$) Distribution}} &
\multirow{2}{*}{\textbf{Dominant Spectral Regime}} &
\multirow{2}{*}{\textbf{Instruments}} \\
\cline{5-7}
& & & & \textbf{Min} & \textbf{Max} & \textbf{Mean} & & \\
\midrule

1 & HBL & 120 & 28.4 & 1.52 & 4.28 & 2.40 &
Broad soft-spectrum distribution &
RO, SA, AS, EI, EX \\

2 & LBL & 55 & 13.0 & 0.60 & 3.75 & 2.10 &
Intermediate-to-hard spectrum &
RO, SA, AS, EI, EX \\

3 & FSRQ & 90 & 21.3 & 0.66 & 2.50 & 1.70 &
Hard-spectrum dominated &
RO, SA, AS, EI, EX \\

\midrule

\multicolumn{2}{c}{\textbf{Spectrally Classified Sources}} &
\textbf{265} & \textbf{62.7} &
-- & -- & -- & -- & -- \\

\multicolumn{2}{c}{\textbf{Other / Unclassified}} &
\textbf{158} & \textbf{37.3} &
-- & -- & -- & -- & -- \\

\midrule

\multicolumn{2}{c}{\textbf{Grand Total}} &
\textbf{423} & \textbf{100} &
-- & -- & -- & -- & -- \\

\bottomrule
\end{tabular}

\vspace{0.2cm}

\begin{minipage}{0.95\textwidth}
\footnotesize
\textbf{Note:} The spectral classifications represent empirical hard X-ray photon-index distributions derived from multiple archival instruments. The terms ``hard-spectrum'', ``intermediate-spectrum'', and ``soft-spectrum'' refer to phenomenological spectral regimes and should not be interpreted as definitive physical state transitions. HBL populations exhibit systematically softer hard X-ray spectra due to synchrotron-tail dominance, whereas LBL and FSRQ populations generally show harder inverse-Compton dominated continua.
\end{minipage}

\end{table*}

\subsection{TevCat}
\begin{figure*}[htbp]
\centering
\includegraphics[width=\textwidth]{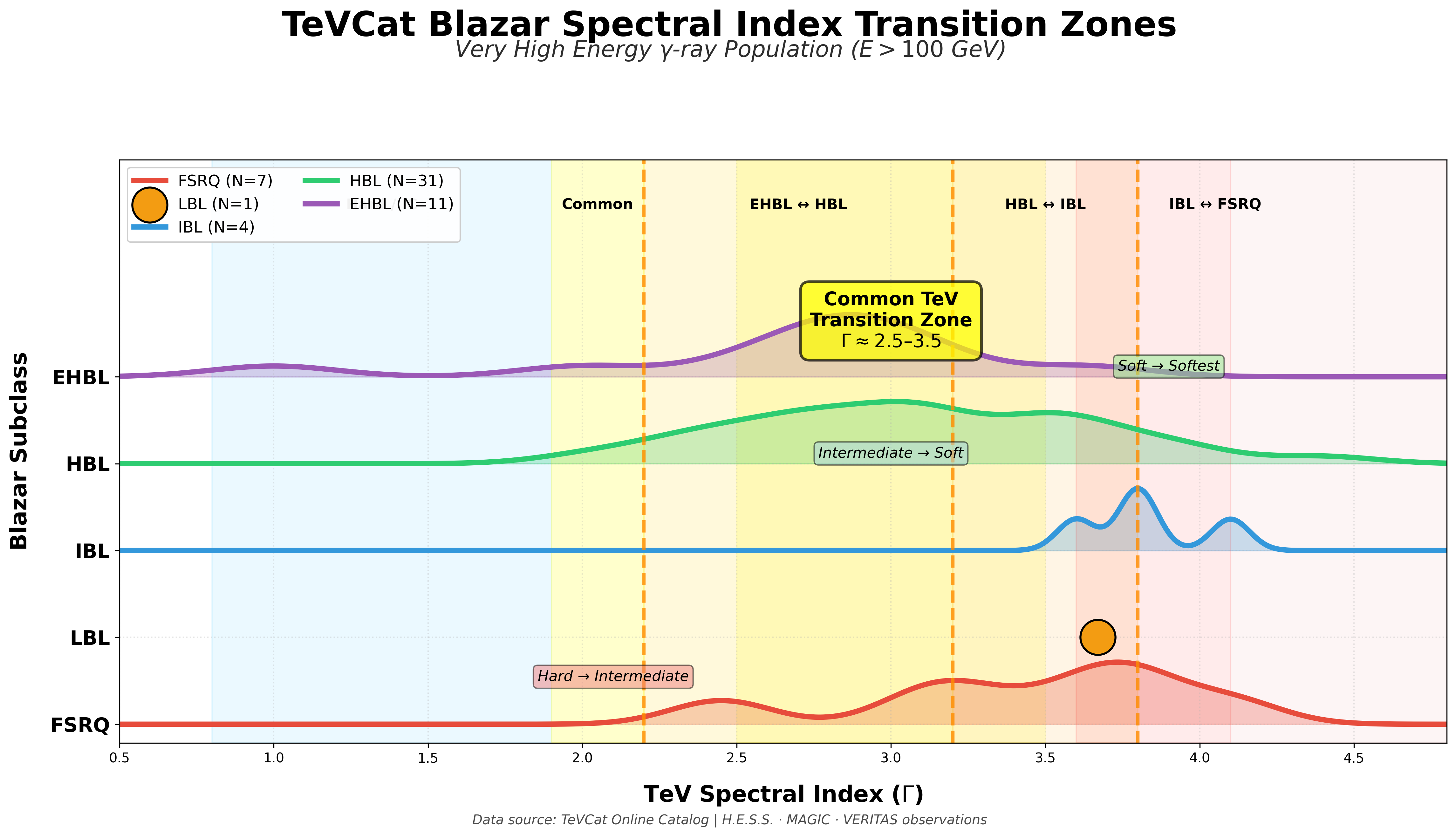}
\caption{TeVCat blazar spectral-index distributions in the very-high-energy ($E > 100$ GeV) $\gamma$-ray regime. The shaded regions indicate broad overlap zones between EHBL, HBL, IBL, LBL, and FSRQ-like populations, with a common TeV transition regime around ($\Gamma \sim 2.5$--$3.5$). The distributions illustrate the gradual progression from hard EHBL/HBL spectra toward softer IBL, LBL, and FSRQ-like regimes. }
\label{fig:tevonly}
\end{figure*}

The Very High Energy (VHE) TeVCat sample further supports the presence of broad spectral-regime overlap across the blazar population. As illustrated in Figure~\ref{fig:tevonly}, the TeV photon-index distributions of EHBL, HBL, IBL, LBL, and FSRQ sources show substantial overlap despite maintaining systematic 
differences in their dominant spectral regimes. The spectral-index versus redshift distribution reveals that FSRQ populations preferentially extend toward higher redshifts 
and generally occupy softer TeV spectral regimes, whereas EHBL and HBL populations remain concentrated at comparatively lower redshifts with harder TeV spectra. Intermediate populations, particularly IBL-like sources, occupy overlapping regions between the classical HBL and FSRQ distributions, indicating 
that the TeV spectral properties form a partially overlapping TeV spectral distributions. The histogram and kernel-density distributions further demonstrate that the TeV photon-index populations are not isolated clusters but instead exhibit partially overlapping spectral regimes, 
particularly within the approximate range $\Gamma_{\gamma,\rm TeV} \sim 2.5$--$3.8$. The corresponding source distributions and representative photon-index ranges are summarized in Table~\ref{tab:tevcat_transition_summary}. These results are consistent with the broader multi-instrument picture developed in this work, where intermediate spectral regime occupation appears to be a common phenomenological property of blazar populations..

\begin{table*}[htbp]
\centering
\renewcommand{\arraystretch}{1.8}

\caption{TeVCat Blazar Populations and Candidate Spectrally Variable Sources Based on Observed TeV Photon-index ($\Gamma$) Distributions.}

\resizebox{\textwidth}{!}{%

\begin{tabular}{p{2.8cm} p{2.5cm} p{3.2cm} p{2.5cm} p{3.3cm} p{5.8cm}}
\hline

\textbf{Category} &
\textbf{Source / Class} &
\textbf{Observed By} &
\textbf{Observed TeV Photon-index Range} &
\textbf{Dominant Spectral Regime} &
\textbf{Physical Interpretation / Importance} \\

\hline

EHBL Population &
EHBL &
H.E.S.S., MAGIC, VERITAS &
$1.00 \leq \Gamma \leq 3.63$ &
EHBL/HBL-like hard-spectrum regime &
Hard synchrotron/SSC-dominated extreme blazar population \\

\hline

HBL Population &
HBL &
H.E.S.S., MAGIC, VERITAS &
$2.00 \leq \Gamma \leq 4.40$ &
Broad HBL-like spectral distribution &
Broad TeV spectral distribution spanning relatively hard and soft states \\

\hline

IBL Population &
IBL &
MAGIC, VERITAS &
$3.60 \leq \Gamma \leq 4.10$ &
Intermediate soft-spectrum regime &
Intermediate-to-soft TeV spectral distribution \\

\hline

FSRQ Population &
FSRQ &
MAGIC, VERITAS, H.E.S.S. &
$2.45 \leq \Gamma \leq 4.10$ &
Soft Compton-dominated regime &
Soft inverse-Compton dominated TeV spectra \\

\hline

Spectrally Variable Candidate &
1ES~0229+200 &
VERITAS, H.E.S.S. &
2.50 &
EHBL/HBL-like hard-spectrum regime &
Extreme hard-spectrum TeV source with broad spectral variability \\

\hline

Spectrally Variable Candidate &
RGB~J0710+591 &
VERITAS &
2.69 &
Intermediate EHBL/HBL-like regime &
Intermediate hard-spectrum behavior between EHBL-like and HBL-like populations \\

\hline

Spectrally Variable Candidate &
1ES~1101$-$232 &
H.E.S.S. &
2.94 &
Intermediate hard-spectrum regime &
Stable intermediate hard-spectrum distribution \\

\hline

Spectrally Variable Candidate &
1ES~0414+009 &
VERITAS, H.E.S.S. &
3.45 &
HBL-like soft-spectrum regime &
Relatively soft TeV spectral distribution compared to typical HBL populations \\

\hline

Spectrally Variable Candidate &
H~1426+428 &
Crimea, VERITAS, MAGIC &
3.50 &
Soft HBL-like regime &
Pronounced soft-spectrum variability within the TeV band \\

\hline

\end{tabular}
}

\vspace{0.2cm}

\begin{minipage}{0.96\textwidth}
\footnotesize
\textbf{Note:}
The TeV photon-index ranges represent empirical spectral distributions derived from archival TeVCat observations and should not be interpreted as definitive subclass boundaries or direct evidence for physical evolutionary transitions. The spectral regimes are phenomenological descriptions of observed TeV spectral behavior and may overlap between different blazar subclasses due to variability, spectral curvature, and instrument-dependent energy coverage.
\end{minipage}

\label{tab:tevcat_transition_summary}

\end{table*}
\subsection{AstroSat and XMM-Newton/NICER Comparative Analysis}

We present the X-ray spectral properties of 10 blazars observed in the $0.3$--$80$~keV energy band using \textit{AstroSat}'s SXT and LAXPC instruments. The sample consists of 21 observations primarily comprising High-frequency peaked BL Lac objects (HBLs) and Intermediate BL Lac objects (IBLs), including: 1ES~0120+340, RGB~J0710+591, 1ES~1101-232, 1ES~1741+196, 1ES~2322-409, Mrk~501, Mrk~421, OJ~287, 1ES~1959+650, and 1ES~1218+304. Spectral indices ($\alpha_{\mathrm{X}}$) were derived from log-parabola fits to the combined broadband spectra. To ensure cross-instrument consistency, we integrated archival \textit{Astrosat}  data for Mrk~421 along with NICER data and OJ~287 with XXM-Newton, facilitating a robust cross-validation of spectral measurements across different X-ray observatories.

The analysis reveals significant variability in the spectral index both across the sample and within individual sources. While the spectral index versus redshift ($z$) distribution displays a weak positive trend with considerable scatter, specific source behaviors provide deeper physical insights. Notably, Mrk~421 exhibits strong stochastic variability; a single observation resolved into 39 time-resolved segments shows spectral fluctuations between $\alpha_{\mathrm{X}} \approx 1.75$ and $2.05$. In contrast, OJ~287 demonstrates three distinct, distinct spectral states (low, intermediate, and high), consistent across both \textit{AstroSat} and \textit{XMM-Newton} epochs. 

Statistical distributions—including Kernel Density Estimation (KDE) and histogram analysis indicate that the majority of spectral index values cluster between $1.8$ and $2.2$, peaking near $2.0$. This concentration reinforces the stability of the transition zone identified in our multi-mission meta-analysis. Furthermore, the identification of 1ES~2322-409 as a high-index outlier ($\alpha \approx 2.34$) suggests a particularly steep particle distribution or a unique accretion state during the observation period. These results highlight \textit{AstroSat}'s multiwavelength capability in resolving complex spectral evolution across the BL Lac--FSRQ continuum.

\subsection{Transition-like Blazar Candidates}
A significant fraction of the analyzed sources occupy intermediate photon-index regimes linking traditionally separated blazar populations. Rather than remaining confined to narrow spectral intervals, several objects exhibit substantial long-term variability that allows them to populate multiple observational subclasses across different epochs. In this work, these objects are conservatively described as candidate spectral-state transition or Transition-like blazar sources in a phenomenological sense. This terminology is intended only to characterize the observed spectral-regime overlap and should not be interpreted as evidence for irreversible physical evolution between established blazar subclasses.

Identified 16 sources showing photon-index excursions of $\Delta\Gamma_{\mathrm{X}} > 0.5$, substantially larger than the typical observational uncertainties, indicating significant long-term spectral variability. These candidates satisfy two observational criteria: (1) occupancy of at least two nominal spectral regimes across multiple epochs, and (2) statistically significant spectral variability beyond instrumental uncertainties. Sources with $\Delta\Gamma_{\mathrm{X}} > 1.5$, including 3C 66A, ON 231, and EXO 1811+3143, are classified as strong-variability candidates, while those with $0.5 < \Delta\Gamma_{\mathrm{X}} < 1.5$ are categorized as moderate-variability sources. Importantly, this classification is purely observational and reflects broad spectral wandering rather than  permanent physical subclass transformation.

The  evidence for such behaviour is found in the long-term Swift-XRT monitoring sample (Table~\ref{tab:swift_revised}; Figure~\ref{fig:giomi}). Sources such as 3C~66A, ON~231, EXO~1811+3143, PKS~2155$-$304, PKS~0235+164, and OJ~287 display large  photon-index excursions spanning intermediate spectral regimes. Several objects repeatedly cross nominal boundaries between harder LBL-/IBL-like and softer HBL-like states over extended observing intervals. Similar overlap behaviour is also present in the NuSTAR hard X-ray sample, where BL~Lac, TXS~0506+056, OJ~287, and 1ES~0229+200 frequently occupy intermediate hard X-ray regimes near $\Gamma_{\mathrm{X}} \sim 1.6$--$2.2$ (Table~\ref{tab:nustar_revised}; Figure~\ref{fig:nublazar_dashboard}). These observations suggest that a substantial fraction of blazars populate broad regions of photon-index space rather than remaining confined to sharply separated subclasses.

The TeVCat analysis extends this phenomenological picture into the very-high-energy regime. Sources including 1ES~0229+200, RGB~J0710+591, 1ES~1101$-$232, 1ES~0414+009, and H~1426+428 occupy intermediate TeV photon-index regimes linking EHBL and HBL like populations. Significant overlap is observed within the approximate range $\Gamma_{\gamma,\rm TeV} \sim 2.5$--$3.5$, indicating that TeV spectral properties also form a continuous observational distribution rather than a set of discrete populations. Although the present analysis does not attempt detailed physical population modeling, the observed overlap behaviour is consistently detected across independent instruments with different energy coverage and selection effects. Histogram, kernel density estimation (KDE), and population-fraction analyses all reveal persistent concentrations near intermediate photon-index values. The recurrence of these overlap regions across multiple datasets argues against their origin being solely due to instrumental systematics or isolated statistical fluctuations. The observed spectral-regime overlap is more plausibly associated with long-term variability in jet emission processes, including changes in particle acceleration efficiency, synchrotron peak location, radiative cooling, jet power, or viewing-angle effects. Consequently, the analyzed sources are best interpreted as objects capable of occupying broad intermediate spectral distributions over time rather than physically evolving from one fixed blazar subclass into another. Taken together, the multi-instrument results support a picture in which blazar spectral populations form a continuous observational distribution shaped by variability and energy-dependent radiative processes, with intermediate spectral-regime occupation emerging as a common phenomenological property across the blazar population.
\subsection{Limitations} 
Several limitations should be considered when interpreting the results presented in this work. First, the analyzed datasets combine observations from instruments with substantially different energy coverage, sensitivity, calibration strategies, and spectral response functions. Consequently, the observed photon-index and hardness-ratio boundaries are instrument dependent and should not be interpreted as universally identical physical thresholds across all wavebands. Second, the archival catalogs used in this study are affected by selection biases inherent to flux-limited surveys. Bright, highly variable, and strongly beamed sources are preferentially detected, particularly in the hard X-ray and TeV regimes. Third, the observations are largely non-simultaneous across different instruments and epochs. Because blazars exhibit strong spectral and flux variability over timescales ranging from hours to years, the measured photon indices may reflect different source states rather than directly comparable physical conditions.
\section{Discussion}
Across more than 32,000 observations spanning multiple X-ray missions, identified a broad intermediate spectral regime ($1.5 \leq \Gamma_{\mathrm{X}} \leq 2.2$) occupied by approximately 26--43\% of the observed populations (\textit{Swift}-XRT: 26\%; \textit{NuSTAR}: 43\%). Despite differences in instrumental energy coverage, the photon-index distributions consistently show enhanced source concentrations near $\Gamma_{\mathrm{X}} \approx 2.0$, suggesting that intermediate spectral states are a persistent observational feature of the blazar population. The precise boundaries of these intermediate regimes vary
slightly between instruments because different X-ray missions sample different regions of the synchrotron and inverse-Compton components of the broadband spectral energy distribution. In the \textit{Swift}-XRT band, the intermediate population is primarily concentrated within $\Gamma_{\rm X}\sim1.5$--$2.0$, whereas the harder \textit{NuSTAR} band extends this range toward $\Gamma_{\rm X}\sim1.6$--$2.2$. Similar intermediate gamma-ray regimes have also been reported in recent studies of
transition-like blazars. For example, \citet{10.1093/mnras/stag542} found that transition blazars preferentially occupy $\gamma$-ray photon indices near
$\Gamma_{\gamma}\sim2.0$--$2.2$, intermediate between classical BL~Lac and FSRQ populations. The consistency between the X-ray and gamma-ray distributions suggests that
intermediate spectral states may represent a common multiwavelength property of several blazar populations rather than isolated features confined to a single energy band.

The \textit{NuSTAR} sample provides particularly clear evidence for this behavior. Approximately 22.5\% of the observations occupy intermediate hard X-ray spectral regimes between classical LBL-like and HBL-like populations. Sources such as BL~Lac, TXS~0506+056, OJ~287, and 1ES~0229+200 repeatedly appear within these intermediate regions, indicating substantial long-term spectral variability. TXS~0506+056 is especially interesting because it has previously been interpreted as a
``masquerading'' BL~Lac object with hidden broad-line emission and an intrinsically radiatively efficient accretion disc \citep{10.1093/mnrasl/slz011}. Its repeated occupation of intermediate hard X-ray regimes further supports the idea that some blazars can display observational characteristics spanning multiple traditional subclasses. The long-term \textit{Swift}-XRT monitoring analysis similarly reveals substantial intra-source spectral variability across the blazar population. Sources such as 3C~66A, ON~231, PKS~2155$-$304, EXO~1811+3143, and OJ~287 show large photon-index excursions extending across multiple nominal
spectral regimes \citep{10.1093/mnras/staf1781, 2023MNRAS.tmp.2362K}. The common \textit{Swift}-XRT photon-index distribution also shows a strong concentration near the ISP regime, indicating that intermediate spectral states are relatively common rather than exceptional. 

The average X-ray photon indices derived in this work are broadly consistent with previous studies of blazar spectral populations. Earlier investigations reported characteristic mean photon indices of $\langle \Gamma_{\rm X} \rangle \approx 2.43 \pm 0.21$ for HBL populations and $\langle \Gamma_{\rm X} \rangle \approx 1.67 \pm 0.13$
for LBL/FSRQ-like sources \citep{comastri1997soft}. Similarly, \citet{2001yCat..33750739D} found average values of $\langle \Gamma_{\rm X} \rangle \approx 2.28$--$2.39$
for HBLs and $\langle \Gamma_{\rm X} \rangle \approx 1.76 \pm 0.06$ for LBL populations, while \citet{Kadler2005} reported typical quasar (FSRQ-like) photon indices near
$\langle \Gamma_{\rm X} \rangle \approx 1.61 \pm 0.20$. Moreover, \citet{giommi2019open} reported characteristic values near $\Gamma_{\rm X}\sim2.0$ for HSP
populations and $\Gamma_{\rm X}\sim1.6$ for LSP populations. These historical results are consistent with our multi-instrument analysis, where HBL/HSP-like sources preferentially occupy softer X-ray spectral regimes, whereas LBL/LSP/FSRQ-like sources remain concentrated toward harder photon-index distributions.

The broad spectral distributions identified in the present work are also consistent with earlier studies of X-ray blazar spectra. \citet{10.1093/mnras/284.3.569} showed that HBLs generally exhibit steeper X-ray spectra ($\langle \alpha_{\rm X,HBL} \rangle = 1.52 \pm 0.06$), while LBLs display flatter distributions
($\langle \alpha_{\rm X,LBL} \rangle = 1.06 \pm 0.09$),consistent with the harder spectra commonly observed in FSRQ-like populations. Motivated by similar spectral intervals, recent simulations by \citet{2025arXiv251025589H} modeled blazar-like spectra using photon indices spanning $\Gamma_{\rm X}=1.5$--$2.3$. Following these empirical trends and  classify the sample into hard-spectrum ($\Gamma<1.5$), intermediate-spectrum
($1.5\leq\Gamma_{\rm X}\leq2.0$), and soft-spectrum ($\Gamma_{\rm X}>2.0$) regimes. Within the 31,068 \textit{Swift}-XRT observations analyzed here,
approximately 55\% occupy the hard-spectrum regime, 26\% lie within the intermediate region, and 19\% populate the soft-spectrum domain.

The hard X-ray \textit{Swift}-BAT and RXTE catalogs extend
this spectral picture beyond the blazar population alone. In
both datasets, Seyfert galaxies, BL~Lac objects, FSRQs, and
NLSy1 galaxies occupy partially overlapping regions of
photon-index space. However, the diagnostic utility of
hardness ratio depends strongly on energy coverage. In the
\textit{Swift}-BAT band, the HR--$\Gamma_{\rm X}$ relation shows
substantial scatter and only a weak anti-correlation, indicating
limited population separation once absorption becomes less
important above $\sim15$~keV. By contrast, the \textit{Chandra}
soft X-ray analysis reveals a clear bimodal hardness-ratio
distribution separating obscured and unobscured populations.
These results demonstrate that hardness ratio is strongly
energy dependent and should not be interpreted as a universal
classifier across all wavebands.\\

The TeVCat analysis provides an important extension of this picture into the very-high-energy (VHE) $\gamma$-ray regime. The TeV photon-index distributions exhibit a gradual progression from EHBL-like hard-spectrum populations toward softer IBL, LBL, and FSRQ-like regimes, broadly consistent with the phenomenological blazar sequence. Nevertheless, substantial overlap remains present between subclasses, particularly within the approximate range $\Gamma_{\gamma,\mathrm{TeV}} \sim 2.5$--$3.5$, where multiple populations occupy partially shared spectral states  Figure~\ref{fig:tevonly}. Sources such as 1ES~0229+200, RGB~J0710+591, 1ES~1101-232, 1ES~0414+009, and H~1426+428 populate intermediate TeV spectral regimes linking EHBL and HBL-like populations \citep{banerjee2023detection,Hota_2024}. Similarly, H~2356$-$309 exhibits a TeV photon index near $\Gamma_{\gamma,\rm TeV} \sim 3.06$, consistent with the EHBL/HBL overlap region identified in the present analysis \citep{2010A&A...516A..56H}. The broad overlap observed across TeV subclasses indicates that blazar spectral properties occupy partially continuous observational distributions rather than forming sharply separated populations. This behavior may reflect gradual variations in particle acceleration efficiency, radiative cooling, inverse-Compton dominance, and external photon-field contributions across different jet environments. However, because TeV photon indices are strongly influenced by variability, spectral curvature, extragalactic background light absorption, and instrument-dependent energy coverage, the inferred transition ranges should be interpreted as empirical observational trends rather than definitive physical subclass boundaries.

The historical VizieR multi-satellite compilation demonstrates that the observed spectral overlap is not unique to modern observatories. Archival observations from ROSAT, ASCA,
EXOSAT, Einstein, and BeppoSAX reveal similarly broad
photon-index distributions across blazar subclasses,
indicating that the observed spectral continuity is a persistent
feature of the blazar population rather than an artifact of a
single mission or survey strategy. In particular, the confirmed
transition blazar S5~1803+784 exhibits a substantial change in
photon index from $\Gamma_{\rm X}=1.45$ during the BeppoSAX epoch
to $\Gamma_{\rm X}=2.42$ during the ROSAT epoch despite nearly
identical X-ray flux levels
($S_{\rm X}\approx0.22$--$0.26$). Such behavior suggests that
significant spectral-state changes can occur without large
variations in observed X-ray flux, possibly reflecting changes
in the relative contributions of synchrotron and inverse-Compton
emission components.

Our AstroSat, XMM-Newton, and NICER analysis further
demonstrates the importance of long-term spectral variability
within individual sources. Mrk~421 shows both the steepest
mean spectrum ($\alpha_{\rm X} = 2.51$) and the largest variability
amplitude ($\Delta\alpha_{\rm X}=0.98$), evolving from
$\alpha_{\rm X}\approx2.9$ to $\alpha_{\rm X}\approx2.4$ across different
epochs. In contrast, RGB~J0710+591 and 1ES~1218+304
maintain comparatively hard spectra near
$\alpha_{\rm X}\approx1.8$, suggesting persistently efficient particle
acceleration. OJ~287 displays three distinct spectral states
($\alpha_{\rm X}=1.86$, 2.16, and 2.55), consistent with episodic
hard-to-soft spectral evolution potentially associated with its
proposed binary supermassive black-hole system
\citep{galaxies10010001}. The absence of any strong
$\alpha_{\rm X}$--redshift correlation indicates that intrinsic jet
physics dominates over cosmological effects in shaping the
observed X-ray spectra.

\begin{figure*}[htbp]
\centering
\includegraphics[width=\textwidth]{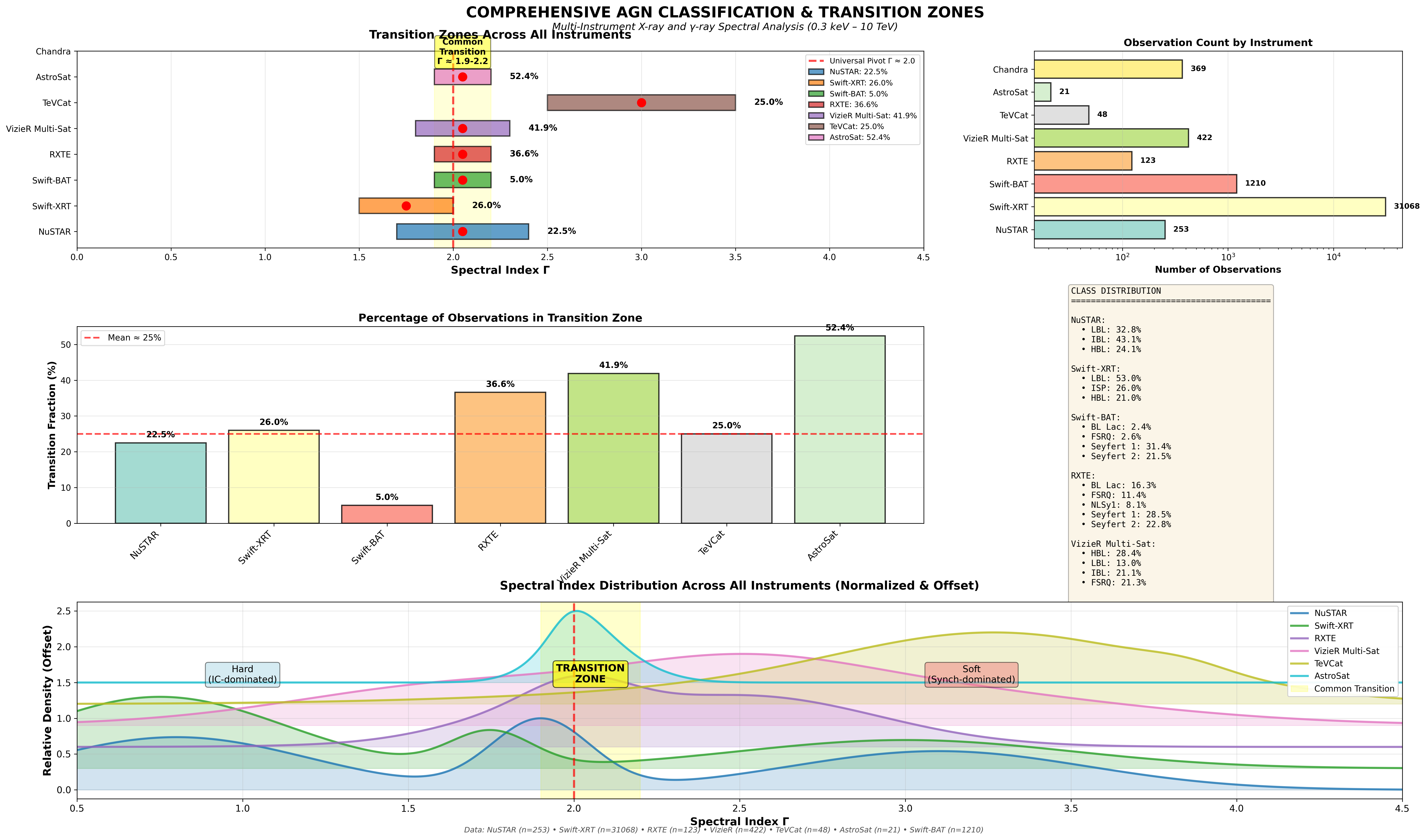}
\caption{Comprehensive comparison of spectral-index distributions across multiple X-ray and $\gamma$-ray instruments. The shaded region ($\Gamma \sim 1.9$--$2.2$) marks a common spectral transition regime between hard Compton-dominated and soft synchrotron-dominated states. The panels show instrument-wise transition fractions, observation statistics, AGN subclass distributions, and normalized spectral-density profiles derived from NuSTAR, Swift-XRT, Swift-BAT, RXTE, VizieR multi-satellite catalogs, TeVCat, AstroSat}
\label{fig:agn_transition}
\end{figure*}

The most compelling evidence for transient spectral evolution is
provided by Mrk~421. Early AstroSat observations are
characterized by steep spectra
($\alpha_{\rm X}\approx2.8$--$3.0$), whereas later epochs show
substantially harder states
($\alpha_{\rm X}\approx2.1$--$2.3$). The broad distribution of NICER
measurements demonstrates that the source repeatedly returns
to both hard and soft spectral states over time, favoring a
scenario involving recurrent particle injection and synchrotron
cooling rather than permanent structural changes within the
jet. This behavior is consistent with the well-known
``harder-when-brighter'' trend observed in TeV blazars
\citep{Abdo_2011, 2015A&A...576A.126A,2016ApJ...824..108L,2016ApJ...819..156B,Hota:2021csa,10.1093/mnras/stae643,AKBAR2025438,Kizhakkekalam:2025moz}.\\
The multi-instrument spectral-index distributions presented in Figure \ref{fig:agn_transition} and summarized in Table \ref{tab:multiinstrument_revised} reveal substantial overlap between traditionally separated blazar populations across different X-ray and TeV $\gamma$-ray energy bands. Rather than occupying sharply distinct photon-index intervals  HBL, IBL, LBL, and FSRQ-like sources populate partially continuous spectral regimes, with the intermediate region concentrated around ($\Gamma_{\rm X} \sim 1.5$--$2.2$). This overlap is consistently observed across NuSTAR, Swift-XRT, ROSAT, VizieR multi-satellite samples, AstroSat, although the precise distributions vary with instrumental energy coverage, source selection effects, and the different physical emission components probed at X-ray . In the TeVCat sample, EHBL and HBL populations predominantly occupy hard TeV spectral regimes, while IBL, LBL, and FSRQ-like sources extend toward softer inverse-Compton-dominated states. The observed concentration within ($2.5 \lesssim \Gamma_{\gamma,\mathrm{TeV}} \lesssim 4$) may therefore reflect a broad phenomenological transition between synchrotron-dominated and inverse-Compton-dominated emission regimes, consistent with gradual variations in jet energetics, particle acceleration efficiency, and radiative cooling processes. These results support a continuum-like interpretation of blazar spectral properties rather than a strictly discrete classification scheme. However, the inferred transition ranges should be regarded as empirical observational trends rather than definitive physical boundaries, since instrumental biases, temporal variability, and spectral-model dependencies may influence the measured photon-index distributions. Moreover, the combined multi-instrument results support a picture in which blazar subclasses occupy broad and partially
overlapping spectral distributions shaped by variability, energy-dependent radiative processes, and long-term changes in jet emission properties. Within this framework, the photon index serves as an effective empirical tracer of spectral-state occupation across multiple wavebands. 

\begin{table*}[htbp]
\centering
\scriptsize
\caption{Empirical spectral-regime distributions and variability characteristics of blazar/AGN populations across multiple X-ray and TeV instruments.}
\label{tab:multiinstrument_revised}

\renewcommand{\arraystretch}{1.15}

\begin{tabular}{l l l r r l l}
\toprule

\textbf{Instrument} &
\textbf{Energy Band} &
\textbf{Population / Sample} &
\textbf{$N_{\rm obs}$} &
\textbf{Fraction (\%)} &
\textbf{Observed Spectral Range} &
\textbf{Dominant Spectral Regime} \\

\midrule

\multirow{3}{*}{NuSTAR} &
\multirow{3}{*}{3--79 keV} &
LBL-like population &
83 &
32.8 &
$\Gamma \lesssim 1.6$ &
Hard hard-X-ray regime \\

& &
Intermediate spectral population &
109 &
43.1 &
$1.6 \lesssim \Gamma \lesssim 2.2$ &
Intermediate hard-X-ray regime \\

& &
HBL-like population &
61 &
24.1 &
$\Gamma \gtrsim 2.2$ &
Soft hard-X-ray regime \\

\midrule

\multirow{3}{*}{Swift-XRT} &
\multirow{3}{*}{0.3--10 keV} &
Hard-spectrum region &
16500 &
53.0 &
$\Gamma < 1.5$ &
Hard soft-X-ray regime \\

& &
Intermediate-spectrum region &
8000 &
26.0 &
$1.5 \leq \Gamma \leq 2.0$ &
Intermediate soft-X-ray regime \\

& &
Soft-spectrum region &
6568 &
21.0 &
$\Gamma > 2.0$ &
Soft soft-X-ray regime \\

\midrule

\multirow{4}{*}{Swift-BAT} &
\multirow{4}{*}{15--195 keV} &
BL Lac &
29 &
2.4 &
$\Gamma \sim 1.45$--$3.24$ &
Broad hard-X-ray spectral distribution \\

& &
FSRQ &
32 &
2.6 &
$\Gamma \sim 1.27$--$2.72$ &
Hard/intermediate spectral regime \\

& &
Seyfert 1 &
380 &
31.4 &
$\Gamma \sim 1.20$--$3.00$ &
Intermediate hard-X-ray regime \\

& &
Seyfert 2 &
260 &
21.5 &
$\Gamma \sim 1.30$--$3.50$ &
Broad intermediate/soft regime \\

\midrule

\multirow{5}{*}{RXTE} &
\multirow{5}{*}{2--10 keV} &
BL Lac &
20 &
16.3 &
$\Gamma \sim 1.83$--$3.20$ &
Broad soft-spectrum distribution \\

& &
FSRQ &
14 &
11.4 &
$\Gamma \sim 1.35$--$2.21$ &
Hard/intermediate spectral regime \\

& &
NLSy1 &
10 &
8.1 &
$\Gamma \sim 2.11$--$3.40$ &
Very soft-spectrum regime \\

& &
Seyfert 1 &
35 &
28.5 &
$\Gamma \sim 1.70$--$2.33$ &
Intermediate-spectrum regime \\

& &
Seyfert 2 &
28 &
22.8 &
$\Gamma \sim 1.32$--$3.30$ &
Broad intermediate/soft regime \\

\midrule

\multirow{4}{*}{VizieR Multi-Satellite} &
\multirow{4}{*}{0.5--10 keV} &
HBL &
120 &
28.4 &
$\Gamma \sim 1.52$--$4.28$ &
Broad soft-spectrum distribution \\

& &
LBL &
55 &
13.0 &
$\Gamma \sim 0.60$--$3.75$ &
Intermediate/hard spectral regime \\

& &
IBL &
89 &
21.1 &
$\Gamma \sim 1.80$--$3.20$ &
Intermediate soft-spectrum regime \\

& &
FSRQ &
90 &
21.3 &
$\Gamma \sim 0.66$--$2.50$ &
Hard/intermediate spectral regime \\

\midrule

\multirow{5}{*}{TeVCat} &
\multirow{5}{*}{$>$100 GeV} &
HBL &
31 &
64.6 &
$\Gamma \sim 2.20$--$4.40$ &
Broad TeV soft-spectrum regime \\

& &
EHBL &
5 &
10.4 &
$\Gamma \sim 1.00$--$3.63$ &
Hard TeV spectral regime \\

& &
IBL &
4 &
8.3 &
$\Gamma \sim 3.60$--$4.10$ &
Soft TeV spectral regime \\

& &
LBL &
1 &
2.1 &
$\Gamma \sim 3.67$ &
Intermediate soft-spectrum regime \\

& &
FSRQ &
7 &
14.6 &
$\Gamma \sim 2.45$--$4.10$ &
Soft inverse-Compton dominated regime \\

\midrule

\multirow{2}{*}{AstroSat / NICER / XMM} &
\multirow{2}{*}{0.2--80 keV} &
HBL-dominated variability sample &
66 &
-- &
$\Gamma \sim 2.7$--$4.0$ &
Soft synchrotron-dominated regime \\

& &
Blazar spectral variability study &
7 sources &
-- &
Source-dependent &
Multi-epoch spectral variability analysis \\

\midrule

\multirow{3}{*}{Chandra} &
\multirow{3}{*}{0.5--7 keV} &
Obscured AGN &
146 &
39.6 &
HR $\sim 0.3$--$1.0$ &
Hard absorbed regime \\

& &
AGN &
85 &
23.0 &
HR $\sim -0.2$--$0.3$ &
Intermediate hardness regime \\

& &
Soft sources &
138 &
37.4 &
HR $\sim -1.0$--$-0.2$ &
Soft X-ray regime \\

\bottomrule
\end{tabular}

\vspace{0.2cm}

\begin{minipage}{0.96\textwidth}
\footnotesize
\textbf{Note:}
The quoted photon-index and hardness-ratio ranges represent empirical observational distributions derived from archival multi-instrument catalogs and targeted variability studies. The spectral-regime terminology is phenomenological and instrument dependent, reflecting differences in energy coverage, dominant emission processes, and spectral variability. These classifications should not be interpreted as strict physical subclass boundaries or definitive evolutionary transitions between blazar populations.
\end{minipage}

\end{table*}

\section{Conclusions}

We have conducted a systematic multi-instrument X-ray and TeV spectral census of blazars using archival observations from \textit{NuSTAR}, \textit{Swift}-XRT, \textit{Swift}-BAT,  \textit{AstroSat}, \textit{XMM-Newton},  \textit{NICER}, \textit{Chandra},  \textit{ROSAT}, and  \textit{TeVCat}. The combined datasets reveal substantial overlap between traditionally defined blazar subclasses across multiple energy bands. A significant fraction of observations occupy intermediate spectral regimes near $\Gamma_{\rm X} \approx 2$, with the precise boundaries varying modestly between instruments because of energy-dependent spectral sampling. The high occupancy of intermediate spectral regimes ($\sim 22$--$43\%$) across instruments) and substantial intra-source variability ($\Delta\Gamma_{\rm X} > 0.5$ in 16 sources) suggest that transition-like blazar behavior may represent a relatively common phenomenological state within the blazar population rather than a rare exception. These intermediate regions are consistently present across soft X-ray, hard X-ray, and TeV observations, supporting the interpretation that blazar spectral properties are distributed continuously rather than forming sharply separated subclasses.

Multi-epoch observations further reveal substantial intra-source spectral variability. Sources such as Mrk~421 exhibit smooth, continuous stochastic spectral evolution, while OJ~287 shows evidence for more discrete state-dependent spectral behavior. Several sources repeatedly occupy intermediate spectral regimes linking EHBL, HBL, IBL, LBL, and FSRQ-like populations, making them strong candidates for transition-like spectral behavior. In this context, the term ``transition-like'' refers to observational occupation of multiple spectral regimes over time and should not be interpreted as definitive evidence for irreversible physical evolution between established blazar subclasses.

The TeVCat and \textit{Chandra} analyses independently support this picture by showing broad overlap between spectral-index and hardness-ratio distributions across different source populations. These results indicate that empirical observational diagnostics, such as photon index and hardness ratio, can effectively trace spectral-regime diversity across multiple wavebands. The concentration of sources within intermediate spectral regimes may reflect gradual shifts in the relative dominance of synchrotron and inverse-Compton emission components driven by long-term changes in jet energetics, particle acceleration efficiency, and radiative cooling processes.

Overall, the combined multi-instrument results favor a phenomenological framework in which blazar subclasses represent overlapping spectral-regime populations shaped by variability, jet emission, and energy-dependent radiative processes rather than strictly discrete source categories. Future coordinated multiwavelength observations combining X-ray, optical, radio, and $\gamma$-ray monitoring will be essential for clarifying how these observed spectral variations relate to the underlying jet physics, accretion state, and particle acceleration processes.

\clearpage 
\clearpage 
\clearpage
\clearpage
\begin{figure*}[htbp]
\centering
\includegraphics[width=\textwidth]{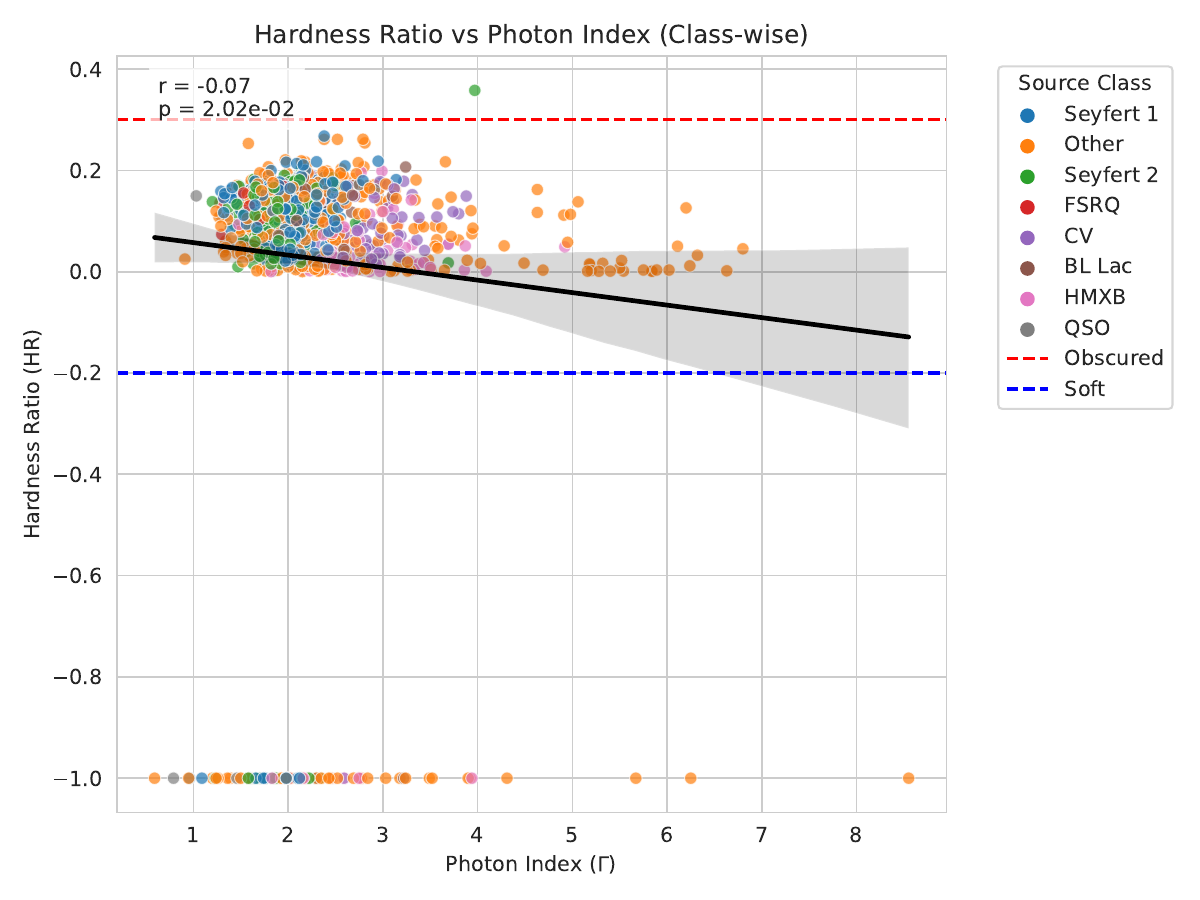}
\caption{ Relationship between hardness ratio (HR) and photon index ($\Gamma_{\mathrm{X}}$) for different source classes, including Seyfert 1, Seyfert 2, blazars (BL Lac and FSRQ), X-ray binaries, QSOs, and other populations. The panels highlight class-dependent clustering and a weak global anti-correlation between HR and $\Gamma_{\mathrm{X}}$ in the Swift-BAT sample.}
\label{fig:hr_gamma}
\end{figure*}
\clearpage
\clearpage
\clearpage
\clearpage
\begin{figure*}[htbp]
\centering
\includegraphics[width=\textwidth]{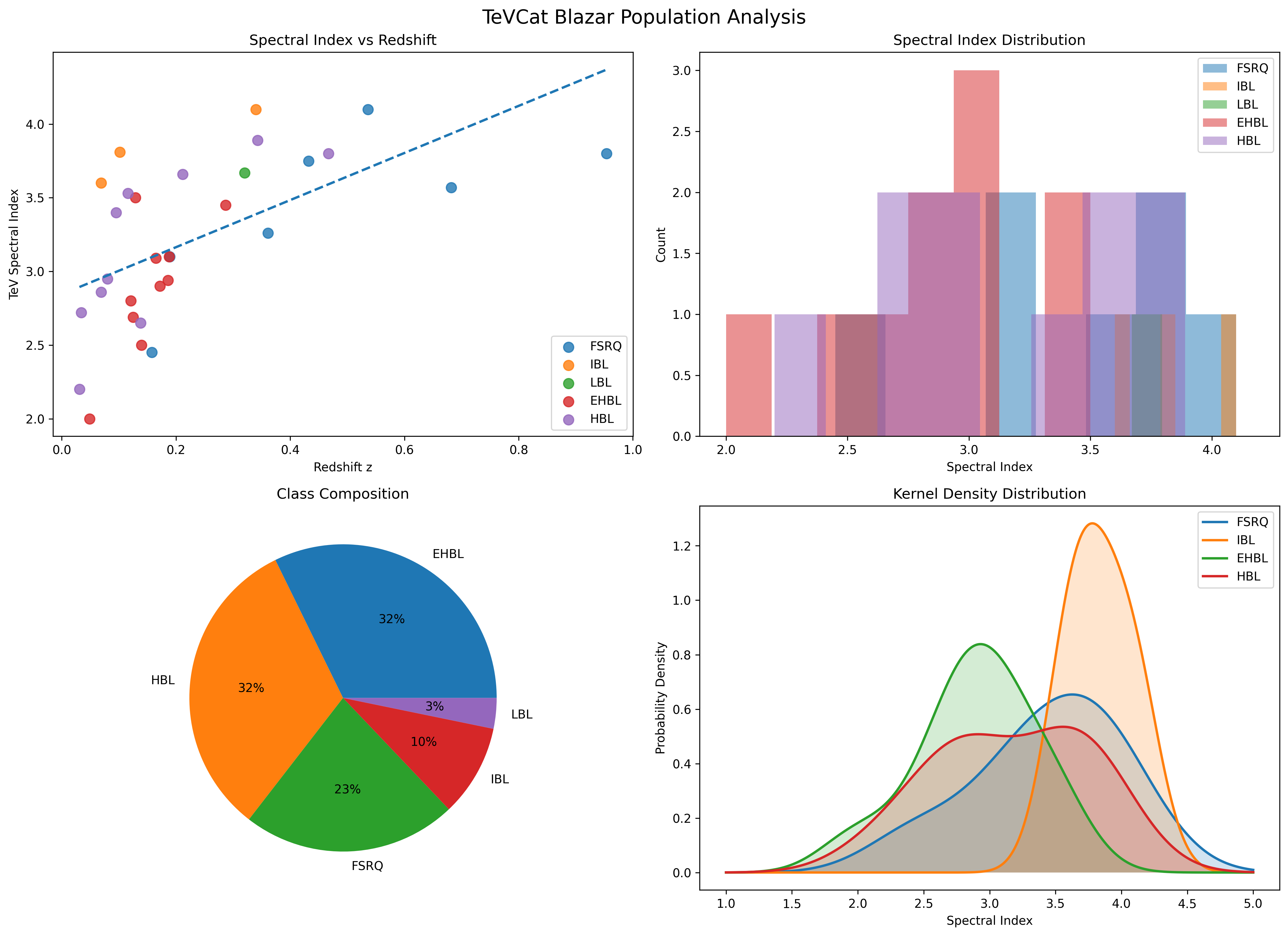}
\caption{ Multi-panel overview of TeVCat blazar populations showing the distributions of TeV spectral index, redshift, class composition, and kernel-density profiles for EHBL, HBL, IBL, LBL, and FSRQ sources. The figure highlights overlapping spectral regimes and class-dependent TeV spectral behavior across the blazar population.}
\label{fig:TEV}
\end{figure*}
\clearpage
\clearpage
\begin{figure*}[htbp]
    \centering
    \includegraphics[width=\textwidth]{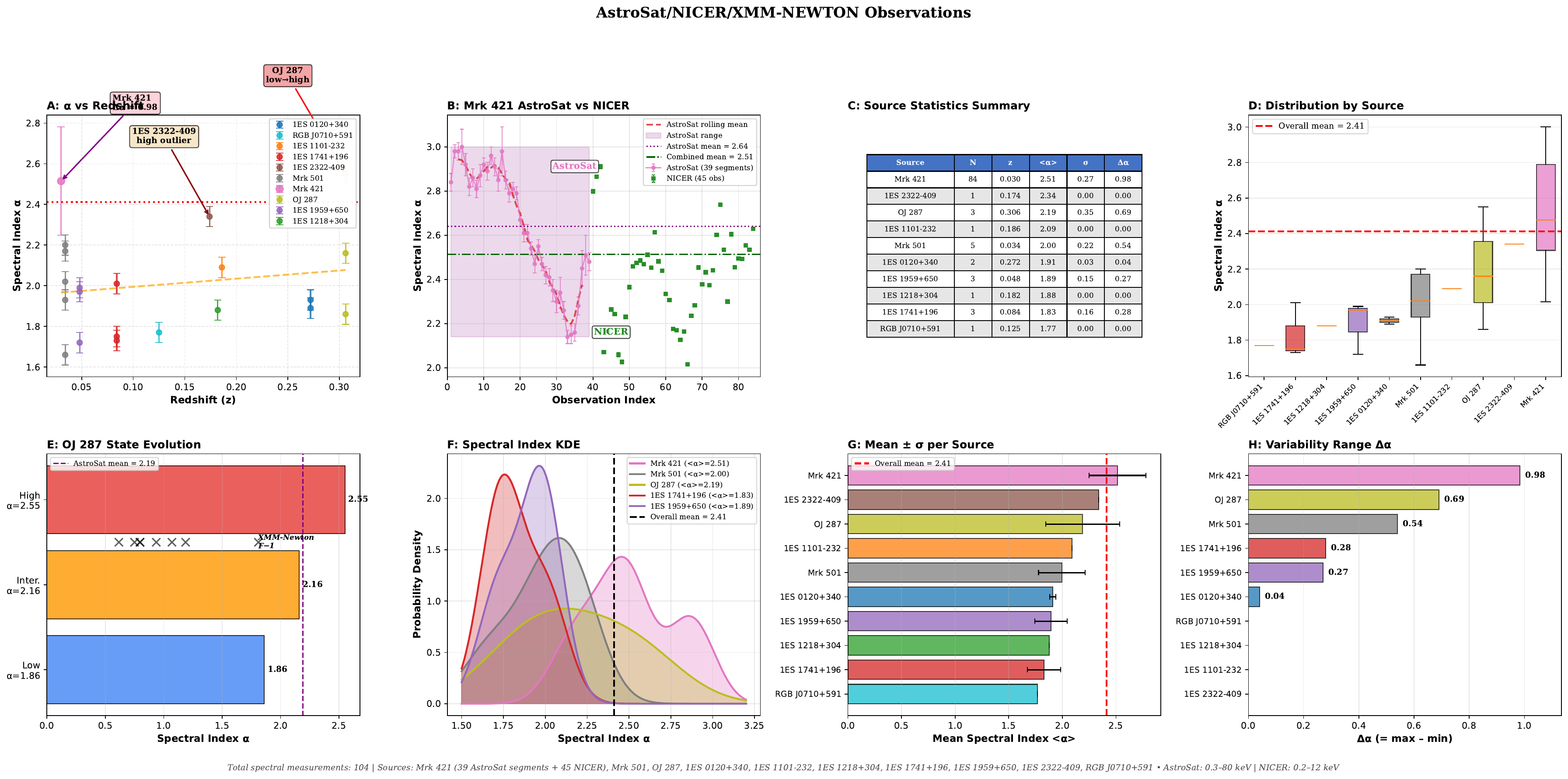}
    \caption{Multi-panel summary of X-ray spectral variability in blazars from AstroSat, NICER, and XMM--Newton observations, showing the dependence of spectral index on redshift, source-wise distributions, state evolution, kernel density distributions, mean spectral indices, and variability ranges across different blazar populations.}
    \label{fig:spectral_index}
\end{figure*}
\clearpage
\clearpage

\newcommand{\aap}{Astronomy \& Astrophysics}
\newcommand{\apj}{The Astrophysical Journal}
\newcommand{\ssr}{Space Science Reviews}
\newcommand{\mnras}{Monthly Notices of the Royal Astronomical Society}
\newcommand{\apjl}{The Astrophysical Journal Letters}
\newcommand{\pasp}{Publications of the Astronomical Society of the Pacific}
\newcommand{\aj}{The Astronomical Journal}
\newcommand{\nat}{Nature}
\newcommand{\araa}{Annual Review of Astronomy and Astrophysics}
\newcommand{\pasj}{Publications of the Astronomical Society of Japan}
\newcommand{\na}{New Astronomy}
\newcommand{\nar}{New Astronomy Reviews}
\newcommand{\jcap}{Journal of Cosmology and Astroparticle Physics}
\newcommand{\physrep}{Physics Reports}
\newcommand{\prd}{Physical Review D}
\newcommand{\prl}{Physical Review Letters}
\newcommand{\rmxaa}{Revista Mexicana de Astronomía y Astrofísica}
\newcommand{\aapr}{Astronomy and Astrophysics Review}
\newcommand{\memsai}{Memorie della Società Astronomica Italiana}
\newcommand{\apss}{Astrophysics and Space Science}

\section*{Acknowledgements}

This research has made use of data from the NuBlazar catalogue \citep{2022MNRAS.514.3179M}, the Swift-XRT blazar monitoring programme \citep{2021MNRAS.507.5690G}, the RXTE/PCA AGN archive, the TeVCat online database, and archival observations from ROSAT, ASCA, BeppoSAX, EXOSAT, Einstein, Chandra, Swift-BAT, AstroSat, XMM-Newton, and NICER.

J.~Tantry gratefully acknowledges financial support from the Indian Space Research Organisation (ISRO), Department of Space, Government of India, through the ISRO--RESPOND programme (Grant No.~DS\_2B-13013(2)/8/2020-Sec.2). The authors also express their gratitude to the Inter-University Centre for Astronomy and Astrophysics (IUCAA), Pune, India, for providing research support and computational facilities.

\bibliographystyle{elsarticle-harv}
\bibliography{example}

\begin{thebibliography}{55}
\expandafter\ifx\csname natexlab\endcsname\relax\def\natexlab#1{#1}\fi
\providecommand{\url}[1]{\texttt{#1}}
\providecommand{\href}[2]{#2}
\providecommand{\path}[1]{#1}
\providecommand{\DOIprefix}{doi:}
\providecommand{\ArXivprefix}{arXiv:}
\providecommand{\URLprefix}{URL: }
\providecommand{\Pubmedprefix}{pmid:}
\providecommand{\doi}[1]{\href{http://dx.doi.org/#1}{\path{#1}}}
\providecommand{\Pubmed}[1]{\href{pmid:#1}{\path{#1}}}
\providecommand{\bibinfo}[2]{#2}
\ifx\xfnm\relax \def\xfnm[#1]{\unskip,\space#1}\fi
\bibitem[{Abdo et~al.(2011)Abdo, Ackermann, Ajello, Baldini, Ballet,
  Barbiellini, Bastieri, Bechtol, Bellazzini, Berenji, Blandford, Bloom,
  Bonamente, Borgland, Bouvier, Bregeon, Brez, Brigida, Bruel, Buehler, Buson,
  Caliandro, Cameron, Cannon, Caraveo, Carrigan, Casandjian, Cavazzuti, Cecchi,
  Çelik, Charles, Chekhtman, Chiang, Ciprini, Claus, Cohen-Tanugi, Conrad,
  Cutini, de~Angelis, de~Palma, Dermer, do~Couto~e Silva, Drell, Dubois,
  Dumora, Escande, Favuzzi, Fegan, Finke, Focke, Fortin, Frailis, Fuhrmann,
  Fukazawa, Fukuyama, Funk, Fusco, Gargano, Gasparrini, Gehrels,
  Georganopoulos, Germani, Giebels, Giglietto, Giommi, Giordano, Giroletti,
  Glanzman, Godfrey, Grenier, Guiriec, Hadasch, Hayashida, Hays, Horan, Hughes,
  Jóhannesson, Johnson, Johnson, Kadler, Kamae, Katagiri, Kataoka,
  Knödlseder, Kuss, Lande, Latronico, Lee, Longo, Loparco, Lott, Lovellette,
  Lubrano, Madejski, Makeev, Max-Moerbeck, Mazziotta, McEnery, Mehault,
  Michelson, Mitthumsiri, Mizuno, Monte, Monzani, Morselli, Moskalenko, Murgia,
  Nakamori, Naumann-Godo, Nishino, Nolan, Norris, Nuss, Ohsugi, Okumura,
  Omodei, Orlando, Ormes, Ozaki, Paneque, Panetta, Parent, Pavlidou, Pearson,
  Pelassa, Pepe, Pesce-Rollins, Pierbattista, Piron, Porter, Rainò, Rando,
  Razzano, Readhead, Reimer, Reimer, Reyes, Richards, Ritz, Roth, Sadrozinski,
  Sanchez, Sander, Sgrò, Siskind, Smith, Spandre, Spinelli, Stawarz,
  Stevenson, Strickman, Suson, Takahashi, Takahashi, Tanaka, Thayer, Thayer,
  Thompson, Tibaldo, Torres, Tosti, Tramacere, Troja, Usher, Vandenbroucke,
  Vasileiou, Vianello, Vilchez, Vitale, Waite, Wang, Wehrle, Winer, Wood, Yang,
  Yatsu, Ylinen, Zensus, Ziegler, Collaboration), Aleksić, Antonelli,
  Antoranz, Backes, Barrio, González, Bednarek, Berdyugin, Berger, Bernardini,
  Biland, Blanch, Bock, Boller, Bonnoli, Bordas, Tridon, Bosch-Ramon, Bose,
  Braun, Bretz, Camara, Carmona, Carosi, Colin, Colombo, Contreras, Cortina,
  Covino, Dazzi, de~Angelis, De~Cea~del Pozo, Mendez, De~Lotto, De~Maria,
  De~Sabata, Ortega, Doert, Domínguez, Prester, Dorner, Doro, Elsaesser,
  Ferenc, Fonseca, Font, López, Garczarczyk, Gaug, Giavitto, Godinovi,
  Hadasch, Herrero, Hildebrand, Höhne-Mönch, Hose, Hrupec, Jogler, Klepser,
  Krähenbühl, Kranich, Krause, Barbera, Leonardo, Lindfors, Lombardi, López,
  Lorenz, Majumdar, Makariev, Maneva, Mankuzhiyil, Mannheim, Maraschi,
  Mariotti, Martínez, Mazin, Meucci, Miranda, Mirzoyan, Miyamoto, Moldón,
  Moralejo, Nieto, Nilsson, Orito, Oya, Paoletti, Paredes, Partini, Pasanen,
  Pauss, Pegna, Perez-Torres, Persic, Peruzzo, Pochon, Prada, Moroni, Prandini,
  Puchades, Puljak, Reichardt, Rhode, Ribó, Rico, Rissi, Rügamer, Saggion,
  Saito, Saito, Salvati, Sánchez-Conde, Satalecka, Scalzotto, Scapin, Schultz,
  Schweizer, Shayduk, Shore, Sierpowska-Bartosik, Sillanpää, Sitarek,
  Sobczynska, Spanier, Spiro, Stamerra, Steinke, Storz, Strah, Struebig, Suric,
  Takalo, Tavecchio, Temnikov, Terzić, Tescaro, Teshima, Vankov, Wagner,
  Weitzel, Zabalza, Zandanel, Zanin, Collaboration), Villata, Raiteri, Aller,
  Aller, Chen, Jordan, Koptelova, Kurtanidze, Lähteenmäki, McBreen, Larionov,
  Lin, Nikolashvili, Reinthal, Angelakis, Capalbi, Carramiñana, Carrasco,
  Cassaro, Cesarini, Falcone, Gurwell, Hovatta, Kovalev, Kovalev, Krichbaum,
  Krimm, Lister, Moody, Maccaferri, Mori, Nestoras, Orlati, Pace, Pagani,
  Pearson, Perri, Piner, Ros, Sadun, Sakamoto, Tammi and Zook}]{Abdo_2011}
\bibinfo{author}{Abdo, A.A.}, \bibinfo{author}{Ackermann, M.},
  \bibinfo{author}{Ajello, M.}, \bibinfo{author}{Baldini, L.},
  \bibinfo{author}{Ballet, J.}, \bibinfo{author}{Barbiellini, G.},
  \bibinfo{author}{Bastieri, D.}, \bibinfo{author}{Bechtol, K.},
  \bibinfo{author}{Bellazzini, R.}, \bibinfo{author}{Berenji, B.},
  \bibinfo{author}{Blandford, R.D.}, \bibinfo{author}{Bloom, E.D.},
  \bibinfo{author}{Bonamente, E.}, \bibinfo{author}{Borgland, A.W.},
  \bibinfo{author}{Bouvier, A.}, \bibinfo{author}{Bregeon, J.},
  \bibinfo{author}{Brez, A.}, \bibinfo{author}{Brigida, M.},
  \bibinfo{author}{Bruel, P.}, \bibinfo{author}{Buehler, R.},
  \bibinfo{author}{Buson, S.}, \bibinfo{author}{Caliandro, G.A.},
  \bibinfo{author}{Cameron, R.A.}, \bibinfo{author}{Cannon, A.},
  \bibinfo{author}{Caraveo, P.A.}, \bibinfo{author}{Carrigan, S.},
  \bibinfo{author}{Casandjian, J.M.}, \bibinfo{author}{Cavazzuti, E.},
  \bibinfo{author}{Cecchi, C.}, \bibinfo{author}{Çelik, Ã.},
  \bibinfo{author}{Charles, E.}, \bibinfo{author}{Chekhtman, A.},
  \bibinfo{author}{Chiang, J.}, \bibinfo{author}{Ciprini, S.},
  \bibinfo{author}{Claus, R.}, \bibinfo{author}{Cohen-Tanugi, J.},
  \bibinfo{author}{Conrad, J.}, \bibinfo{author}{Cutini, S.},
  \bibinfo{author}{de~Angelis, A.}, \bibinfo{author}{de~Palma, F.},
  \bibinfo{author}{Dermer, C.D.}, \bibinfo{author}{do~Couto~e Silva, E.},
  \bibinfo{author}{Drell, P.S.}, \bibinfo{author}{Dubois, R.},
  \bibinfo{author}{Dumora, D.}, \bibinfo{author}{Escande, L.},
  \bibinfo{author}{Favuzzi, C.}, \bibinfo{author}{Fegan, S.J.},
  \bibinfo{author}{Finke, J.}, \bibinfo{author}{Focke, W.B.},
  \bibinfo{author}{Fortin, P.}, \bibinfo{author}{Frailis, M.},
  \bibinfo{author}{Fuhrmann, L.}, \bibinfo{author}{Fukazawa, Y.},
  \bibinfo{author}{Fukuyama, T.}, \bibinfo{author}{Funk, S.},
  \bibinfo{author}{Fusco, P.}, \bibinfo{author}{Gargano, F.},
  \bibinfo{author}{Gasparrini, D.}, \bibinfo{author}{Gehrels, N.},
  \bibinfo{author}{Georganopoulos, M.}, \bibinfo{author}{Germani, S.},
  \bibinfo{author}{Giebels, B.}, \bibinfo{author}{Giglietto, N.},
  \bibinfo{author}{Giommi, P.}, \bibinfo{author}{Giordano, F.},
  \bibinfo{author}{Giroletti, M.}, \bibinfo{author}{Glanzman, T.},
  \bibinfo{author}{Godfrey, G.}, \bibinfo{author}{Grenier, I.A.},
  \bibinfo{author}{Guiriec, S.}, \bibinfo{author}{Hadasch, D.},
  \bibinfo{author}{Hayashida, M.}, \bibinfo{author}{Hays, E.},
  \bibinfo{author}{Horan, D.}, \bibinfo{author}{Hughes, R.E.},
  \bibinfo{author}{Jóhannesson, G.}, \bibinfo{author}{Johnson, A.S.},
  \bibinfo{author}{Johnson, W.N.}, \bibinfo{author}{Kadler, M.},
  \bibinfo{author}{Kamae, T.}, \bibinfo{author}{Katagiri, H.},
  \bibinfo{author}{Kataoka, J.}, \bibinfo{author}{Knödlseder, J.},
  \bibinfo{author}{Kuss, M.}, \bibinfo{author}{Lande, J.},
  \bibinfo{author}{Latronico, L.}, \bibinfo{author}{Lee, S.H.},
  \bibinfo{author}{Longo, F.}, \bibinfo{author}{Loparco, F.},
  \bibinfo{author}{Lott, B.}, \bibinfo{author}{Lovellette, M.N.},
  \bibinfo{author}{Lubrano, P.}, \bibinfo{author}{Madejski, G.M.},
  \bibinfo{author}{Makeev, A.}, \bibinfo{author}{Max-Moerbeck, W.},
  \bibinfo{author}{Mazziotta, M.N.}, \bibinfo{author}{McEnery, J.E.},
  \bibinfo{author}{Mehault, J.}, \bibinfo{author}{Michelson, P.F.},
  \bibinfo{author}{Mitthumsiri, W.}, \bibinfo{author}{Mizuno, T.},
  \bibinfo{author}{Monte, C.}, \bibinfo{author}{Monzani, M.E.},
  \bibinfo{author}{Morselli, A.}, \bibinfo{author}{Moskalenko, I.V.},
  \bibinfo{author}{Murgia, S.}, \bibinfo{author}{Nakamori, T.},
  \bibinfo{author}{Naumann-Godo, M.}, \bibinfo{author}{Nishino, S.},
  \bibinfo{author}{Nolan, P.L.}, \bibinfo{author}{Norris, J.P.},
  \bibinfo{author}{Nuss, E.}, \bibinfo{author}{Ohsugi, T.},
  \bibinfo{author}{Okumura, A.}, \bibinfo{author}{Omodei, N.},
  \bibinfo{author}{Orlando, E.}, \bibinfo{author}{Ormes, J.F.},
  \bibinfo{author}{Ozaki, M.}, \bibinfo{author}{Paneque, D.},
  \bibinfo{author}{Panetta, J.H.}, \bibinfo{author}{Parent, D.},
  \bibinfo{author}{Pavlidou, V.}, \bibinfo{author}{Pearson, T.J.},
  \bibinfo{author}{Pelassa, V.}, \bibinfo{author}{Pepe, M.},
  \bibinfo{author}{Pesce-Rollins, M.}, \bibinfo{author}{Pierbattista, M.},
  \bibinfo{author}{Piron, F.}, \bibinfo{author}{Porter, T.A.},
  \bibinfo{author}{Rainò, S.}, \bibinfo{author}{Rando, R.},
  \bibinfo{author}{Razzano, M.}, \bibinfo{author}{Readhead, A.},
  \bibinfo{author}{Reimer, A.}, \bibinfo{author}{Reimer, O.},
  \bibinfo{author}{Reyes, L.C.}, \bibinfo{author}{Richards, J.L.},
  \bibinfo{author}{Ritz, S.}, \bibinfo{author}{Roth, M.},
  \bibinfo{author}{Sadrozinski, H.F.W.}, \bibinfo{author}{Sanchez, D.},
  \bibinfo{author}{Sander, A.}, \bibinfo{author}{Sgrò, C.},
  \bibinfo{author}{Siskind, E.J.}, \bibinfo{author}{Smith, P.D.},
  \bibinfo{author}{Spandre, G.}, \bibinfo{author}{Spinelli, P.},
  \bibinfo{author}{Stawarz, Å.}, \bibinfo{author}{Stevenson, M.},
  \bibinfo{author}{Strickman, M.S.}, \bibinfo{author}{Suson, D.J.},
  \bibinfo{author}{Takahashi, H.}, \bibinfo{author}{Takahashi, T.},
  \bibinfo{author}{Tanaka, T.}, \bibinfo{author}{Thayer, J.G.},
  \bibinfo{author}{Thayer, J.B.}, \bibinfo{author}{Thompson, D.J.},
  \bibinfo{author}{Tibaldo, L.}, \bibinfo{author}{Torres, D.F.},
  \bibinfo{author}{Tosti, G.}, \bibinfo{author}{Tramacere, A.},
  \bibinfo{author}{Troja, E.}, \bibinfo{author}{Usher, T.L.},
  \bibinfo{author}{Vandenbroucke, J.}, \bibinfo{author}{Vasileiou, V.},
  \bibinfo{author}{Vianello, G.}, \bibinfo{author}{Vilchez, N.},
  \bibinfo{author}{Vitale, V.}, \bibinfo{author}{Waite, A.P.},
  \bibinfo{author}{Wang, P.}, \bibinfo{author}{Wehrle, A.E.},
  \bibinfo{author}{Winer, B.L.}, \bibinfo{author}{Wood, K.S.},
  \bibinfo{author}{Yang, Z.}, \bibinfo{author}{Yatsu, Y.},
  \bibinfo{author}{Ylinen, T.}, \bibinfo{author}{Zensus, J.A.},
  \bibinfo{author}{Ziegler, M.}, \bibinfo{author}{Collaboration), T.F.L.},
  \bibinfo{author}{Aleksić, J.}, \bibinfo{author}{Antonelli, L.A.},
  \bibinfo{author}{Antoranz, P.}, \bibinfo{author}{Backes, M.},
  \bibinfo{author}{Barrio, J.A.}, \bibinfo{author}{González, J.B.},
  \bibinfo{author}{Bednarek, W.}, \bibinfo{author}{Berdyugin, A.},
  \bibinfo{author}{Berger, K.}, \bibinfo{author}{Bernardini, E.},
  \bibinfo{author}{Biland, A.}, \bibinfo{author}{Blanch, O.},
  \bibinfo{author}{Bock, R.K.}, \bibinfo{author}{Boller, A.},
  \bibinfo{author}{Bonnoli, G.}, \bibinfo{author}{Bordas, P.},
  \bibinfo{author}{Tridon, D.B.}, \bibinfo{author}{Bosch-Ramon, V.},
  \bibinfo{author}{Bose, D.}, \bibinfo{author}{Braun, I.},
  \bibinfo{author}{Bretz, T.}, \bibinfo{author}{Camara, M.},
  \bibinfo{author}{Carmona, E.}, \bibinfo{author}{Carosi, A.},
  \bibinfo{author}{Colin, P.}, \bibinfo{author}{Colombo, E.},
  \bibinfo{author}{Contreras, J.L.}, \bibinfo{author}{Cortina, J.},
  \bibinfo{author}{Covino, S.}, \bibinfo{author}{Dazzi, F.},
  \bibinfo{author}{de~Angelis, A.}, \bibinfo{author}{De~Cea~del Pozo, E.},
  \bibinfo{author}{Mendez, C.D.}, \bibinfo{author}{De~Lotto, B.},
  \bibinfo{author}{De~Maria, M.}, \bibinfo{author}{De~Sabata, F.},
  \bibinfo{author}{Ortega, A.D.}, \bibinfo{author}{Doert, M.},
  \bibinfo{author}{Domínguez, A.}, \bibinfo{author}{Prester, D.D.},
  \bibinfo{author}{Dorner, D.}, \bibinfo{author}{Doro, M.},
  \bibinfo{author}{Elsaesser, D.}, \bibinfo{author}{Ferenc, D.},
  \bibinfo{author}{Fonseca, M.V.}, \bibinfo{author}{Font, L.},
  \bibinfo{author}{López, R.J.G.}, \bibinfo{author}{Garczarczyk, M.},
  \bibinfo{author}{Gaug, M.}, \bibinfo{author}{Giavitto, G.},
  \bibinfo{author}{Godinovi, N.}, \bibinfo{author}{Hadasch, D.},
  \bibinfo{author}{Herrero, A.}, \bibinfo{author}{Hildebrand, D.},
  \bibinfo{author}{Höhne-Mönch, D.}, \bibinfo{author}{Hose, J.},
  \bibinfo{author}{Hrupec, D.}, \bibinfo{author}{Jogler, T.},
  \bibinfo{author}{Klepser, S.}, \bibinfo{author}{Krähenbühl, T.},
  \bibinfo{author}{Kranich, D.}, \bibinfo{author}{Krause, J.},
  \bibinfo{author}{Barbera, A.L.}, \bibinfo{author}{Leonardo, E.},
  \bibinfo{author}{Lindfors, E.}, \bibinfo{author}{Lombardi, S.},
  \bibinfo{author}{López, M.}, \bibinfo{author}{Lorenz, E.},
  \bibinfo{author}{Majumdar, P.}, \bibinfo{author}{Makariev, E.},
  \bibinfo{author}{Maneva, G.}, \bibinfo{author}{Mankuzhiyil, N.},
  \bibinfo{author}{Mannheim, K.}, \bibinfo{author}{Maraschi, L.},
  \bibinfo{author}{Mariotti, M.}, \bibinfo{author}{Martínez, M.},
  \bibinfo{author}{Mazin, D.}, \bibinfo{author}{Meucci, M.},
  \bibinfo{author}{Miranda, J.M.}, \bibinfo{author}{Mirzoyan, R.},
  \bibinfo{author}{Miyamoto, H.}, \bibinfo{author}{Moldón, J.},
  \bibinfo{author}{Moralejo, A.}, \bibinfo{author}{Nieto, D.},
  \bibinfo{author}{Nilsson, K.}, \bibinfo{author}{Orito, R.},
  \bibinfo{author}{Oya, I.}, \bibinfo{author}{Paoletti, R.},
  \bibinfo{author}{Paredes, J.M.}, \bibinfo{author}{Partini, S.},
  \bibinfo{author}{Pasanen, M.}, \bibinfo{author}{Pauss, F.},
  \bibinfo{author}{Pegna, R.G.}, \bibinfo{author}{Perez-Torres, M.A.},
  \bibinfo{author}{Persic, M.}, \bibinfo{author}{Peruzzo, J.},
  \bibinfo{author}{Pochon, J.}, \bibinfo{author}{Prada, F.},
  \bibinfo{author}{Moroni, P.G.P.}, \bibinfo{author}{Prandini, E.},
  \bibinfo{author}{Puchades, N.}, \bibinfo{author}{Puljak, I.},
  \bibinfo{author}{Reichardt, T.}, \bibinfo{author}{Rhode, W.},
  \bibinfo{author}{Ribó, M.}, \bibinfo{author}{Rico, J.},
  \bibinfo{author}{Rissi, M.}, \bibinfo{author}{Rügamer, S.},
  \bibinfo{author}{Saggion, A.}, \bibinfo{author}{Saito, K.},
  \bibinfo{author}{Saito, T.Y.}, \bibinfo{author}{Salvati, M.},
  \bibinfo{author}{Sánchez-Conde, M.}, \bibinfo{author}{Satalecka, K.},
  \bibinfo{author}{Scalzotto, V.}, \bibinfo{author}{Scapin, V.},
  \bibinfo{author}{Schultz, C.}, \bibinfo{author}{Schweizer, T.},
  \bibinfo{author}{Shayduk, M.}, \bibinfo{author}{Shore, S.N.},
  \bibinfo{author}{Sierpowska-Bartosik, A.}, \bibinfo{author}{Sillanpää, A.},
  \bibinfo{author}{Sitarek, J.}, \bibinfo{author}{Sobczynska, D.},
  \bibinfo{author}{Spanier, F.}, \bibinfo{author}{Spiro, S.},
  \bibinfo{author}{Stamerra, A.}, \bibinfo{author}{Steinke, B.},
  \bibinfo{author}{Storz, J.}, \bibinfo{author}{Strah, N.},
  \bibinfo{author}{Struebig, J.C.}, \bibinfo{author}{Suric, T.},
  \bibinfo{author}{Takalo, L.O.}, \bibinfo{author}{Tavecchio, F.},
  \bibinfo{author}{Temnikov, P.}, \bibinfo{author}{Terzić, T.},
  \bibinfo{author}{Tescaro, D.}, \bibinfo{author}{Teshima, M.},
  \bibinfo{author}{Vankov, H.}, \bibinfo{author}{Wagner, R.M.},
  \bibinfo{author}{Weitzel, Q.}, \bibinfo{author}{Zabalza, V.},
  \bibinfo{author}{Zandanel, F.}, \bibinfo{author}{Zanin, R.},
  \bibinfo{author}{Collaboration), T.M.}, \bibinfo{author}{Villata, M.},
  \bibinfo{author}{Raiteri, C.}, \bibinfo{author}{Aller, H.D.},
  \bibinfo{author}{Aller, M.F.}, \bibinfo{author}{Chen, W.P.},
  \bibinfo{author}{Jordan, B.}, \bibinfo{author}{Koptelova, E.},
  \bibinfo{author}{Kurtanidze, O.M.}, \bibinfo{author}{Lähteenmäki, A.},
  \bibinfo{author}{McBreen, B.}, \bibinfo{author}{Larionov, V.M.},
  \bibinfo{author}{Lin, C.S.}, \bibinfo{author}{Nikolashvili, M.G.},
  \bibinfo{author}{Reinthal, R.}, \bibinfo{author}{Angelakis, E.},
  \bibinfo{author}{Capalbi, M.}, \bibinfo{author}{Carramiñana, A.},
  \bibinfo{author}{Carrasco, L.}, \bibinfo{author}{Cassaro, P.},
  \bibinfo{author}{Cesarini, A.}, \bibinfo{author}{Falcone, A.},
  \bibinfo{author}{Gurwell, M.A.}, \bibinfo{author}{Hovatta, T.},
  \bibinfo{author}{Kovalev, Y.A.}, \bibinfo{author}{Kovalev, Y.Y.},
  \bibinfo{author}{Krichbaum, T.P.}, \bibinfo{author}{Krimm, H.A.},
  \bibinfo{author}{Lister, M.L.}, \bibinfo{author}{Moody, J.W.},
  \bibinfo{author}{Maccaferri, G.}, \bibinfo{author}{Mori, Y.},
  \bibinfo{author}{Nestoras, I.}, \bibinfo{author}{Orlati, A.},
  \bibinfo{author}{Pace, C.}, \bibinfo{author}{Pagani, C.},
  \bibinfo{author}{Pearson, R.}, \bibinfo{author}{Perri, M.},
  \bibinfo{author}{Piner, B.G.}, \bibinfo{author}{Ros, E.},
  \bibinfo{author}{Sadun, A.C.}, \bibinfo{author}{Sakamoto, T.},
  \bibinfo{author}{Tammi, J.}, \bibinfo{author}{Zook, A.},
  \bibinfo{year}{2011}.
\newblock \bibinfo{title}{Fermi large area telescope observations of markarian
  421: The missing piece of its spectral energy distribution}.
\newblock \bibinfo{journal}{The Astrophysical Journal} \bibinfo{volume}{736},
  \bibinfo{pages}{131}.
\newblock \URLprefix \url{https://doi.org/10.1088/0004-637X/736/2/131},
  \DOIprefix\doi{10.1088/0004-637X/736/2/131}.
\bibitem[{{Ajello} et~al.(2020){Ajello}, {Angioni}, {Axelsson}, {Ballet},
  {Barbiellini}, {Bastieri}, {Becerra Gonzalez}, {Bellazzini}, {Bissaldi},
  {Bloom}, {Bonino}, {Bottacini}, {Bruel}, {Buson}, {Cafardo}, {Cameron},
  {Cavazzuti}, {Chen}, {Cheung}, {Ciprini}, {Costantin}, {Cutini}, {D'Ammando},
  {de la Torre Luque}, {de Menezes}, {de Palma}, {Desai}, {Di Lalla}, {Di
  Venere}, {Dom{\'\i}nguez}, {Dirirsa}, {Ferrara}, {Finke}, {Franckowiak},
  {Fukazawa}, {Funk}, {Fusco}, {Gargano}, {Garrappa}, {Gasparrini},
  {Giglietto}, {Giordano}, {Giroletti}, {Green}, {Grenier}, {Guiriec},
  {Harita}, {Hays}, {Horan}, {Itoh}, {J{\'o}hannesson}, {Kovac'evic'},
  {Krauss}, {Kreter}, {Kuss}, {Larsson}, {Leto}, {Li}, {Liodakis}, {Longo},
  {Loparco}, {Lott}, {Lovellette}, {Lubrano}, {Madejski}, {Maldera},
  {Manfreda}, {Mart{\'\i}-Devesa}, {Massaro}, {Mazziotta}, {Mereu}, {Meyer},
  {Migliori}, {Mirabal}, {Mizuno}, {Monzani}, {Morselli}, {Moskalenko},
  {Negro}, {Nemmen}, {Nuss}, {Ojha}, {Ojha}, {Omodei}, {Orienti}, {Orlando},
  {Ormes}, {Paliya}, {Pei}, {Pe{\~n}a-Herazo}, {Persic}, {Pesce-Rollins},
  {Petrov}, {Piron}, {Poon}, {Principe}, {Rain{\`o}}, {Rando}, {Rani},
  {Razzano}, {Razzaque}, {Reimer}, {Reimer}, {Schinzel}, {Serini}, {Sgr{\`o}},
  {Siskind}, {Spandre}, {Spinelli}, {Suson}, {Tachibana}, {Thompson}, {Torres},
  {Torresi}, {Troja}, {Valverde}, {van Zyl} and
  {Yassine}}]{2020ApJ...892..105A}
\bibinfo{author}{{Ajello}, M.}, \bibinfo{author}{{Angioni}, R.},
  \bibinfo{author}{{Axelsson}, M.}, \bibinfo{author}{{Ballet}, J.},
  \bibinfo{author}{{Barbiellini}, G.}, \bibinfo{author}{{Bastieri}, D.},
  \bibinfo{author}{{Becerra Gonzalez}, J.}, \bibinfo{author}{{Bellazzini}, R.},
  \bibinfo{author}{{Bissaldi}, E.}, \bibinfo{author}{{Bloom}, E.D.},
  \bibinfo{author}{{Bonino}, R.}, \bibinfo{author}{{Bottacini}, E.},
  \bibinfo{author}{{Bruel}, P.}, \bibinfo{author}{{Buson}, S.},
  \bibinfo{author}{{Cafardo}, F.}, \bibinfo{author}{{Cameron}, R.A.},
  \bibinfo{author}{{Cavazzuti}, E.}, \bibinfo{author}{{Chen}, S.},
  \bibinfo{author}{{Cheung}, C.C.}, \bibinfo{author}{{Ciprini}, S.},
  \bibinfo{author}{{Costantin}, D.}, \bibinfo{author}{{Cutini}, S.},
  \bibinfo{author}{{D'Ammando}, F.}, \bibinfo{author}{{de la Torre Luque}, P.},
  \bibinfo{author}{{de Menezes}, R.}, \bibinfo{author}{{de Palma}, F.},
  \bibinfo{author}{{Desai}, A.}, \bibinfo{author}{{Di Lalla}, N.},
  \bibinfo{author}{{Di Venere}, L.}, \bibinfo{author}{{Dom{\'\i}nguez}, A.},
  \bibinfo{author}{{Dirirsa}, F.F.}, \bibinfo{author}{{Ferrara}, E.C.},
  \bibinfo{author}{{Finke}, J.}, \bibinfo{author}{{Franckowiak}, A.},
  \bibinfo{author}{{Fukazawa}, Y.}, \bibinfo{author}{{Funk}, S.},
  \bibinfo{author}{{Fusco}, P.}, \bibinfo{author}{{Gargano}, F.},
  \bibinfo{author}{{Garrappa}, S.}, \bibinfo{author}{{Gasparrini}, D.},
  \bibinfo{author}{{Giglietto}, N.}, \bibinfo{author}{{Giordano}, F.},
  \bibinfo{author}{{Giroletti}, M.}, \bibinfo{author}{{Green}, D.},
  \bibinfo{author}{{Grenier}, I.A.}, \bibinfo{author}{{Guiriec}, S.},
  \bibinfo{author}{{Harita}, S.}, \bibinfo{author}{{Hays}, E.},
  \bibinfo{author}{{Horan}, D.}, \bibinfo{author}{{Itoh}, R.},
  \bibinfo{author}{{J{\'o}hannesson}, G.}, \bibinfo{author}{{Kovac'evic'}, M.},
  \bibinfo{author}{{Krauss}, F.}, \bibinfo{author}{{Kreter}, M.},
  \bibinfo{author}{{Kuss}, M.}, \bibinfo{author}{{Larsson}, S.},
  \bibinfo{author}{{Leto}, C.}, \bibinfo{author}{{Li}, J.},
  \bibinfo{author}{{Liodakis}, I.}, \bibinfo{author}{{Longo}, F.},
  \bibinfo{author}{{Loparco}, F.}, \bibinfo{author}{{Lott}, B.},
  \bibinfo{author}{{Lovellette}, M.N.}, \bibinfo{author}{{Lubrano}, P.},
  \bibinfo{author}{{Madejski}, G.M.}, \bibinfo{author}{{Maldera}, S.},
  \bibinfo{author}{{Manfreda}, A.}, \bibinfo{author}{{Mart{\'\i}-Devesa}, G.},
  \bibinfo{author}{{Massaro}, F.}, \bibinfo{author}{{Mazziotta}, M.N.},
  \bibinfo{author}{{Mereu}, I.}, \bibinfo{author}{{Meyer}, M.},
  \bibinfo{author}{{Migliori}, G.}, \bibinfo{author}{{Mirabal}, N.},
  \bibinfo{author}{{Mizuno}, T.}, \bibinfo{author}{{Monzani}, M.E.},
  \bibinfo{author}{{Morselli}, A.}, \bibinfo{author}{{Moskalenko}, I.V.},
  \bibinfo{author}{{Negro}, M.}, \bibinfo{author}{{Nemmen}, R.},
  \bibinfo{author}{{Nuss}, E.}, \bibinfo{author}{{Ojha}, L.S.},
  \bibinfo{author}{{Ojha}, R.}, \bibinfo{author}{{Omodei}, N.},
  \bibinfo{author}{{Orienti}, M.}, \bibinfo{author}{{Orlando}, E.},
  \bibinfo{author}{{Ormes}, J.F.}, \bibinfo{author}{{Paliya}, V.S.},
  \bibinfo{author}{{Pei}, Z.}, \bibinfo{author}{{Pe{\~n}a-Herazo}, H.},
  \bibinfo{author}{{Persic}, M.}, \bibinfo{author}{{Pesce-Rollins}, M.},
  \bibinfo{author}{{Petrov}, L.}, \bibinfo{author}{{Piron}, F.},
  \bibinfo{author}{{Poon}, H.}, \bibinfo{author}{{Principe}, G.},
  \bibinfo{author}{{Rain{\`o}}, S.}, \bibinfo{author}{{Rando}, R.},
  \bibinfo{author}{{Rani}, B.}, \bibinfo{author}{{Razzano}, M.},
  \bibinfo{author}{{Razzaque}, S.}, \bibinfo{author}{{Reimer}, A.},
  \bibinfo{author}{{Reimer}, O.}, \bibinfo{author}{{Schinzel}, F.K.},
  \bibinfo{author}{{Serini}, D.}, \bibinfo{author}{{Sgr{\`o}}, C.},
  \bibinfo{author}{{Siskind}, E.J.}, \bibinfo{author}{{Spandre}, G.},
  \bibinfo{author}{{Spinelli}, P.}, \bibinfo{author}{{Suson}, D.J.},
  \bibinfo{author}{{Tachibana}, Y.}, \bibinfo{author}{{Thompson}, D.J.},
  \bibinfo{author}{{Torres}, D.F.}, \bibinfo{author}{{Torresi}, E.},
  \bibinfo{author}{{Troja}, E.}, \bibinfo{author}{{Valverde}, J.},
  \bibinfo{author}{{van Zyl}, P.}, \bibinfo{author}{{Yassine}, M.},
  \bibinfo{year}{2020}.
\newblock \bibinfo{title}{{The Fourth Catalog of Active Galactic Nuclei
  Detected by the Fermi Large Area Telescope}}.
\newblock \bibinfo{journal}{\apj} \bibinfo{volume}{892}, \bibinfo{pages}{105}.
\newblock \DOIprefix\doi{10.3847/1538-4357/ab791e},
  \href{http://arxiv.org/abs/1905.10771}{{\tt arXiv:1905.10771}}.
\bibitem[{Akbar et~al.(2025)Akbar, Shah, Misra, Iqbal and
  Tantry}]{AKBAR2025438}
\bibinfo{author}{Akbar, S.}, \bibinfo{author}{Shah, Z.},
  \bibinfo{author}{Misra, R.}, \bibinfo{author}{Iqbal, N.},
  \bibinfo{author}{Tantry, J.}, \bibinfo{year}{2025}.
\newblock \bibinfo{title}{Probing spectral evolution and intrinsic variability
  of mkn 421: A multi-epoch astrosat study of x-ray spectra}.
\newblock \bibinfo{journal}{Journal of High Energy Astrophysics}
  \bibinfo{volume}{45}, \bibinfo{pages}{438--455}.
\newblock \URLprefix
  \url{https://www.sciencedirect.com/science/article/pii/S2214404825000096},
  \DOIprefix\doi{https://doi.org/10.1016/j.jheap.2025.01.009}.
\bibitem[{{Aleksi{\'c}} et~al.(2015){Aleksi{\'c}}, {Ansoldi}, {Antonelli},
  {Antoranz}, {Babic}, {Bangale}, {Barres de Almeida}, {Barrio}, {Becerra
  Gonz{\'a}lez}, {Bednarek}, {Berger}, {Bernardini}, {Biland}, {Blanch},
  {Bock}, {Bonnefoy}, {Bonnoli}, {Borracci}, {Bretz}, {Carmona}, {Carosi},
  {Carreto Fidalgo}, {Colin}, {Colombo}, {Contreras}, {Cortina}, {Covino}, {Da
  Vela}, {Dazzi}, {De Angelis}, {De Caneva}, {De Lotto}, {Delgado Mendez},
  {Doert}, {Dom{\'\i}nguez}, {Dominis Prester}, {Dorner}, {Doro}, {Einecke},
  {Eisenacher}, {Elsaesser}, {Farina}, {Ferenc}, {Fonseca}, {Font}, {Frantzen},
  {Fruck}, {Garc{\'\i}a L{\'o}pez}, {Garczarczyk}, {Garrido Terrats}, {Gaug},
  {Giavitto}, {Godinovi{\'c}}, {Gonz{\'a}lez Mu{\~n}oz}, {Gozzini}, {Hadamek},
  {Hadasch}, {Herrero}, {Hildebrand}, {Hose}, {Hrupec}, {Idec}, {Kadenius},
  {Kellermann}, {Knoetig}, {Krause}, {Kushida}, {La Barbera}, {Lelas},
  {Lewandowska}, {Lindfors}, {Longo}, {Lombardi}, {L{\'o}pez},
  {L{\'o}pez-Coto}, {L{\'o}pez-Oramas}, {Lorenz}, {Lozano}, {Makariev},
  {Mallot}, {Maneva}, {Mankuzhiyil}, {Mannheim}, {Maraschi}, {Marcote},
  {Mariotti}, {Mart{\'\i}nez}, {Mazin}, {Menzel}, {Meucci}, {Miranda},
  {Mirzoyan}, {Moralejo}, {Munar-Adrover}, {Nakajima}, {Niedzwiecki},
  {Nilsson}, {Nowak}, {Orito}, {Overkemping}, {Paiano}, {Palatiello},
  {Paneque}, {Paoletti}, {Paredes}, {Paredes-Fortuny}, {Partini}, {Persic},
  {Prada}, {Prada Moroni}, {Prandini}, {Preziuso}, {Puljak}, {Reinthal},
  {Rhode}, {Rib{\'o}}, {Rico}, {RodriguezGarcia}, {R{\"u}gamer}, {Saggion},
  {Saito}, {Salvati}, {Satalecka}, {Scalzotto}, {Scapin}, {Schultz},
  {Schweizer}, {Shore}, {Sillanp{\"a}{\"a}}, {Sitarek}, {Snidaric},
  {Sobczynska}, {Spanier}, {Stamatescu}, {Stamerra}, {Steinbring}, {Storz},
  {Sun}, {Suri{\'c}}, {Takalo}, {Tavecchio}, {Temnikov}, {Terzi{\'c}},
  {Tescaro}, {Teshima}, {Thaele}, {Tibolla}, {Torres}, {Toyama}, {Treves},
  {Uellenbeck}, {Vogler}, {Wagner}, {Zandanel}, {Zanin}, {MAGIC Collaboration},
  {Archambault}, {Behera}, {Beilicke}, {Benbow}, {Bird}, {Buckley}, {Bugaev},
  {Cerruti}, {Chen}, {Ciupik}, {Collins-Hughes}, {Cui}, {Dumm}, {Eisch},
  {Falcone}, {Federici}, {Feng}, {Finley}, {Fleischhack}, {Fortin}, {Fortson},
  {Furniss}, {Griffin}, {Griffiths}, {Grube}, {Gyuk}, {Hanna}, {Holder},
  {Hughes}, {Humensky}, {Johnson}, {Kaaret}, {Kertzman}, {Khassen}, {Kieda},
  {Krawczynski}, {Krennrich}, {Kumar}, {Lang}, {Maier}, {McArthur}, {Meagher},
  {Moriarty} and {Mukherjee}}]{2015A&A...576A.126A}
\bibinfo{author}{{Aleksi{\'c}}, J.}, \bibinfo{author}{{Ansoldi}, S.},
  \bibinfo{author}{{Antonelli}, L.A.}, \bibinfo{author}{{Antoranz}, P.},
  \bibinfo{author}{{Babic}, A.}, \bibinfo{author}{{Bangale}, P.},
  \bibinfo{author}{{Barres de Almeida}, U.}, \bibinfo{author}{{Barrio}, J.A.},
  \bibinfo{author}{{Becerra Gonz{\'a}lez}, J.}, \bibinfo{author}{{Bednarek},
  W.}, \bibinfo{author}{{Berger}, K.}, \bibinfo{author}{{Bernardini}, E.},
  \bibinfo{author}{{Biland}, A.}, \bibinfo{author}{{Blanch}, O.},
  \bibinfo{author}{{Bock}, R.K.}, \bibinfo{author}{{Bonnefoy}, S.},
  \bibinfo{author}{{Bonnoli}, G.}, \bibinfo{author}{{Borracci}, F.},
  \bibinfo{author}{{Bretz}, T.}, \bibinfo{author}{{Carmona}, E.},
  \bibinfo{author}{{Carosi}, A.}, \bibinfo{author}{{Carreto Fidalgo}, D.},
  \bibinfo{author}{{Colin}, P.}, \bibinfo{author}{{Colombo}, E.},
  \bibinfo{author}{{Contreras}, J.L.}, \bibinfo{author}{{Cortina}, J.},
  \bibinfo{author}{{Covino}, S.}, \bibinfo{author}{{Da Vela}, P.},
  \bibinfo{author}{{Dazzi}, F.}, \bibinfo{author}{{De Angelis}, A.},
  \bibinfo{author}{{De Caneva}, G.}, \bibinfo{author}{{De Lotto}, B.},
  \bibinfo{author}{{Delgado Mendez}, C.}, \bibinfo{author}{{Doert}, M.},
  \bibinfo{author}{{Dom{\'\i}nguez}, A.}, \bibinfo{author}{{Dominis Prester},
  D.}, \bibinfo{author}{{Dorner}, D.}, \bibinfo{author}{{Doro}, M.},
  \bibinfo{author}{{Einecke}, S.}, \bibinfo{author}{{Eisenacher}, D.},
  \bibinfo{author}{{Elsaesser}, D.}, \bibinfo{author}{{Farina}, E.},
  \bibinfo{author}{{Ferenc}, D.}, \bibinfo{author}{{Fonseca}, M.V.},
  \bibinfo{author}{{Font}, L.}, \bibinfo{author}{{Frantzen}, K.},
  \bibinfo{author}{{Fruck}, C.}, \bibinfo{author}{{Garc{\'\i}a L{\'o}pez},
  R.J.}, \bibinfo{author}{{Garczarczyk}, M.}, \bibinfo{author}{{Garrido
  Terrats}, D.}, \bibinfo{author}{{Gaug}, M.}, \bibinfo{author}{{Giavitto},
  G.}, \bibinfo{author}{{Godinovi{\'c}}, N.}, \bibinfo{author}{{Gonz{\'a}lez
  Mu{\~n}oz}, A.}, \bibinfo{author}{{Gozzini}, S.R.},
  \bibinfo{author}{{Hadamek}, A.}, \bibinfo{author}{{Hadasch}, D.},
  \bibinfo{author}{{Herrero}, A.}, \bibinfo{author}{{Hildebrand}, D.},
  \bibinfo{author}{{Hose}, J.}, \bibinfo{author}{{Hrupec}, D.},
  \bibinfo{author}{{Idec}, W.}, \bibinfo{author}{{Kadenius}, V.},
  \bibinfo{author}{{Kellermann}, H.}, \bibinfo{author}{{Knoetig}, M.L.},
  \bibinfo{author}{{Krause}, J.}, \bibinfo{author}{{Kushida}, J.},
  \bibinfo{author}{{La Barbera}, A.}, \bibinfo{author}{{Lelas}, D.},
  \bibinfo{author}{{Lewandowska}, N.}, \bibinfo{author}{{Lindfors}, E.},
  \bibinfo{author}{{Longo}, F.}, \bibinfo{author}{{Lombardi}, S.},
  \bibinfo{author}{{L{\'o}pez}, M.}, \bibinfo{author}{{L{\'o}pez-Coto}, R.},
  \bibinfo{author}{{L{\'o}pez-Oramas}, A.}, \bibinfo{author}{{Lorenz}, E.},
  \bibinfo{author}{{Lozano}, I.}, \bibinfo{author}{{Makariev}, M.},
  \bibinfo{author}{{Mallot}, K.}, \bibinfo{author}{{Maneva}, G.},
  \bibinfo{author}{{Mankuzhiyil}, N.}, \bibinfo{author}{{Mannheim}, K.},
  \bibinfo{author}{{Maraschi}, L.}, \bibinfo{author}{{Marcote}, B.},
  \bibinfo{author}{{Mariotti}, M.}, \bibinfo{author}{{Mart{\'\i}nez}, M.},
  \bibinfo{author}{{Mazin}, D.}, \bibinfo{author}{{Menzel}, U.},
  \bibinfo{author}{{Meucci}, M.}, \bibinfo{author}{{Miranda}, J.M.},
  \bibinfo{author}{{Mirzoyan}, R.}, \bibinfo{author}{{Moralejo}, A.},
  \bibinfo{author}{{Munar-Adrover}, P.}, \bibinfo{author}{{Nakajima}, D.},
  \bibinfo{author}{{Niedzwiecki}, A.}, \bibinfo{author}{{Nilsson}, K.},
  \bibinfo{author}{{Nowak}, N.}, \bibinfo{author}{{Orito}, R.},
  \bibinfo{author}{{Overkemping}, A.}, \bibinfo{author}{{Paiano}, S.},
  \bibinfo{author}{{Palatiello}, M.}, \bibinfo{author}{{Paneque}, D.},
  \bibinfo{author}{{Paoletti}, R.}, \bibinfo{author}{{Paredes}, J.M.},
  \bibinfo{author}{{Paredes-Fortuny}, X.}, \bibinfo{author}{{Partini}, S.},
  \bibinfo{author}{{Persic}, M.}, \bibinfo{author}{{Prada}, F.},
  \bibinfo{author}{{Prada Moroni}, P.G.}, \bibinfo{author}{{Prandini}, E.},
  \bibinfo{author}{{Preziuso}, S.}, \bibinfo{author}{{Puljak}, I.},
  \bibinfo{author}{{Reinthal}, R.}, \bibinfo{author}{{Rhode}, W.},
  \bibinfo{author}{{Rib{\'o}}, M.}, \bibinfo{author}{{Rico}, J.},
  \bibinfo{author}{{RodriguezGarcia}, J.}, \bibinfo{author}{{R{\"u}gamer}, S.},
  \bibinfo{author}{{Saggion}, A.}, \bibinfo{author}{{Saito}, K.},
  \bibinfo{author}{{Salvati}, M.}, \bibinfo{author}{{Satalecka}, K.},
  \bibinfo{author}{{Scalzotto}, V.}, \bibinfo{author}{{Scapin}, V.},
  \bibinfo{author}{{Schultz}, C.}, \bibinfo{author}{{Schweizer}, T.},
  \bibinfo{author}{{Shore}, S.N.}, \bibinfo{author}{{Sillanp{\"a}{\"a}}, A.},
  \bibinfo{author}{{Sitarek}, J.}, \bibinfo{author}{{Snidaric}, I.},
  \bibinfo{author}{{Sobczynska}, D.}, \bibinfo{author}{{Spanier}, F.},
  \bibinfo{author}{{Stamatescu}, V.}, \bibinfo{author}{{Stamerra}, A.},
  \bibinfo{author}{{Steinbring}, T.}, \bibinfo{author}{{Storz}, J.},
  \bibinfo{author}{{Sun}, S.}, \bibinfo{author}{{Suri{\'c}}, T.},
  \bibinfo{author}{{Takalo}, L.}, \bibinfo{author}{{Tavecchio}, F.},
  \bibinfo{author}{{Temnikov}, P.}, \bibinfo{author}{{Terzi{\'c}}, T.},
  \bibinfo{author}{{Tescaro}, D.}, \bibinfo{author}{{Teshima}, M.},
  \bibinfo{author}{{Thaele}, J.}, \bibinfo{author}{{Tibolla}, O.},
  \bibinfo{author}{{Torres}, D.F.}, \bibinfo{author}{{Toyama}, T.},
  \bibinfo{author}{{Treves}, A.}, \bibinfo{author}{{Uellenbeck}, M.},
  \bibinfo{author}{{Vogler}, P.}, \bibinfo{author}{{Wagner}, R.M.},
  \bibinfo{author}{{Zandanel}, F.}, \bibinfo{author}{{Zanin}, R.},
  \bibinfo{author}{{MAGIC Collaboration}}, \bibinfo{author}{{Archambault}, S.},
  \bibinfo{author}{{Behera}, B.}, \bibinfo{author}{{Beilicke}, M.},
  \bibinfo{author}{{Benbow}, W.}, \bibinfo{author}{{Bird}, R.},
  \bibinfo{author}{{Buckley}, J.H.}, \bibinfo{author}{{Bugaev}, V.},
  \bibinfo{author}{{Cerruti}, M.}, \bibinfo{author}{{Chen}, X.},
  \bibinfo{author}{{Ciupik}, L.}, \bibinfo{author}{{Collins-Hughes}, E.},
  \bibinfo{author}{{Cui}, W.}, \bibinfo{author}{{Dumm}, J.},
  \bibinfo{author}{{Eisch}, J.D.}, \bibinfo{author}{{Falcone}, A.},
  \bibinfo{author}{{Federici}, S.}, \bibinfo{author}{{Feng}, Q.},
  \bibinfo{author}{{Finley}, J.P.}, \bibinfo{author}{{Fleischhack}, H.},
  \bibinfo{author}{{Fortin}, P.}, \bibinfo{author}{{Fortson}, L.},
  \bibinfo{author}{{Furniss}, A.}, \bibinfo{author}{{Griffin}, S.},
  \bibinfo{author}{{Griffiths}, S.T.}, \bibinfo{author}{{Grube}, J.},
  \bibinfo{author}{{Gyuk}, G.}, \bibinfo{author}{{Hanna}, D.},
  \bibinfo{author}{{Holder}, J.}, \bibinfo{author}{{Hughes}, G.},
  \bibinfo{author}{{Humensky}, T.B.}, \bibinfo{author}{{Johnson}, C.A.},
  \bibinfo{author}{{Kaaret}, P.}, \bibinfo{author}{{Kertzman}, M.},
  \bibinfo{author}{{Khassen}, Y.}, \bibinfo{author}{{Kieda}, D.},
  \bibinfo{author}{{Krawczynski}, H.}, \bibinfo{author}{{Krennrich}, F.},
  \bibinfo{author}{{Kumar}, S.}, \bibinfo{author}{{Lang}, M.J.},
  \bibinfo{author}{{Maier}, G.}, \bibinfo{author}{{McArthur}, S.},
  \bibinfo{author}{{Meagher}, K.}, \bibinfo{author}{{Moriarty}, P.},
  \bibinfo{author}{{Mukherjee}, R.}, \bibinfo{year}{2015}.
\newblock \bibinfo{title}{{The 2009 multiwavelength campaign on Mrk 421:
  Variability and correlation studies}}.
\newblock \bibinfo{journal}{\aap} \bibinfo{volume}{576}, \bibinfo{pages}{A126}.
\newblock \DOIprefix\doi{10.1051/0004-6361/201424216},
  \href{http://arxiv.org/abs/1502.02650}{{\tt arXiv:1502.02650}}.
\bibitem[{{Balokovi{\'c}} et~al.(2016){Balokovi{\'c}}, {Paneque}, {Madejski},
  {Furniss}, {Chiang}, {Ajello}, {Alexander}, {Barret}, {Blandford}, {Boggs},
  {Christensen}, {Craig}, {Forster}, {Giommi}, {Grefenstette}, {Hailey},
  {Harrison}, {Hornstrup}, {Kitaguchi}, {Koglin}, {Madsen}, {Mao}, {Miyasaka},
  {Mori}, {Perri}, {Pivovaroff}, {Puccetti}, {Rana}, {Stern}, {Tagliaferri},
  {Urry}, {Westergaard}, {Zhang}, {Zoglauer}, {NuSTAR Team}, {Archambault},
  {Archer}, {Barnacka}, {Benbow}, {Bird}, {Buckley}, {Bugaev}, {Cerruti},
  {Chen}, {Ciupik}, {Connolly}, {Cui}, {Dickinson}, {Dumm}, {Eisch}, {Falcone},
  {Feng}, {Finley}, {Fleischhack}, {Fortson}, {Griffin}, {Griffiths}, {Grube},
  {Gyuk}, {Huetten}, {H{\r{a}}kansson}, {Holder}, {Humensky}, {Johnson},
  {Kaaret}, {Kertzman}, {Khassen}, {Kieda}, {Krause}, {Krennrich}, {Lang},
  {Maier}, {McArthur}, {Meagher}, {Moriarty}, {Nelson}, {Nieto}, {Ong}, {Park},
  {Pohl}, {Popkow}, {Pueschel}, {Reynolds}, {Richards}, {Roache}, {Santander},
  {Sembroski}, {Shahinyan}, {Smith}, {Staszak}, {Telezhinsky}, {Todd}, {Tucci},
  {Tyler}, {Vincent}, {Weinstein}, {Wilhelm}, {Williams}, {Zitzer}, {VERITAS
  Collaboration}, {Ahnen}, {Ansoldi}, {Antonelli}, {Antoranz}, {Babic},
  {Banerjee}, {Bangale}, {Barres de Almeida}, {Barrio}, {Becerra Gonz{\'a}lez},
  {Bednarek}, {Bernardini}, {Biasuzzi}, {Biland}, {Blanch}, {Bonnefoy},
  {Bonnoli}, {Borracci}, {Bretz}, {Carmona}, {Carosi}, {Chatterjee}, {Clavero},
  {Colin}, {Colombo}, {Contreras}, {Cortina}, {Covino}, {Da Vela}, {Dazzi}, {De
  Angelis}, {De Lotto}, {de O{\~n}a Wilhelmi}, {Delgado Mendez}, {Di Pierro},
  {Dominis Prester}, {Dorner}, {Doro}, {Einecke}, {Elsaesser},
  {Fern{\'a}ndez-Barral}, {Fidalgo}, {Fonseca}, {Font}, {Frantzen}, {Fruck},
  {Galindo}, {Garc{\'\i}a L{\'o}pez}, {Garczarczyk}, {Garrido Terrats}, {Gaug},
  {Giammaria}, {Glawion (Eisenacher}, {Godinovi{\'c}}, {Gonz{\'a}lez
  Mu{\~n}oz}, {Guberman}, {Hahn}, {Hanabata}, {Hayashida}, {Herrera}, {Hose},
  {Hrupec}, {Hughes}, {Idec}, {Kodani}, {Konno}, {Kubo}, {Kushida}, {La
  Barbera}, {Lelas}, {Lindfors}, {Lombardi}, {Longo}, {L{\'o}pez},
  {L{\'o}pez-Coto}, {L{\'o}pez-Oramas}, {Lorenz}, {Majumdar}, {Makariev},
  {Mallot}, {Maneva}, {Manganaro}, {Mannheim}, {Maraschi}, {Marcote},
  {Mariotti}, {Mart{\'\i}nez}, {Mazin}, {Menzel}, {Miranda}, {Mirzoyan},
  {Moralejo}, {Moretti}, {Nakajima}, {Neustroev}, {Niedzwiecki}, {Nievas
  Rosillo}, {Nilsson}, {Nishijima} and {Noda}}]{2016ApJ...819..156B}
\bibinfo{author}{{Balokovi{\'c}}, M.}, \bibinfo{author}{{Paneque}, D.},
  \bibinfo{author}{{Madejski}, G.}, \bibinfo{author}{{Furniss}, A.},
  \bibinfo{author}{{Chiang}, J.}, \bibinfo{author}{{Ajello}, M.},
  \bibinfo{author}{{Alexander}, D.M.}, \bibinfo{author}{{Barret}, D.},
  \bibinfo{author}{{Blandford}, R.D.}, \bibinfo{author}{{Boggs}, S.E.},
  \bibinfo{author}{{Christensen}, F.E.}, \bibinfo{author}{{Craig}, W.W.},
  \bibinfo{author}{{Forster}, K.}, \bibinfo{author}{{Giommi}, P.},
  \bibinfo{author}{{Grefenstette}, B.}, \bibinfo{author}{{Hailey}, C.},
  \bibinfo{author}{{Harrison}, F.A.}, \bibinfo{author}{{Hornstrup}, A.},
  \bibinfo{author}{{Kitaguchi}, T.}, \bibinfo{author}{{Koglin}, J.E.},
  \bibinfo{author}{{Madsen}, K.K.}, \bibinfo{author}{{Mao}, P.H.},
  \bibinfo{author}{{Miyasaka}, H.}, \bibinfo{author}{{Mori}, K.},
  \bibinfo{author}{{Perri}, M.}, \bibinfo{author}{{Pivovaroff}, M.J.},
  \bibinfo{author}{{Puccetti}, S.}, \bibinfo{author}{{Rana}, V.},
  \bibinfo{author}{{Stern}, D.}, \bibinfo{author}{{Tagliaferri}, G.},
  \bibinfo{author}{{Urry}, C.M.}, \bibinfo{author}{{Westergaard}, N.J.},
  \bibinfo{author}{{Zhang}, W.W.}, \bibinfo{author}{{Zoglauer}, A.},
  \bibinfo{author}{{NuSTAR Team}}, \bibinfo{author}{{Archambault}, S.},
  \bibinfo{author}{{Archer}, A.}, \bibinfo{author}{{Barnacka}, A.},
  \bibinfo{author}{{Benbow}, W.}, \bibinfo{author}{{Bird}, R.},
  \bibinfo{author}{{Buckley}, J.H.}, \bibinfo{author}{{Bugaev}, V.},
  \bibinfo{author}{{Cerruti}, M.}, \bibinfo{author}{{Chen}, X.},
  \bibinfo{author}{{Ciupik}, L.}, \bibinfo{author}{{Connolly}, M.P.},
  \bibinfo{author}{{Cui}, W.}, \bibinfo{author}{{Dickinson}, H.J.},
  \bibinfo{author}{{Dumm}, J.}, \bibinfo{author}{{Eisch}, J.D.},
  \bibinfo{author}{{Falcone}, A.}, \bibinfo{author}{{Feng}, Q.},
  \bibinfo{author}{{Finley}, J.P.}, \bibinfo{author}{{Fleischhack}, H.},
  \bibinfo{author}{{Fortson}, L.}, \bibinfo{author}{{Griffin}, S.},
  \bibinfo{author}{{Griffiths}, S.T.}, \bibinfo{author}{{Grube}, J.},
  \bibinfo{author}{{Gyuk}, G.}, \bibinfo{author}{{Huetten}, M.},
  \bibinfo{author}{{H{\r{a}}kansson}, N.}, \bibinfo{author}{{Holder}, J.},
  \bibinfo{author}{{Humensky}, T.B.}, \bibinfo{author}{{Johnson}, C.A.},
  \bibinfo{author}{{Kaaret}, P.}, \bibinfo{author}{{Kertzman}, M.},
  \bibinfo{author}{{Khassen}, Y.}, \bibinfo{author}{{Kieda}, D.},
  \bibinfo{author}{{Krause}, M.}, \bibinfo{author}{{Krennrich}, F.},
  \bibinfo{author}{{Lang}, M.J.}, \bibinfo{author}{{Maier}, G.},
  \bibinfo{author}{{McArthur}, S.}, \bibinfo{author}{{Meagher}, K.},
  \bibinfo{author}{{Moriarty}, P.}, \bibinfo{author}{{Nelson}, T.},
  \bibinfo{author}{{Nieto}, D.}, \bibinfo{author}{{Ong}, R.A.},
  \bibinfo{author}{{Park}, N.}, \bibinfo{author}{{Pohl}, M.},
  \bibinfo{author}{{Popkow}, A.}, \bibinfo{author}{{Pueschel}, E.},
  \bibinfo{author}{{Reynolds}, P.T.}, \bibinfo{author}{{Richards}, G.T.},
  \bibinfo{author}{{Roache}, E.}, \bibinfo{author}{{Santander}, M.},
  \bibinfo{author}{{Sembroski}, G.H.}, \bibinfo{author}{{Shahinyan}, K.},
  \bibinfo{author}{{Smith}, A.W.}, \bibinfo{author}{{Staszak}, D.},
  \bibinfo{author}{{Telezhinsky}, I.}, \bibinfo{author}{{Todd}, N.W.},
  \bibinfo{author}{{Tucci}, J.V.}, \bibinfo{author}{{Tyler}, J.},
  \bibinfo{author}{{Vincent}, S.}, \bibinfo{author}{{Weinstein}, A.},
  \bibinfo{author}{{Wilhelm}, A.}, \bibinfo{author}{{Williams}, D.A.},
  \bibinfo{author}{{Zitzer}, B.}, \bibinfo{author}{{VERITAS Collaboration}},
  \bibinfo{author}{{Ahnen}, M.L.}, \bibinfo{author}{{Ansoldi}, S.},
  \bibinfo{author}{{Antonelli}, L.A.}, \bibinfo{author}{{Antoranz}, P.},
  \bibinfo{author}{{Babic}, A.}, \bibinfo{author}{{Banerjee}, B.},
  \bibinfo{author}{{Bangale}, P.}, \bibinfo{author}{{Barres de Almeida}, U.},
  \bibinfo{author}{{Barrio}, J.A.}, \bibinfo{author}{{Becerra Gonz{\'a}lez},
  J.}, \bibinfo{author}{{Bednarek}, W.}, \bibinfo{author}{{Bernardini}, E.},
  \bibinfo{author}{{Biasuzzi}, B.}, \bibinfo{author}{{Biland}, A.},
  \bibinfo{author}{{Blanch}, O.}, \bibinfo{author}{{Bonnefoy}, S.},
  \bibinfo{author}{{Bonnoli}, G.}, \bibinfo{author}{{Borracci}, F.},
  \bibinfo{author}{{Bretz}, T.}, \bibinfo{author}{{Carmona}, E.},
  \bibinfo{author}{{Carosi}, A.}, \bibinfo{author}{{Chatterjee}, A.},
  \bibinfo{author}{{Clavero}, R.}, \bibinfo{author}{{Colin}, P.},
  \bibinfo{author}{{Colombo}, E.}, \bibinfo{author}{{Contreras}, J.L.},
  \bibinfo{author}{{Cortina}, J.}, \bibinfo{author}{{Covino}, S.},
  \bibinfo{author}{{Da Vela}, P.}, \bibinfo{author}{{Dazzi}, F.},
  \bibinfo{author}{{De Angelis}, A.}, \bibinfo{author}{{De Lotto}, B.},
  \bibinfo{author}{{de O{\~n}a Wilhelmi}, E.}, \bibinfo{author}{{Delgado
  Mendez}, C.}, \bibinfo{author}{{Di Pierro}, F.}, \bibinfo{author}{{Dominis
  Prester}, D.}, \bibinfo{author}{{Dorner}, D.}, \bibinfo{author}{{Doro}, M.},
  \bibinfo{author}{{Einecke}, S.}, \bibinfo{author}{{Elsaesser}, D.},
  \bibinfo{author}{{Fern{\'a}ndez-Barral}, A.}, \bibinfo{author}{{Fidalgo},
  D.}, \bibinfo{author}{{Fonseca}, M.V.}, \bibinfo{author}{{Font}, L.},
  \bibinfo{author}{{Frantzen}, K.}, \bibinfo{author}{{Fruck}, C.},
  \bibinfo{author}{{Galindo}, D.}, \bibinfo{author}{{Garc{\'\i}a L{\'o}pez},
  R.J.}, \bibinfo{author}{{Garczarczyk}, M.}, \bibinfo{author}{{Garrido
  Terrats}, D.}, \bibinfo{author}{{Gaug}, M.}, \bibinfo{author}{{Giammaria},
  P.}, \bibinfo{author}{{Glawion (Eisenacher}, D.},
  \bibinfo{author}{{Godinovi{\'c}}, N.}, \bibinfo{author}{{Gonz{\'a}lez
  Mu{\~n}oz}, A.}, \bibinfo{author}{{Guberman}, D.}, \bibinfo{author}{{Hahn},
  A.}, \bibinfo{author}{{Hanabata}, Y.}, \bibinfo{author}{{Hayashida}, M.},
  \bibinfo{author}{{Herrera}, J.}, \bibinfo{author}{{Hose}, J.},
  \bibinfo{author}{{Hrupec}, D.}, \bibinfo{author}{{Hughes}, G.},
  \bibinfo{author}{{Idec}, W.}, \bibinfo{author}{{Kodani}, K.},
  \bibinfo{author}{{Konno}, Y.}, \bibinfo{author}{{Kubo}, H.},
  \bibinfo{author}{{Kushida}, J.}, \bibinfo{author}{{La Barbera}, A.},
  \bibinfo{author}{{Lelas}, D.}, \bibinfo{author}{{Lindfors}, E.},
  \bibinfo{author}{{Lombardi}, S.}, \bibinfo{author}{{Longo}, F.},
  \bibinfo{author}{{L{\'o}pez}, M.}, \bibinfo{author}{{L{\'o}pez-Coto}, R.},
  \bibinfo{author}{{L{\'o}pez-Oramas}, A.}, \bibinfo{author}{{Lorenz}, E.},
  \bibinfo{author}{{Majumdar}, P.}, \bibinfo{author}{{Makariev}, M.},
  \bibinfo{author}{{Mallot}, K.}, \bibinfo{author}{{Maneva}, G.},
  \bibinfo{author}{{Manganaro}, M.}, \bibinfo{author}{{Mannheim}, K.},
  \bibinfo{author}{{Maraschi}, L.}, \bibinfo{author}{{Marcote}, B.},
  \bibinfo{author}{{Mariotti}, M.}, \bibinfo{author}{{Mart{\'\i}nez}, M.},
  \bibinfo{author}{{Mazin}, D.}, \bibinfo{author}{{Menzel}, U.},
  \bibinfo{author}{{Miranda}, J.M.}, \bibinfo{author}{{Mirzoyan}, R.},
  \bibinfo{author}{{Moralejo}, A.}, \bibinfo{author}{{Moretti}, E.},
  \bibinfo{author}{{Nakajima}, D.}, \bibinfo{author}{{Neustroev}, V.},
  \bibinfo{author}{{Niedzwiecki}, A.}, \bibinfo{author}{{Nievas Rosillo}, M.},
  \bibinfo{author}{{Nilsson}, K.}, \bibinfo{author}{{Nishijima}, K.},
  \bibinfo{author}{{Noda}, K.}, \bibinfo{year}{2016}.
\newblock \bibinfo{title}{{Multiwavelength Study of Quiescent States of Mrk 421
  with Unprecedented Hard X-Ray Coverage Provided by NuSTAR in 2013}}.
\newblock \bibinfo{journal}{\apj} \bibinfo{volume}{819}, \bibinfo{pages}{156}.
\newblock \DOIprefix\doi{10.3847/0004-637X/819/2/156},
  \href{http://arxiv.org/abs/1512.02235}{{\tt arXiv:1512.02235}}.
\bibitem[{Banerjee et~al.(2023)Banerjee, Sharma, Mandal, Das, Bhatta and
  Bose}]{banerjee2023detection}
\bibinfo{author}{Banerjee, A.}, \bibinfo{author}{Sharma, A.},
  \bibinfo{author}{Mandal, A.}, \bibinfo{author}{Das, A.K.},
  \bibinfo{author}{Bhatta, G.}, \bibinfo{author}{Bose, D.},
  \bibinfo{year}{2023}.
\newblock \bibinfo{title}{Detection of periodicity in the gamma-ray light curve
  of the bl lac 4fgl j2202. 7+ 4216}.
\newblock \bibinfo{journal}{Monthly Notices of the Royal Astronomical Society:
  Letters} \bibinfo{volume}{523}, \bibinfo{pages}{L52--L57}.
\bibitem[{Baumgartner et~al.(2013)Baumgartner, Tueller, Markwardt, Skinner,
  Barthelmy, Mushotzky, Evans and Gehrels}]{Baumgartner_2013}
\bibinfo{author}{Baumgartner, W.H.}, \bibinfo{author}{Tueller, J.},
  \bibinfo{author}{Markwardt, C.B.}, \bibinfo{author}{Skinner, G.K.},
  \bibinfo{author}{Barthelmy, S.}, \bibinfo{author}{Mushotzky, R.F.},
  \bibinfo{author}{Evans, P.A.}, \bibinfo{author}{Gehrels, N.},
  \bibinfo{year}{2013}.
\newblock \bibinfo{title}{The 70 month swift-bat all-sky hard x-ray survey}.
\newblock \bibinfo{journal}{The Astrophysical Journal Supplement Series}
  \bibinfo{volume}{207}, \bibinfo{pages}{19}.
\newblock \URLprefix \url{https://doi.org/10.1088/0067-0049/207/2/19},
  \DOIprefix\doi{10.1088/0067-0049/207/2/19}.
\bibitem[{{Bianchin} et~al.(2009){Bianchin}, {Foschini}, {Ghisellini},
  {Tagliaferri}, {Tavecchio}, {Treves}, {Di Cocco}, {Gliozzi}, {Pian},
  {Sambruna} and {Wolter}}]{2009A&A...496..423B}
\bibinfo{author}{{Bianchin}, V.}, \bibinfo{author}{{Foschini}, L.},
  \bibinfo{author}{{Ghisellini}, G.}, \bibinfo{author}{{Tagliaferri}, G.},
  \bibinfo{author}{{Tavecchio}, F.}, \bibinfo{author}{{Treves}, A.},
  \bibinfo{author}{{Di Cocco}, G.}, \bibinfo{author}{{Gliozzi}, M.},
  \bibinfo{author}{{Pian}, E.}, \bibinfo{author}{{Sambruna}, R.M.},
  \bibinfo{author}{{Wolter}, A.}, \bibinfo{year}{2009}.
\newblock \bibinfo{title}{{The changing look of PKS 2149-306}}.
\newblock \bibinfo{journal}{\aap} \bibinfo{volume}{496},
  \bibinfo{pages}{423--428}.
\newblock \DOIprefix\doi{10.1051/0004-6361/200811128},
  \href{http://arxiv.org/abs/0902.1789}{{\tt arXiv:0902.1789}}.
\bibitem[{Cavaliere and D'Elia(2002)}]{Cavaliere_2002}
\bibinfo{author}{Cavaliere, A.}, \bibinfo{author}{D'Elia, V.},
  \bibinfo{year}{2002}.
\newblock \bibinfo{title}{The blazar main sequence}.
\newblock \bibinfo{journal}{The Astrophysical Journal} \bibinfo{volume}{571},
  \bibinfo{pages}{226}.
\newblock \URLprefix \url{https://doi.org/10.1086/339778},
  \DOIprefix\doi{10.1086/339778}.
\bibitem[{Comastri et~al.(1997)Comastri, Fossati, Ghisellini and
  Molendi}]{comastri1997soft}
\bibinfo{author}{Comastri, A.}, \bibinfo{author}{Fossati, G.},
  \bibinfo{author}{Ghisellini, G.}, \bibinfo{author}{Molendi, S.},
  \bibinfo{year}{1997}.
\newblock \bibinfo{title}{On the soft x-ray spectra of $\gamma$-loud blazars}.
\newblock \bibinfo{journal}{The Astrophysical Journal} \bibinfo{volume}{480},
  \bibinfo{pages}{534--546}.
\bibitem[{D'Elia et~al.(2015)D'Elia, Padovani, Giommi and
  Turriziani}]{10.1093/mnras/stv573}
\bibinfo{author}{D'Elia, V.}, \bibinfo{author}{Padovani, P.},
  \bibinfo{author}{Giommi, P.}, \bibinfo{author}{Turriziani, S.},
  \bibinfo{year}{2015}.
\newblock \bibinfo{title}{Are many radio-selected bl lacs radio quasars in
  disguise?}
\newblock \bibinfo{journal}{Monthly Notices of the Royal Astronomical Society}
  \bibinfo{volume}{449}, \bibinfo{pages}{3517--3521}.
\newblock \URLprefix \url{https://doi.org/10.1093/mnras/stv573},
  \DOIprefix\doi{10.1093/mnras/stv573},
  \href{http://arxiv.org/abs/https://academic.oup.com/mnras/article-pdf/449/4/3517/18510031/stv573.pdf}{{\tt
  arXiv:https://academic.oup.com/mnras/article-pdf/449/4/3517/18510031/stv573.pdf}}.
\bibitem[{Dermer et~al.(2009)Dermer, Finke, Krug and Böttcher}]{Dermer_2009}
\bibinfo{author}{Dermer, C.D.}, \bibinfo{author}{Finke, J.D.},
  \bibinfo{author}{Krug, H.}, \bibinfo{author}{Böttcher, M.},
  \bibinfo{year}{2009}.
\newblock \bibinfo{title}{Gamma-ray studies of blazars: Synchro-compton
  analysis of flat spectrum radio quasars}.
\newblock \bibinfo{journal}{The Astrophysical Journal} \bibinfo{volume}{692},
  \bibinfo{pages}{32}.
\newblock \URLprefix \url{https://dx.doi.org/10.1088/0004-637X/692/1/32},
  \DOIprefix\doi{10.1088/0004-637X/692/1/32}.
\bibitem[{Ding et~al.(2023)Ding, Gu, Tang, Geng, Chen and Guo}]{Ding_2023}
\bibinfo{author}{Ding, N.}, \bibinfo{author}{Gu, Q.}, \bibinfo{author}{Tang,
  Y.}, \bibinfo{author}{Geng, X.}, \bibinfo{author}{Chen, Y.},
  \bibinfo{author}{Guo, X.}, \bibinfo{year}{2023}.
\newblock \bibinfo{title}{A special state transition in the blazar ot 081:
  Implication for the unified state transition paradigm of different-scale
  black hole systems}.
\newblock \bibinfo{journal}{The Astrophysical Journal} \bibinfo{volume}{944},
  \bibinfo{pages}{12}.
\newblock \URLprefix \url{https://doi.org/10.3847/1538-4357/acae97},
  \DOIprefix\doi{10.3847/1538-4357/acae97}.
\bibitem[{{Donato} et~al.(2001){Donato}, {Ghisellini}, {Tagliaferri} and
  {Fossati}}]{2001yCat..33750739D}
\bibinfo{author}{{Donato}, D.}, \bibinfo{author}{{Ghisellini}, G.},
  \bibinfo{author}{{Tagliaferri}, G.}, \bibinfo{author}{{Fossati}, G.},
  \bibinfo{year}{2001}.
\newblock \bibinfo{title}{{VizieR Online Data Catalog: Hard X-ray properties of
  blazars (Donato+, 2001)}}.
\newblock \bibinfo{journal}{VizieR Online Data Catalog}
  \DOIprefix\doi{10.26093/cds/vizier.33750739}.
\bibitem[{{Dondi} and {Ghisellini}(1995)}]{1995MNRAS.273..583D}
\bibinfo{author}{{Dondi}, L.}, \bibinfo{author}{{Ghisellini}, G.},
  \bibinfo{year}{1995}.
\newblock \bibinfo{title}{{Gamma-ray-loud blazars and beaming}}.
\newblock \bibinfo{journal}{\mnras} \bibinfo{volume}{273},
  \bibinfo{pages}{583--595}.
\newblock \DOIprefix\doi{10.1093/mnras/273.3.583}.
\bibitem[{{Finke}(2013)}]{2013ApJ...763..134F}
\bibinfo{author}{{Finke}, J.D.}, \bibinfo{year}{2013}.
\newblock \bibinfo{title}{{Compton Dominance and the Blazar Sequence}}.
\newblock \bibinfo{journal}{\apj} \bibinfo{volume}{763}, \bibinfo{pages}{134}.
\newblock \DOIprefix\doi{10.1088/0004-637X/763/2/134},
  \href{http://arxiv.org/abs/1212.0869}{{\tt arXiv:1212.0869}}.
\bibitem[{Fossati et~al.(1998)Fossati, Maraschi, Celotti, Comastri and
  Ghisellini}]{10.1046/j.1365-8711.1998.01828.x}
\bibinfo{author}{Fossati, G.}, \bibinfo{author}{Maraschi, L.},
  \bibinfo{author}{Celotti, A.}, \bibinfo{author}{Comastri, A.},
  \bibinfo{author}{Ghisellini, G.}, \bibinfo{year}{1998}.
\newblock \bibinfo{title}{A unifying view of the spectral energy distributions
  of blazars}.
\newblock \bibinfo{journal}{Monthly Notices of the Royal Astronomical Society}
  \bibinfo{volume}{299}, \bibinfo{pages}{433--448}.
\newblock \URLprefix \url{https://doi.org/10.1046/j.1365-8711.1998.01828.x},
  \DOIprefix\doi{10.1046/j.1365-8711.1998.01828.x},
  \href{http://arxiv.org/abs/https://academic.oup.com/mnras/article-pdf/299/2/433/3340292/299-2-433.pdf}{{\tt
  arXiv:https://academic.oup.com/mnras/article-pdf/299/2/433/3340292/299-2-433.pdf}}.
\bibitem[{Ghisellini et~al.(2017)Ghisellini, Righi, Costamante and
  Tavecchio}]{10.1093/mnras/stx806}
\bibinfo{author}{Ghisellini, G.}, \bibinfo{author}{Righi, C.},
  \bibinfo{author}{Costamante, L.}, \bibinfo{author}{Tavecchio, F.},
  \bibinfo{year}{2017}.
\newblock \bibinfo{title}{The fermi blazar sequence}.
\newblock \bibinfo{journal}{Monthly Notices of the Royal Astronomical Society}
  \bibinfo{volume}{469}, \bibinfo{pages}{255--266}.
\newblock \URLprefix \url{https://doi.org/10.1093/mnras/stx806},
  \DOIprefix\doi{10.1093/mnras/stx806},
  \href{http://arxiv.org/abs/https://academic.oup.com/mnras/article-pdf/469/1/255/17503062/stx806.pdf}{{\tt
  arXiv:https://academic.oup.com/mnras/article-pdf/469/1/255/17503062/stx806.pdf}}.
\bibitem[{{Ghisellini} et~al.(2013){Ghisellini}, {Tavecchio}, {Foschini},
  {Bonnoli} and {Tagliaferri}}]{2013MNRAS.432L..66G}
\bibinfo{author}{{Ghisellini}, G.}, \bibinfo{author}{{Tavecchio}, F.},
  \bibinfo{author}{{Foschini}, L.}, \bibinfo{author}{{Bonnoli}, G.},
  \bibinfo{author}{{Tagliaferri}, G.}, \bibinfo{year}{2013}.
\newblock \bibinfo{title}{{The red blazar PMN J2345-1555 becomes blue.}}
\newblock \bibinfo{journal}{\mnras} \bibinfo{volume}{432},
  \bibinfo{pages}{L66--L70}.
\newblock \DOIprefix\doi{10.1093/mnrasl/slt041},
  \href{http://arxiv.org/abs/1302.4444}{{\tt arXiv:1302.4444}}.
\bibitem[{Ghisellini et~al.(2012)Ghisellini, Tavecchio, Foschini, Sbarrato,
  Ghirlanda and Maraschi}]{10.1111/j.1365-2966.2012.21554.x}
\bibinfo{author}{Ghisellini, G.}, \bibinfo{author}{Tavecchio, F.},
  \bibinfo{author}{Foschini, L.}, \bibinfo{author}{Sbarrato, T.},
  \bibinfo{author}{Ghirlanda, G.}, \bibinfo{author}{Maraschi, L.},
  \bibinfo{year}{2012}.
\newblock \bibinfo{title}{Blue fermi flat spectrum radio quasars}.
\newblock \bibinfo{journal}{Monthly Notices of the Royal Astronomical Society}
  \bibinfo{volume}{425}, \bibinfo{pages}{1371--1379}.
\newblock \URLprefix \url{https://doi.org/10.1111/j.1365-2966.2012.21554.x},
  \DOIprefix\doi{10.1111/j.1365-2966.2012.21554.x},
  \href{http://arxiv.org/abs/https://academic.oup.com/mnras/article-pdf/425/2/1371/4027112/425-2-1371.pdf}{{\tt
  arXiv:https://academic.oup.com/mnras/article-pdf/425/2/1371/4027112/425-2-1371.pdf}}.
\bibitem[{Giommi et~al.(2019)Giommi, Brandt, de~Almeida, Pollock, Arneodo,
  Chang, Civitarese, De~Angelis, D'Elia, Vera et~al.}]{giommi2019open}
\bibinfo{author}{Giommi, P.}, \bibinfo{author}{Brandt, C.},
  \bibinfo{author}{de~Almeida, U.B.}, \bibinfo{author}{Pollock, A.},
  \bibinfo{author}{Arneodo, F.}, \bibinfo{author}{Chang, Y.},
  \bibinfo{author}{Civitarese, O.}, \bibinfo{author}{De~Angelis, M.},
  \bibinfo{author}{D'Elia, V.}, \bibinfo{author}{Vera, J.D.R.}, et~al.,
  \bibinfo{year}{2019}.
\newblock \bibinfo{title}{Open universe for blazars: a new generation of
  astronomical products based on 14 years of swift-xrt data}.
\newblock \bibinfo{journal}{Astronomy \& Astrophysics} \bibinfo{volume}{631},
  \bibinfo{pages}{A116}.
\bibitem[{Giommi et~al.(2012a)Giommi, Padovani, Polenta, Turriziani, D'Elia and
  Piranomonte}]{Giommi:2011sn}
\bibinfo{author}{Giommi, P.}, \bibinfo{author}{Padovani, P.},
  \bibinfo{author}{Polenta, G.}, \bibinfo{author}{Turriziani, S.},
  \bibinfo{author}{D'Elia, V.}, \bibinfo{author}{Piranomonte, S.},
  \bibinfo{year}{2012}a.
\newblock \bibinfo{title}{{A simplified view of blazars: clearing the fog
  around long-standing selection effects}}.
\newblock \bibinfo{journal}{Mon. Not. Roy. Astron. Soc.} \bibinfo{volume}{420},
  \bibinfo{pages}{2899}.
\newblock \DOIprefix\doi{10.1111/j.1365-2966.2011.20044.x},
  \href{http://arxiv.org/abs/1110.4706}{{\tt arXiv:1110.4706}}.
\bibitem[{Giommi et~al.(2012b)Giommi, Padovani, Polenta, Turriziani, D'Elia
  and Piranomonte}]{10.1111/j.1365-2966.2011.20044.x}
\bibinfo{author}{Giommi, P.}, \bibinfo{author}{Padovani, P.},
  \bibinfo{author}{Polenta, G.}, \bibinfo{author}{Turriziani, S.},
  \bibinfo{author}{D'Elia, V.}, \bibinfo{author}{Piranomonte, S.},
  \bibinfo{year}{2012}b.
\newblock \bibinfo{title}{A simplified view of blazars: clearing the fog around
  long-standing selection effects}.
\newblock \bibinfo{journal}{Monthly Notices of the Royal Astronomical Society}
  \bibinfo{volume}{420}, \bibinfo{pages}{2899--2911}.
\newblock \URLprefix \url{https://doi.org/10.1111/j.1365-2966.2011.20044.x},
  \DOIprefix\doi{10.1111/j.1365-2966.2011.20044.x},
  \href{http://arxiv.org/abs/https://academic.oup.com/mnras/article-pdf/420/4/2899/2933812/mnras0420-2899.pdf}{{\tt
  arXiv:https://academic.oup.com/mnras/article-pdf/420/4/2899/2933812/mnras0420-2899.pdf}}.
\bibitem[{{Giommi} et~al.(2021){Giommi}, {Perri}, {Capalbi}, {D'Elia}, {Barres
  de Almeida}, {Brandt}, {Pollock}, {Arneodo}, {Di Giovanni}, {Chang},
  {Civitarese}, {De Angelis}, {Leto}, {Verrecchia}, {Ricard}, {Di Pippo},
  {Middei}, {Penacchioni}, {Ruffini}, {Sahakyan}, {Israyelyan} and
  {Turriziani}}]{2021MNRAS.507.5690G}
\bibinfo{author}{{Giommi}, P.}, \bibinfo{author}{{Perri}, M.},
  \bibinfo{author}{{Capalbi}, M.}, \bibinfo{author}{{D'Elia}, V.},
  \bibinfo{author}{{Barres de Almeida}, U.}, \bibinfo{author}{{Brandt}, C.H.},
  \bibinfo{author}{{Pollock}, A.M.T.}, \bibinfo{author}{{Arneodo}, F.},
  \bibinfo{author}{{Di Giovanni}, A.}, \bibinfo{author}{{Chang}, Y.L.},
  \bibinfo{author}{{Civitarese}, O.}, \bibinfo{author}{{De Angelis}, M.},
  \bibinfo{author}{{Leto}, C.}, \bibinfo{author}{{Verrecchia}, F.},
  \bibinfo{author}{{Ricard}, N.}, \bibinfo{author}{{Di Pippo}, S.},
  \bibinfo{author}{{Middei}, R.}, \bibinfo{author}{{Penacchioni}, A.V.},
  \bibinfo{author}{{Ruffini}, R.}, \bibinfo{author}{{Sahakyan}, N.},
  \bibinfo{author}{{Israyelyan}, D.}, \bibinfo{author}{{Turriziani}, S.},
  \bibinfo{year}{2021}.
\newblock \bibinfo{title}{{X-ray spectra, light curves and SEDs of blazars
  frequently observed by Swift}}.
\newblock \bibinfo{journal}{\mnras} \bibinfo{volume}{507},
  \bibinfo{pages}{5690--5702}.
\newblock \DOIprefix\doi{10.1093/mnras/stab2425},
  \href{http://arxiv.org/abs/2108.07255}{{\tt arXiv:2108.07255}}.
\bibitem[{Gokus et~al.(2024)Gokus, Wilms, Kadler, Dorner, Nowak, Kreikenbohm,
  Leiter, Bretz, Schleicher, Markowitz, Pottschmidt, Mannheim, Kreykenbohm,
  Langejahn, McBride, Beuchert, Dauser, Kreter, Abhir, Baack, Balbo, Biland,
  Brand, Buss, Eisenberger, Elsaesser, Günther, Hildebrand, Linhoff, Paravac,
  Rhode, Sliusar, Hasan and Walter}]{10.1093/mnras/stae643}
\bibinfo{author}{Gokus, A.}, \bibinfo{author}{Wilms, J.},
  \bibinfo{author}{Kadler, M.}, \bibinfo{author}{Dorner, D.},
  \bibinfo{author}{Nowak, M.A.}, \bibinfo{author}{Kreikenbohm, A.},
  \bibinfo{author}{Leiter, K.}, \bibinfo{author}{Bretz, T.},
  \bibinfo{author}{Schleicher, B.}, \bibinfo{author}{Markowitz, A.G.},
  \bibinfo{author}{Pottschmidt, K.}, \bibinfo{author}{Mannheim, K.},
  \bibinfo{author}{Kreykenbohm, I.}, \bibinfo{author}{Langejahn, M.},
  \bibinfo{author}{McBride, F.}, \bibinfo{author}{Beuchert, T.},
  \bibinfo{author}{Dauser, T.}, \bibinfo{author}{Kreter, M.},
  \bibinfo{author}{Abhir, J.}, \bibinfo{author}{Baack, D.},
  \bibinfo{author}{Balbo, M.}, \bibinfo{author}{Biland, A.},
  \bibinfo{author}{Brand, K.}, \bibinfo{author}{Buss, J.},
  \bibinfo{author}{Eisenberger, L.}, \bibinfo{author}{Elsaesser, D.},
  \bibinfo{author}{Günther, P.}, \bibinfo{author}{Hildebrand, D.},
  \bibinfo{author}{Linhoff, M.}, \bibinfo{author}{Paravac, A.},
  \bibinfo{author}{Rhode, W.}, \bibinfo{author}{Sliusar, V.},
  \bibinfo{author}{Hasan, S.}, \bibinfo{author}{Walter, R.},
  \bibinfo{year}{2024}.
\newblock \bibinfo{title}{Rapid variability of markarian 421 during extreme
  flaring as seen through the eyes of xmm–newton}.
\newblock \bibinfo{journal}{Monthly Notices of the Royal Astronomical Society}
  \bibinfo{volume}{529}, \bibinfo{pages}{1450--1462}.
\newblock \URLprefix \url{https://doi.org/10.1093/mnras/stae643},
  \DOIprefix\doi{10.1093/mnras/stae643},
  \href{http://arxiv.org/abs/https://academic.oup.com/mnras/article-pdf/529/2/1450/56948581/stae643.pdf}{{\tt
  arXiv:https://academic.oup.com/mnras/article-pdf/529/2/1450/56948581/stae643.pdf}}.
\bibitem[{{H.~E.~S.~S. Collaboration} et~al.(2010){H.~E.~S.~S. Collaboration},
  {Abramowski}, {Acero}, {Aharonian}, {Akhperjanian}, {Anton}, {Barres de
  Almeida}, {Bazer-Bachi}, {Becherini}, {Behera}, {Benbow}, {Bernl{\"o}hr},
  {Bochow}, {Boisson}, {Bolmont}, {Borrel}, {Brucker}, {Brun}, {Brun},
  {B{\"u}hler}, {Bulik}, {B{\"u}sching}, {Boutelier}, {Chadwick},
  {Charbonnier}, {Chaves}, {Cheesebrough}, {Conrad}, {Chounet}, {Clapson},
  {Coignet}, {Costamante}, {Dalton}, {Daniel}, {Davids}, {Degrange}, {Deil},
  {Dickinson}, {Djannati-Ata{\"\i}}, {Domainko}, {O'C. Drury}, {Dubois},
  {Dubus}, {Dyks}, {Dyrda}, {Egberts}, {Eger}, {Espigat}, {Fallon}, {Farnier},
  {Fegan}, {Feinstein}, {Fernandes}, {Fiasson}, {F{\"o}rster}, {Fontaine},
  {F{\"u}{\ss}ling}, {Gabici}, {Gallant}, {G{\'e}rard}, {Gerbig}, {Giebels},
  {Glicenstein}, {Gl{\"u}ck}, {Goret}, {G{\"o}ring}, {Hampf}, {Hauser},
  {Heinz}, {Heinzelmann}, {Henri}, {Hermann}, {Hinton}, {Hoffmann}, {Hofmann},
  {Hofverberg}, {Holleran}, {Hoppe}, {Horns}, {Jacholkowska}, {de Jager},
  {Jahn}, {Jung}, {Katarzy{\'n}ski}, {Katz}, {Kaufmann}, {Kerschhaggl},
  {Khangulyan}, {Kh{\'e}lifi}, {Keogh}, {Klochkov}, {Klu{\v{z}}niak},
  {Kneiske}, {Komin}, {Kosack}, {Kossakowski}, {Lamanna}, {Lenain}, {Lohse},
  {Lu}, {Marandon}, {Marcowith}, {Masbou}, {Maurin}, {McComb}, {Medina},
  {M{\'e}hault}, {Moderski}, {Moulin}, {Naumann-Godo}, {de Naurois}, {Nedbal},
  {Nekrassov}, {Nguyen}, {Nicholas}, {Niemiec}, {Nolan}, {Ohm}, {Olive}, {de
  O{\~n}a Wilhelmi}, {Opitz}, {Orford}, {Ostrowski}, {Panter}, {Paz Arribas},
  {Pedaletti}, {Pelletier}, {Petrucci}, {Pita}, {P{\"u}hlhofer}, {Punch},
  {Quirrenbach}, {Raubenheimer}, {Raue}, {Rayner}, {Reimer}, {Renaud}, {de Los
  Reyes}, {Rieger}, {Ripken}, {Rob}, {Rosier-Lees}, {Rowell}, {Rudak},
  {Rulten}, {Ruppel}, {Ryde}, {Sahakian}, {Santangelo}, {Schlickeiser},
  {Sch{\"o}ck}, {Sch{\"o}nwald}, {Schwanke}, {Schwarzburg}, {Schwemmer},
  {Shalchi}, {Sushch}, {Sikora}, {Skilton}, {Sol}, {Stawarz}, {Steenkamp},
  {Stegmann}, {Stinzing}, {Szostek}, {Tam}, {Tavernet}, {Terrier}, {Tibolla},
  {Tluczykont}, {Valerius}, {van Eldik}, {Vasileiadis}, {Venter}, {Venter},
  {Vialle}, {Viana}, {Vincent}, {Vivier}, {V{\"o}lk}, {Volpe}, {Vorobiov},
  {Wagner}, {Ward}, {Zdziarski}, {Zech} and {Zechlin}}]{2010A&A...516A..56H}
\bibinfo{author}{{H.~E.~S.~S. Collaboration}}, \bibinfo{author}{{Abramowski},
  A.}, \bibinfo{author}{{Acero}, F.}, \bibinfo{author}{{Aharonian}, F.},
  \bibinfo{author}{{Akhperjanian}, A.G.}, \bibinfo{author}{{Anton}, G.},
  \bibinfo{author}{{Barres de Almeida}, U.}, \bibinfo{author}{{Bazer-Bachi},
  A.R.}, \bibinfo{author}{{Becherini}, Y.}, \bibinfo{author}{{Behera}, B.},
  \bibinfo{author}{{Benbow}, W.}, \bibinfo{author}{{Bernl{\"o}hr}, K.},
  \bibinfo{author}{{Bochow}, A.}, \bibinfo{author}{{Boisson}, C.},
  \bibinfo{author}{{Bolmont}, J.}, \bibinfo{author}{{Borrel}, V.},
  \bibinfo{author}{{Brucker}, J.}, \bibinfo{author}{{Brun}, F.},
  \bibinfo{author}{{Brun}, P.}, \bibinfo{author}{{B{\"u}hler}, R.},
  \bibinfo{author}{{Bulik}, T.}, \bibinfo{author}{{B{\"u}sching}, I.},
  \bibinfo{author}{{Boutelier}, T.}, \bibinfo{author}{{Chadwick}, P.M.},
  \bibinfo{author}{{Charbonnier}, A.}, \bibinfo{author}{{Chaves}, R.C.G.},
  \bibinfo{author}{{Cheesebrough}, A.}, \bibinfo{author}{{Conrad}, J.},
  \bibinfo{author}{{Chounet}, L.M.}, \bibinfo{author}{{Clapson}, A.C.},
  \bibinfo{author}{{Coignet}, G.}, \bibinfo{author}{{Costamante}, L.},
  \bibinfo{author}{{Dalton}, M.}, \bibinfo{author}{{Daniel}, M.K.},
  \bibinfo{author}{{Davids}, I.D.}, \bibinfo{author}{{Degrange}, B.},
  \bibinfo{author}{{Deil}, C.}, \bibinfo{author}{{Dickinson}, H.J.},
  \bibinfo{author}{{Djannati-Ata{\"\i}}, A.}, \bibinfo{author}{{Domainko}, W.},
  \bibinfo{author}{{O'C. Drury}, L.}, \bibinfo{author}{{Dubois}, F.},
  \bibinfo{author}{{Dubus}, G.}, \bibinfo{author}{{Dyks}, J.},
  \bibinfo{author}{{Dyrda}, M.}, \bibinfo{author}{{Egberts}, K.},
  \bibinfo{author}{{Eger}, P.}, \bibinfo{author}{{Espigat}, P.},
  \bibinfo{author}{{Fallon}, L.}, \bibinfo{author}{{Farnier}, C.},
  \bibinfo{author}{{Fegan}, S.}, \bibinfo{author}{{Feinstein}, F.},
  \bibinfo{author}{{Fernandes}, M.V.}, \bibinfo{author}{{Fiasson}, A.},
  \bibinfo{author}{{F{\"o}rster}, A.}, \bibinfo{author}{{Fontaine}, G.},
  \bibinfo{author}{{F{\"u}{\ss}ling}, M.}, \bibinfo{author}{{Gabici}, S.},
  \bibinfo{author}{{Gallant}, Y.A.}, \bibinfo{author}{{G{\'e}rard}, L.},
  \bibinfo{author}{{Gerbig}, D.}, \bibinfo{author}{{Giebels}, B.},
  \bibinfo{author}{{Glicenstein}, J.F.}, \bibinfo{author}{{Gl{\"u}ck}, B.},
  \bibinfo{author}{{Goret}, P.}, \bibinfo{author}{{G{\"o}ring}, D.},
  \bibinfo{author}{{Hampf}, D.}, \bibinfo{author}{{Hauser}, M.},
  \bibinfo{author}{{Heinz}, S.}, \bibinfo{author}{{Heinzelmann}, G.},
  \bibinfo{author}{{Henri}, G.}, \bibinfo{author}{{Hermann}, G.},
  \bibinfo{author}{{Hinton}, J.A.}, \bibinfo{author}{{Hoffmann}, A.},
  \bibinfo{author}{{Hofmann}, W.}, \bibinfo{author}{{Hofverberg}, P.},
  \bibinfo{author}{{Holleran}, M.}, \bibinfo{author}{{Hoppe}, S.},
  \bibinfo{author}{{Horns}, D.}, \bibinfo{author}{{Jacholkowska}, A.},
  \bibinfo{author}{{de Jager}, O.C.}, \bibinfo{author}{{Jahn}, C.},
  \bibinfo{author}{{Jung}, I.}, \bibinfo{author}{{Katarzy{\'n}ski}, K.},
  \bibinfo{author}{{Katz}, U.}, \bibinfo{author}{{Kaufmann}, S.},
  \bibinfo{author}{{Kerschhaggl}, M.}, \bibinfo{author}{{Khangulyan}, D.},
  \bibinfo{author}{{Kh{\'e}lifi}, B.}, \bibinfo{author}{{Keogh}, D.},
  \bibinfo{author}{{Klochkov}, D.}, \bibinfo{author}{{Klu{\v{z}}niak}, W.},
  \bibinfo{author}{{Kneiske}, T.}, \bibinfo{author}{{Komin}, N.},
  \bibinfo{author}{{Kosack}, K.}, \bibinfo{author}{{Kossakowski}, R.},
  \bibinfo{author}{{Lamanna}, G.}, \bibinfo{author}{{Lenain}, J.P.},
  \bibinfo{author}{{Lohse}, T.}, \bibinfo{author}{{Lu}, C.C.},
  \bibinfo{author}{{Marandon}, V.}, \bibinfo{author}{{Marcowith}, A.},
  \bibinfo{author}{{Masbou}, J.}, \bibinfo{author}{{Maurin}, D.},
  \bibinfo{author}{{McComb}, T.J.L.}, \bibinfo{author}{{Medina}, M.C.},
  \bibinfo{author}{{M{\'e}hault}, J.}, \bibinfo{author}{{Moderski}, R.},
  \bibinfo{author}{{Moulin}, E.}, \bibinfo{author}{{Naumann-Godo}, M.},
  \bibinfo{author}{{de Naurois}, M.}, \bibinfo{author}{{Nedbal}, D.},
  \bibinfo{author}{{Nekrassov}, D.}, \bibinfo{author}{{Nguyen}, N.},
  \bibinfo{author}{{Nicholas}, B.}, \bibinfo{author}{{Niemiec}, J.},
  \bibinfo{author}{{Nolan}, S.J.}, \bibinfo{author}{{Ohm}, S.},
  \bibinfo{author}{{Olive}, J.F.}, \bibinfo{author}{{de O{\~n}a Wilhelmi}, E.},
  \bibinfo{author}{{Opitz}, B.}, \bibinfo{author}{{Orford}, K.J.},
  \bibinfo{author}{{Ostrowski}, M.}, \bibinfo{author}{{Panter}, M.},
  \bibinfo{author}{{Paz Arribas}, M.}, \bibinfo{author}{{Pedaletti}, G.},
  \bibinfo{author}{{Pelletier}, G.}, \bibinfo{author}{{Petrucci}, P.O.},
  \bibinfo{author}{{Pita}, S.}, \bibinfo{author}{{P{\"u}hlhofer}, G.},
  \bibinfo{author}{{Punch}, M.}, \bibinfo{author}{{Quirrenbach}, A.},
  \bibinfo{author}{{Raubenheimer}, B.C.}, \bibinfo{author}{{Raue}, M.},
  \bibinfo{author}{{Rayner}, S.M.}, \bibinfo{author}{{Reimer}, O.},
  \bibinfo{author}{{Renaud}, M.}, \bibinfo{author}{{de Los Reyes}, R.},
  \bibinfo{author}{{Rieger}, F.}, \bibinfo{author}{{Ripken}, J.},
  \bibinfo{author}{{Rob}, L.}, \bibinfo{author}{{Rosier-Lees}, S.},
  \bibinfo{author}{{Rowell}, G.}, \bibinfo{author}{{Rudak}, B.},
  \bibinfo{author}{{Rulten}, C.B.}, \bibinfo{author}{{Ruppel}, J.},
  \bibinfo{author}{{Ryde}, F.}, \bibinfo{author}{{Sahakian}, V.},
  \bibinfo{author}{{Santangelo}, A.}, \bibinfo{author}{{Schlickeiser}, R.},
  \bibinfo{author}{{Sch{\"o}ck}, F.M.}, \bibinfo{author}{{Sch{\"o}nwald}, A.},
  \bibinfo{author}{{Schwanke}, U.}, \bibinfo{author}{{Schwarzburg}, S.},
  \bibinfo{author}{{Schwemmer}, S.}, \bibinfo{author}{{Shalchi}, A.},
  \bibinfo{author}{{Sushch}, I.}, \bibinfo{author}{{Sikora}, M.},
  \bibinfo{author}{{Skilton}, J.L.}, \bibinfo{author}{{Sol}, H.},
  \bibinfo{author}{{Stawarz}, {\L}.}, \bibinfo{author}{{Steenkamp}, R.},
  \bibinfo{author}{{Stegmann}, C.}, \bibinfo{author}{{Stinzing}, F.},
  \bibinfo{author}{{Szostek}, A.}, \bibinfo{author}{{Tam}, P.H.},
  \bibinfo{author}{{Tavernet}, J.P.}, \bibinfo{author}{{Terrier}, R.},
  \bibinfo{author}{{Tibolla}, O.}, \bibinfo{author}{{Tluczykont}, M.},
  \bibinfo{author}{{Valerius}, K.}, \bibinfo{author}{{van Eldik}, C.},
  \bibinfo{author}{{Vasileiadis}, G.}, \bibinfo{author}{{Venter}, C.},
  \bibinfo{author}{{Venter}, L.}, \bibinfo{author}{{Vialle}, J.P.},
  \bibinfo{author}{{Viana}, A.}, \bibinfo{author}{{Vincent}, P.},
  \bibinfo{author}{{Vivier}, M.}, \bibinfo{author}{{V{\"o}lk}, H.J.},
  \bibinfo{author}{{Volpe}, F.}, \bibinfo{author}{{Vorobiov}, S.},
  \bibinfo{author}{{Wagner}, S.J.}, \bibinfo{author}{{Ward}, M.},
  \bibinfo{author}{{Zdziarski}, A.A.}, \bibinfo{author}{{Zech}, A.},
  \bibinfo{author}{{Zechlin}, H.S.}, \bibinfo{year}{2010}.
\newblock \bibinfo{title}{{Multi-wavelength observations of H 2356-309}}.
\newblock \bibinfo{journal}{\aap} \bibinfo{volume}{516}, \bibinfo{pages}{A56}.
\newblock \DOIprefix\doi{10.1051/0004-6361/201014321},
  \href{http://arxiv.org/abs/1004.2089}{{\tt arXiv:1004.2089}}.
\bibitem[{{Haemmerich} et~al.(2025){Haemmerich}, {Gokus}, {McBride}, {Weber},
  {Marcotulli}, {Zainab}, {Collmar}, {Salvato}, {Wolf}, {Sbarrato},
  {Belladitta}, {Buchner}, {Saeedi}, {Dauner}, {Lorenz}, {Koenig}, {Kirsch},
  {Berger}, {Bahic}, {Tubin-Arenas}, {Krumpe}, {Homan}, {Markowitz}, {Benke},
  {Roesch}, {Rajasekar Kavitha}, {Tambe}, {Kadler}, {Ros}, {Ojha} and
  {Wilms}}]{2025arXiv251025589H}
\bibinfo{author}{{Haemmerich}, S.}, \bibinfo{author}{{Gokus}, A.},
  \bibinfo{author}{{McBride}, F.}, \bibinfo{author}{{Weber}, P.},
  \bibinfo{author}{{Marcotulli}, L.}, \bibinfo{author}{{Zainab}, A.},
  \bibinfo{author}{{Collmar}, W.}, \bibinfo{author}{{Salvato}, M.},
  \bibinfo{author}{{Wolf}, J.}, \bibinfo{author}{{Sbarrato}, T.},
  \bibinfo{author}{{Belladitta}, S.}, \bibinfo{author}{{Buchner}, J.},
  \bibinfo{author}{{Saeedi}, S.}, \bibinfo{author}{{Dauner}, L.},
  \bibinfo{author}{{Lorenz}, M.}, \bibinfo{author}{{Koenig}, O.},
  \bibinfo{author}{{Kirsch}, C.}, \bibinfo{author}{{Berger}, K.},
  \bibinfo{author}{{Bahic}, S.}, \bibinfo{author}{{Tubin-Arenas}, D.},
  \bibinfo{author}{{Krumpe}, M.}, \bibinfo{author}{{Homan}, D.},
  \bibinfo{author}{{Markowitz}, A.}, \bibinfo{author}{{Benke}, P.},
  \bibinfo{author}{{Roesch}, F.}, \bibinfo{author}{{Rajasekar Kavitha}, P.},
  \bibinfo{author}{{Tambe}, H.}, \bibinfo{author}{{Kadler}, M.},
  \bibinfo{author}{{Ros}, E.}, \bibinfo{author}{{Ojha}, R.},
  \bibinfo{author}{{Wilms}, J.}, \bibinfo{year}{2025}.
\newblock \bibinfo{title}{{BlazEr1: The eROSITA Blazar Catalog. Blazars and
  Blazar Candidates in the First eROSITA Survey}}.
\newblock \bibinfo{journal}{arXiv e-prints} ,
  \bibinfo{pages}{arXiv:2510.25589}\DOIprefix\doi{10.48550/arXiv.2510.25589},
  \href{http://arxiv.org/abs/2510.25589}{{\tt arXiv:2510.25589}}.
\bibitem[{{Harrison} et~al.(2013){Harrison}, {Craig}, {Christensen}, {Hailey},
  {Zhang}, {Boggs}, {Stern}, {Cook}, {Forster}, {Giommi}, {Grefenstette},
  {Kim}, {Kitaguchi}, {Koglin}, {Madsen}, {Mao}, {Miyasaka}, {Mori}, {Perri},
  {Pivovaroff}, {Puccetti}, {Rana}, {Westergaard}, {Willis}, {Zoglauer}, {An},
  {Bachetti}, {Barri{\`e}re}, {Bellm}, {Bhalerao}, {Brejnholt}, {Fuerst},
  {Liebe}, {Markwardt}, {Nynka}, {Vogel}, {Walton}, {Wik}, {Alexander},
  {Cominsky}, {Hornschemeier}, {Hornstrup}, {Kaspi}, {Madejski}, {Matt},
  {Molendi}, {Smith}, {Tomsick}, {Ajello}, {Ballantyne}, {Balokovi{\'c}},
  {Barret}, {Bauer}, {Blandford}, {Brandt}, {Brenneman}, {Chiang},
  {Chakrabarty}, {Chenevez}, {Comastri}, {Dufour}, {Elvis}, {Fabian}, {Farrah},
  {Fryer}, {Gotthelf}, {Grindlay}, {Helfand}, {Krivonos}, {Meier}, {Miller},
  {Natalucci}, {Ogle}, {Ofek}, {Ptak}, {Reynolds}, {Rigby}, {Tagliaferri},
  {Thorsett}, {Treister} and {Urry}}]{2013ApJ...770..103H}
\bibinfo{author}{{Harrison}, F.A.}, \bibinfo{author}{{Craig}, W.W.},
  \bibinfo{author}{{Christensen}, F.E.}, \bibinfo{author}{{Hailey}, C.J.},
  \bibinfo{author}{{Zhang}, W.W.}, \bibinfo{author}{{Boggs}, S.E.},
  \bibinfo{author}{{Stern}, D.}, \bibinfo{author}{{Cook}, W.R.},
  \bibinfo{author}{{Forster}, K.}, \bibinfo{author}{{Giommi}, P.},
  \bibinfo{author}{{Grefenstette}, B.W.}, \bibinfo{author}{{Kim}, Y.},
  \bibinfo{author}{{Kitaguchi}, T.}, \bibinfo{author}{{Koglin}, J.E.},
  \bibinfo{author}{{Madsen}, K.K.}, \bibinfo{author}{{Mao}, P.H.},
  \bibinfo{author}{{Miyasaka}, H.}, \bibinfo{author}{{Mori}, K.},
  \bibinfo{author}{{Perri}, M.}, \bibinfo{author}{{Pivovaroff}, M.J.},
  \bibinfo{author}{{Puccetti}, S.}, \bibinfo{author}{{Rana}, V.R.},
  \bibinfo{author}{{Westergaard}, N.J.}, \bibinfo{author}{{Willis}, J.},
  \bibinfo{author}{{Zoglauer}, A.}, \bibinfo{author}{{An}, H.},
  \bibinfo{author}{{Bachetti}, M.}, \bibinfo{author}{{Barri{\`e}re}, N.M.},
  \bibinfo{author}{{Bellm}, E.C.}, \bibinfo{author}{{Bhalerao}, V.},
  \bibinfo{author}{{Brejnholt}, N.F.}, \bibinfo{author}{{Fuerst}, F.},
  \bibinfo{author}{{Liebe}, C.C.}, \bibinfo{author}{{Markwardt}, C.B.},
  \bibinfo{author}{{Nynka}, M.}, \bibinfo{author}{{Vogel}, J.K.},
  \bibinfo{author}{{Walton}, D.J.}, \bibinfo{author}{{Wik}, D.R.},
  \bibinfo{author}{{Alexander}, D.M.}, \bibinfo{author}{{Cominsky}, L.R.},
  \bibinfo{author}{{Hornschemeier}, A.E.}, \bibinfo{author}{{Hornstrup}, A.},
  \bibinfo{author}{{Kaspi}, V.M.}, \bibinfo{author}{{Madejski}, G.M.},
  \bibinfo{author}{{Matt}, G.}, \bibinfo{author}{{Molendi}, S.},
  \bibinfo{author}{{Smith}, D.M.}, \bibinfo{author}{{Tomsick}, J.A.},
  \bibinfo{author}{{Ajello}, M.}, \bibinfo{author}{{Ballantyne}, D.R.},
  \bibinfo{author}{{Balokovi{\'c}}, M.}, \bibinfo{author}{{Barret}, D.},
  \bibinfo{author}{{Bauer}, F.E.}, \bibinfo{author}{{Blandford}, R.D.},
  \bibinfo{author}{{Brandt}, W.N.}, \bibinfo{author}{{Brenneman}, L.W.},
  \bibinfo{author}{{Chiang}, J.}, \bibinfo{author}{{Chakrabarty}, D.},
  \bibinfo{author}{{Chenevez}, J.}, \bibinfo{author}{{Comastri}, A.},
  \bibinfo{author}{{Dufour}, F.}, \bibinfo{author}{{Elvis}, M.},
  \bibinfo{author}{{Fabian}, A.C.}, \bibinfo{author}{{Farrah}, D.},
  \bibinfo{author}{{Fryer}, C.L.}, \bibinfo{author}{{Gotthelf}, E.V.},
  \bibinfo{author}{{Grindlay}, J.E.}, \bibinfo{author}{{Helfand}, D.J.},
  \bibinfo{author}{{Krivonos}, R.}, \bibinfo{author}{{Meier}, D.L.},
  \bibinfo{author}{{Miller}, J.M.}, \bibinfo{author}{{Natalucci}, L.},
  \bibinfo{author}{{Ogle}, P.}, \bibinfo{author}{{Ofek}, E.O.},
  \bibinfo{author}{{Ptak}, A.}, \bibinfo{author}{{Reynolds}, S.P.},
  \bibinfo{author}{{Rigby}, J.R.}, \bibinfo{author}{{Tagliaferri}, G.},
  \bibinfo{author}{{Thorsett}, S.E.}, \bibinfo{author}{{Treister}, E.},
  \bibinfo{author}{{Urry}, C.M.}, \bibinfo{year}{2013}.
\newblock \bibinfo{title}{{The Nuclear Spectroscopic Telescope Array (NuSTAR)
  High-energy X-Ray Mission}}.
\newblock \bibinfo{journal}{\apj} \bibinfo{volume}{770}, \bibinfo{pages}{103}.
\newblock \DOIprefix\doi{10.1088/0004-637X/770/2/103},
  \href{http://arxiv.org/abs/1301.7307}{{\tt arXiv:1301.7307}}.
\bibitem[{Hota et~al.(2024)Hota, Khatoon, Misra and Pradhan}]{Hota_2024}
\bibinfo{author}{Hota, J.}, \bibinfo{author}{Khatoon, R.},
  \bibinfo{author}{Misra, R.}, \bibinfo{author}{Pradhan, A.C.},
  \bibinfo{year}{2024}.
\newblock \bibinfo{title}{Multiwavelength study of extreme high-energy peaked
  bl lac source 1es 0229+200 using ultraviolet, x-ray, and γ-ray
  observations}.
\newblock \bibinfo{journal}{The Astrophysical Journal} \bibinfo{volume}{976},
  \bibinfo{pages}{69}.
\newblock \URLprefix \url{https://doi.org/10.3847/1538-4357/ad8085},
  \DOIprefix\doi{10.3847/1538-4357/ad8085}.
\bibitem[{Hota et~al.(2021)Hota, Shah, Khatoon, Misra, Pradhan and
  Gogoi}]{Hota:2021csa}
\bibinfo{author}{Hota, J.}, \bibinfo{author}{Shah, Z.},
  \bibinfo{author}{Khatoon, R.}, \bibinfo{author}{Misra, R.},
  \bibinfo{author}{Pradhan, A.C.}, \bibinfo{author}{Gogoi, R.},
  \bibinfo{year}{2021}.
\newblock \bibinfo{title}{{Understanding the X-ray spectral curvature of
  Mkn\,421 using broad-band AstroSat observations}}.
\newblock \bibinfo{journal}{Mon. Not. Roy. Astron. Soc.} \bibinfo{volume}{508},
  \bibinfo{pages}{5921--5934}.
\newblock \DOIprefix\doi{10.1093/mnras/stab2903},
  \href{http://arxiv.org/abs/2110.03344}{{\tt arXiv:2110.03344}}.
\bibitem[{Kadler(2005)}]{Kadler2005}
\bibinfo{author}{Kadler, M.}, \bibinfo{year}{2005}.
\newblock \bibinfo{title}{The X-ray Spectral Evolution of Blazars}.
\newblock Ph.D. thesis. Rheinische Friedrich-Wilhelms-Universität Bonn.
  \bibinfo{address}{Bonn, Germany}.
\bibitem[{{Kang} et~al.(2023){Kang}, {Zheng} and {Wu}}]{2023MNRAS.tmp.2362K}
\bibinfo{author}{{Kang}, S.J.}, \bibinfo{author}{{Zheng}, Y.G.},
  \bibinfo{author}{{Wu}, Q.}, \bibinfo{year}{2023}.
\newblock \bibinfo{title}{{Hunting for the candidates of misclassified sources
  in LSP BL Lacs using Machine learning}}.
\newblock \bibinfo{journal}{\mnras} \DOIprefix\doi{10.1093/mnras/stad2456},
  \href{http://arxiv.org/abs/2308.05794}{{\tt arXiv:2308.05794}}.
\bibitem[{Kizhakkekalam et~al.(2025)Kizhakkekalam, Bhatta, K and
  Adhikari}]{Kizhakkekalam:2025moz}
\bibinfo{author}{Kizhakkekalam, S.}, \bibinfo{author}{Bhatta, G.},
  \bibinfo{author}{K, N.P.}, \bibinfo{author}{Adhikari, T.P.},
  \bibinfo{year}{2025}.
\newblock \bibinfo{title}{Nicer perspective on tev blazar
  mrk{\textasciitilde}421: X-ray variability and particle acceleration}.
\newblock \bibinfo{journal}{arXiv e-prints}
  \href{http://arxiv.org/abs/2512.08531}{{\tt arXiv:2512.08531}}.
\bibitem[{{Lewis} et~al.(2016){Lewis}, {Becker} and
  {Finke}}]{2016ApJ...824..108L}
\bibinfo{author}{{Lewis}, T.R.}, \bibinfo{author}{{Becker}, P.A.},
  \bibinfo{author}{{Finke}, J.D.}, \bibinfo{year}{2016}.
\newblock \bibinfo{title}{{Time-dependent Electron Acceleration in Blazar
  Transients: X-Ray Time Lags and Spectral Formation}}.
\newblock \bibinfo{journal}{\apj} \bibinfo{volume}{824}, \bibinfo{pages}{108}.
\newblock \DOIprefix\doi{10.3847/0004-637X/824/2/108},
  \href{http://arxiv.org/abs/1603.07386}{{\tt arXiv:1603.07386}}.
\bibitem[{{Middei} et~al.(2022){Middei}, {Giommi}, {Perri}, {Turriziani},
  {Sahakyan}, {Chang}, {Leto} and {Verrecchia}}]{2022MNRAS.514.3179M}
\bibinfo{author}{{Middei}, R.}, \bibinfo{author}{{Giommi}, P.},
  \bibinfo{author}{{Perri}, M.}, \bibinfo{author}{{Turriziani}, S.},
  \bibinfo{author}{{Sahakyan}, N.}, \bibinfo{author}{{Chang}, Y.L.},
  \bibinfo{author}{{Leto}, C.}, \bibinfo{author}{{Verrecchia}, F.},
  \bibinfo{year}{2022}.
\newblock \bibinfo{title}{{The first hard X-ray spectral catalogue of Blazars
  observed by NuSTAR}}.
\newblock \bibinfo{journal}{\mnras} \bibinfo{volume}{514},
  \bibinfo{pages}{3179--3190}.
\newblock \DOIprefix\doi{10.1093/mnras/stac1185},
  \href{http://arxiv.org/abs/2205.05089}{{\tt arXiv:2205.05089}}.
\bibitem[{{Miyaji} et~al.(2024){Miyaji}, {Bravo-Navarro}, {D{\'\i}az Tello},
  {Krumpe}, {Herrera-Endoqui}, {Ikeda}, {Takagi}, {Oi}, {Shogaki}, {Matsuura},
  {Kim}, {Malkan}, {Hwang}, {Kim}, {Ishigaki}, {Hanami}, {Kim}, {Ohyama},
  {Goto} and {Matsuhara}}]{2024A&A...689A..83M}
\bibinfo{author}{{Miyaji}, T.}, \bibinfo{author}{{Bravo-Navarro}, B.A.},
  \bibinfo{author}{{D{\'\i}az Tello}, J.}, \bibinfo{author}{{Krumpe}, M.},
  \bibinfo{author}{{Herrera-Endoqui}, M.}, \bibinfo{author}{{Ikeda}, H.},
  \bibinfo{author}{{Takagi}, T.}, \bibinfo{author}{{Oi}, N.},
  \bibinfo{author}{{Shogaki}, A.}, \bibinfo{author}{{Matsuura}, S.},
  \bibinfo{author}{{Kim}, H.}, \bibinfo{author}{{Malkan}, M.A.},
  \bibinfo{author}{{Hwang}, H.S.}, \bibinfo{author}{{Kim}, T.},
  \bibinfo{author}{{Ishigaki}, T.}, \bibinfo{author}{{Hanami}, H.},
  \bibinfo{author}{{Kim}, S.J.}, \bibinfo{author}{{Ohyama}, Y.},
  \bibinfo{author}{{Goto}, T.}, \bibinfo{author}{{Matsuhara}, H.},
  \bibinfo{year}{2024}.
\newblock \bibinfo{title}{{Chandra Survey in the AKARI deep field at the North
  Ecliptic Pole: Optical and near-infrared identifications of X-ray sources}}.
\newblock \bibinfo{journal}{\aap} \bibinfo{volume}{689}, \bibinfo{pages}{A83}.
\newblock \DOIprefix\doi{10.1051/0004-6361/202450453},
  \href{http://arxiv.org/abs/2407.13864}{{\tt arXiv:2407.13864}}.
\bibitem[{Navaneeth et~al.(2025)Navaneeth, Bhatta and
  Kizhakkekalam}]{10.1093/mnras/staf1781}
\bibinfo{author}{Navaneeth, P.K.}, \bibinfo{author}{Bhatta, G.},
  \bibinfo{author}{Kizhakkekalam, S.}, \bibinfo{year}{2025}.
\newblock \bibinfo{title}{Transitional spectral behaviour in blazar oj 287: A
  comprehensive multi-epoch sed study}.
\newblock \bibinfo{journal}{Monthly Notices of the Royal Astronomical Society}
  \bibinfo{volume}{544}, \bibinfo{pages}{2455--2466}.
\newblock \URLprefix \url{https://doi.org/10.1093/mnras/staf1781},
  \DOIprefix\doi{10.1093/mnras/staf1781},
  \href{http://arxiv.org/abs/https://academic.oup.com/mnras/article-pdf/544/2/2455/64718390/staf1781.pdf}{{\tt
  arXiv:https://academic.oup.com/mnras/article-pdf/544/2/2455/64718390/staf1781.pdf}}.
\bibitem[{Padovani et~al.(1997)Padovani, Giommi and
  Fiore}]{10.1093/mnras/284.3.569}
\bibinfo{author}{Padovani, P.}, \bibinfo{author}{Giommi, P.},
  \bibinfo{author}{Fiore, F.}, \bibinfo{year}{1997}.
\newblock \bibinfo{title}{Are the x-ray spectra of flat-spectrum radio quasars
  and bl lacertae objects different?}
\newblock \bibinfo{journal}{Monthly Notices of the Royal Astronomical Society}
  \bibinfo{volume}{284}, \bibinfo{pages}{569--575}.
\newblock \DOIprefix\doi{10.1093/mnras/284.3.569}.
\bibitem[{Padovani et~al.(2019a)Padovani, Oikonomou, Petropoulou, Giommi and
  Resconi}]{2019MNRAS.484L.104P}
\bibinfo{author}{Padovani, P.}, \bibinfo{author}{Oikonomou, F.},
  \bibinfo{author}{Petropoulou, M.}, \bibinfo{author}{Giommi, P.},
  \bibinfo{author}{Resconi, E.}, \bibinfo{year}{2019}a.
\newblock \bibinfo{title}{Txs 0506+056, the first cosmic neutrino source, is
  not a bl lac}.
\newblock \bibinfo{journal}{MNRAS Letters} \bibinfo{volume}{484},
  \bibinfo{pages}{L104--L108}.
\newblock \DOIprefix\doi{10.1093/mnrasl/slz011}.
\bibitem[{Padovani et~al.(2019b)Padovani, Oikonomou, Petropoulou, Giommi and
  Resconi}]{10.1093/mnrasl/slz011}
\bibinfo{author}{Padovani, P.}, \bibinfo{author}{Oikonomou, F.},
  \bibinfo{author}{Petropoulou, M.}, \bibinfo{author}{Giommi, P.},
  \bibinfo{author}{Resconi, E.}, \bibinfo{year}{2019}b.
\newblock \bibinfo{title}{Txs 0506+056, the first cosmic neutrino source, is
  not a bl lac}.
\newblock \bibinfo{journal}{MNRAS Letters} \bibinfo{volume}{484},
  \bibinfo{pages}{L104--L108}.
\newblock \DOIprefix\doi{10.1093/mnrasl/slz011}.
\bibitem[{Pasham and Wevers(2019)}]{Pasham_2019}
\bibinfo{author}{Pasham, D.R.}, \bibinfo{author}{Wevers, T.},
  \bibinfo{year}{2019}.
\newblock \bibinfo{title}{Gaia19bsj/at2019evq: A "changing-look" quasar at
  a redshift of 1.3}.
\newblock \bibinfo{journal}{Research Notes of the AAS} \bibinfo{volume}{3},
  \bibinfo{pages}{92}.
\newblock \URLprefix \url{https://doi.org/10.3847/2515-5172/ab304a},
  \DOIprefix\doi{10.3847/2515-5172/ab304a}.
\bibitem[{Peña-Herazo et~al.(2021)Peña-Herazo, Massaro, Gu, Paggi, Landoni,
  D'Abrusco, Ricci, Masetti and Chavushyan}]{Peña-Herazo_2021}
\bibinfo{author}{Peña-Herazo, H.A.}, \bibinfo{author}{Massaro, F.},
  \bibinfo{author}{Gu, M.}, \bibinfo{author}{Paggi, A.},
  \bibinfo{author}{Landoni, M.}, \bibinfo{author}{D'Abrusco, R.},
  \bibinfo{author}{Ricci, F.}, \bibinfo{author}{Masetti, N.},
  \bibinfo{author}{Chavushyan, V.}, \bibinfo{year}{2021}.
\newblock \bibinfo{title}{An optical overview of blazars with lamost. i.
  hunting changing-look blazars and new redshift estimates}.
\newblock \bibinfo{journal}{The Astronomical Journal} \bibinfo{volume}{161},
  \bibinfo{pages}{196}.
\newblock \URLprefix \url{https://dx.doi.org/10.3847/1538-3881/abe41d},
  \DOIprefix\doi{10.3847/1538-3881/abe41d}.
\bibitem[{{Ren} et~al.(2024){Ren}, {Zhou}, {Zheng}, {Kang} and
  {Wu}}]{2024A&A...685A.140R}
\bibinfo{author}{{Ren}, S.S.}, \bibinfo{author}{{Zhou}, R.X.},
  \bibinfo{author}{{Zheng}, Y.G.}, \bibinfo{author}{{Kang}, S.J.},
  \bibinfo{author}{{Wu}, Q.}, \bibinfo{year}{2024}.
\newblock \bibinfo{title}{{The Fermi-LAT view of the changing-look blazar OQ
  334}}.
\newblock \bibinfo{journal}{\aap} \bibinfo{volume}{685}, \bibinfo{pages}{A140}.
\newblock \DOIprefix\doi{10.1051/0004-6361/202347312},
  \href{http://arxiv.org/abs/2402.17099}{{\tt arXiv:2402.17099}}.
\bibitem[{Rivers et~al.(2013)Rivers, Markowitz and Rothschild}]{Rivers_2013}
\bibinfo{author}{Rivers, E.}, \bibinfo{author}{Markowitz, A.},
  \bibinfo{author}{Rothschild, R.}, \bibinfo{year}{2013}.
\newblock \bibinfo{title}{Full spectral survey of active galactic nuclei in the
  rossi x-ray timing explorer archive}.
\newblock \bibinfo{journal}{The Astrophysical Journal} \bibinfo{volume}{772},
  \bibinfo{pages}{114}.
\newblock \URLprefix \url{https://doi.org/10.1088/0004-637X/772/2/114},
  \DOIprefix\doi{10.1088/0004-637X/772/2/114}.
\bibitem[{Tantry et~al.(2025)Tantry, Sharma, Shah, Iqbal and
  Bose}]{TANTRY2025100372}
\bibinfo{author}{Tantry, J.}, \bibinfo{author}{Sharma, A.},
  \bibinfo{author}{Shah, Z.}, \bibinfo{author}{Iqbal, N.},
  \bibinfo{author}{Bose, D.}, \bibinfo{year}{2025}.
\newblock \bibinfo{title}{Study of multi-wavelength variability, emission
  mechanism and quasi-periodic oscillation for transition blazar s5 1803+784}.
\newblock \bibinfo{journal}{Journal of High Energy Astrophysics}
  \bibinfo{volume}{47}, \bibinfo{pages}{100372}.
\newblock \URLprefix
  \url{https://www.sciencedirect.com/science/article/pii/S2214404825000539},
  \DOIprefix\doi{https://doi.org/10.1016/j.jheap.2025.100372}.
\bibitem[{Urry and Padovani(1995)}]{urry1995unified}
\bibinfo{author}{Urry, C.M.}, \bibinfo{author}{Padovani, P.},
  \bibinfo{year}{1995}.
\newblock \bibinfo{title}{Unified schemes for radio-loud active galactic
  nuclei}.
\newblock \bibinfo{journal}{Publications of the Astronomical Society of the
  Pacific} \bibinfo{volume}{107}, \bibinfo{pages}{803}.
\bibitem[{Valtonen et~al.(2022)Valtonen, Dey, Gopakumar, Zola, Komossa,
  Pursimo, Gomez, Hudec, Jermak and Berdyugin}]{galaxies10010001}
\bibinfo{author}{Valtonen, M.J.}, \bibinfo{author}{Dey, L.},
  \bibinfo{author}{Gopakumar, A.}, \bibinfo{author}{Zola, S.},
  \bibinfo{author}{Komossa, S.}, \bibinfo{author}{Pursimo, T.},
  \bibinfo{author}{Gomez, J.L.}, \bibinfo{author}{Hudec, R.},
  \bibinfo{author}{Jermak, H.}, \bibinfo{author}{Berdyugin, A.V.},
  \bibinfo{year}{2022}.
\newblock \bibinfo{title}{Promise of persistent multi-messenger astronomy with
  the blazar oj 287}.
\newblock \bibinfo{journal}{Galaxies} \bibinfo{volume}{10}.
\newblock \URLprefix \url{https://www.mdpi.com/2075-4434/10/1/1},
  \DOIprefix\doi{10.3390/galaxies10010001}.
\bibitem[{Vermeulen et~al.(1995)Vermeulen, Ogle, Tran, Browne, Cohen, Readhead,
  Taylor and Goodrich}]{Vermeulen_1995}
\bibinfo{author}{Vermeulen, R.C.}, \bibinfo{author}{Ogle, P.M.},
  \bibinfo{author}{Tran, H.D.}, \bibinfo{author}{Browne, I.W.A.},
  \bibinfo{author}{Cohen, M.H.}, \bibinfo{author}{Readhead, A.C.S.},
  \bibinfo{author}{Taylor, G.B.}, \bibinfo{author}{Goodrich, R.W.},
  \bibinfo{year}{1995}.
\newblock \bibinfo{title}{When is bl lac not a bl lac?}
\newblock \bibinfo{journal}{The Astrophysical Journal} \bibinfo{volume}{452},
  \bibinfo{pages}{L5}.
\newblock \URLprefix \url{https://doi.org/10.1086/309716},
  \DOIprefix\doi{10.1086/309716}.
\bibitem[{{Wierzcholska} and {Siejkowski}(2025)}]{2025A&A...703A.150W}
\bibinfo{author}{{Wierzcholska}, A.}, \bibinfo{author}{{Siejkowski}, H.},
  \bibinfo{year}{2025}.
\newblock \bibinfo{title}{{Disentangling two spectral components in the X-ray
  emission of the blazar 1ES 0229+200}}.
\newblock \bibinfo{journal}{\aap} \bibinfo{volume}{703}, \bibinfo{pages}{A150}.
\newblock \DOIprefix\doi{10.1051/0004-6361/202555323},
  \href{http://arxiv.org/abs/2510.15465}{{\tt arXiv:2510.15465}}.
\bibitem[{Xiao et~al.(2022)Xiao, Fan, Ouyang, Hu, Chen, Fu and
  Zhang}]{Xiao_2022}
\bibinfo{author}{Xiao, H.}, \bibinfo{author}{Fan, J.}, \bibinfo{author}{Ouyang,
  Z.}, \bibinfo{author}{Hu, L.}, \bibinfo{author}{Chen, G.},
  \bibinfo{author}{Fu, L.}, \bibinfo{author}{Zhang, S.}, \bibinfo{year}{2022}.
\newblock \bibinfo{title}{An extensive study of blazar broad emission line:
  Changing-look blazars and the baldwin effect}.
\newblock \bibinfo{journal}{The Astrophysical Journal} \bibinfo{volume}{936},
  \bibinfo{pages}{146}.
\newblock \URLprefix \url{https://dx.doi.org/10.3847/1538-4357/ac887f},
  \DOIprefix\doi{10.3847/1538-4357/ac887f}.
\bibitem[{{Yuan} and {Narayan}(2014)}]{2014ARA&A..52..529Y}
\bibinfo{author}{{Yuan}, F.}, \bibinfo{author}{{Narayan}, R.},
  \bibinfo{year}{2014}.
\newblock \bibinfo{title}{{Hot Accretion Flows Around Black Holes}}.
\newblock \bibinfo{journal}{\araa} \bibinfo{volume}{52},
  \bibinfo{pages}{529--588}.
\newblock \DOIprefix\doi{10.1146/annurev-astro-082812-141003},
  \href{http://arxiv.org/abs/1401.0586}{{\tt arXiv:1401.0586}}.
\bibitem[{{Yuan} and {Fan}(2014)}]{2014Ap&SS.352..207Y}
\bibinfo{author}{{Yuan}, Y.}, \bibinfo{author}{{Fan}, J.},
  \bibinfo{year}{2014}.
\newblock \bibinfo{title}{{X-ray spectral indices of the Fermi/LAT blazars}}.
\newblock \bibinfo{journal}{\apss} \bibinfo{volume}{352},
  \bibinfo{pages}{207--214}.
\newblock \DOIprefix\doi{10.1007/s10509-014-1878-y}.
\bibitem[{Zhang et~al.(2025)Zhang, Kokubo, Doi, Hagio, Seki, Takahashi, Murata,
  Matsubayashi, Isogai, Kawabata, Sasada, Niwano, Hashizume, Shidatsu, Higuchi,
  Imazawa, Joshima, Sako, Hayatsu, Yatsu, Iwakiri and Kubo}]{Zhang_2025}
\bibinfo{author}{Zhang, T.}, \bibinfo{author}{Kokubo, M.},
  \bibinfo{author}{Doi, M.}, \bibinfo{author}{Hagio, H.},
  \bibinfo{author}{Seki, H.}, \bibinfo{author}{Takahashi, I.},
  \bibinfo{author}{Murata, K.L.}, \bibinfo{author}{Matsubayashi, K.},
  \bibinfo{author}{Isogai, K.}, \bibinfo{author}{Kawabata, K.},
  \bibinfo{author}{Sasada, M.}, \bibinfo{author}{Niwano, M.},
  \bibinfo{author}{Hashizume, M.}, \bibinfo{author}{Shidatsu, M.},
  \bibinfo{author}{Higuchi, N.}, \bibinfo{author}{Imazawa, R.},
  \bibinfo{author}{Joshima, S.}, \bibinfo{author}{Sako, S.},
  \bibinfo{author}{Hayatsu, S.}, \bibinfo{author}{Yatsu, Y.},
  \bibinfo{author}{Iwakiri, W.}, \bibinfo{author}{Kubo, Y.},
  \bibinfo{year}{2025}.
\newblock \bibinfo{title}{Multiband optical photometric and spectroscopic
  monitoring of the 2024 flare event in transition blazar op313}.
\newblock \bibinfo{journal}{The Astrophysical Journal} \bibinfo{volume}{994},
  \bibinfo{pages}{99}.
\newblock \URLprefix \url{https://doi.org/10.3847/1538-4357/ae0c95},
  \DOIprefix\doi{10.3847/1538-4357/ae0c95}.
\bibitem[{Zhang et~al.(2021)Zhang, Gupta, Gaur, Wiita, An, Lu, Fan and
  Xu}]{Zhang_2021}
\bibinfo{author}{Zhang, Z.}, \bibinfo{author}{Gupta, A.C.},
  \bibinfo{author}{Gaur, H.}, \bibinfo{author}{Wiita, P.J.},
  \bibinfo{author}{An, T.}, \bibinfo{author}{Lu, Y.}, \bibinfo{author}{Fan,
  S.}, \bibinfo{author}{Xu, H.}, \bibinfo{year}{2021}.
\newblock \bibinfo{title}{X-ray intraday variability of the tev blazar pks
  2155–304 with suzaku during 2005–2014}.
\newblock \bibinfo{journal}{The Astrophysical Journal} \bibinfo{volume}{909},
  \bibinfo{pages}{103}.
\newblock \URLprefix \url{https://dx.doi.org/10.3847/1538-4357/abdd38},
  \DOIprefix\doi{10.3847/1538-4357/abdd38}.
\bibitem[{Zhao et~al.(2026)Zhao, Zhou, Zheng and Kang}]{10.1093/mnras/stag542}
\bibinfo{author}{Zhao, J.W.}, \bibinfo{author}{Zhou, R.X.},
  \bibinfo{author}{Zheng, Y.G.}, \bibinfo{author}{Kang, S.J.},
  \bibinfo{year}{2026}.
\newblock \bibinfo{title}{The gamma-ray photon spectral index view of
  transition blazars}.
\newblock \bibinfo{journal}{Monthly Notices of the Royal Astronomical Society}
  , \bibinfo{pages}{stag542}\URLprefix
  \url{https://doi.org/10.1093/mnras/stag542},
  \DOIprefix\doi{10.1093/mnras/stag542},
  \href{http://arxiv.org/abs/https://academic.oup.com/mnras/advance-article-pdf/doi/10.1093/mnras/stag542/67654577/stag542.pdf}{{\tt
  arXiv:https://academic.oup.com/mnras/advance-article-pdf/doi/10.1093/mnras/stag542/67654577/stag542.pdf}}.

\end{thebibliography}

\end{document}